\documentclass[twocolumn,secnumarabic,amssymb, nobibnotes, superscriptaddress, nofootinbib, aps, prd]{revtex4-2}
\usepackage{natbib}
\usepackage{xcolor}

\usepackage{hyperref}     
\usepackage{amsmath}
\usepackage{adjustbox}
\usepackage{graphicx}   
\usepackage{amsfonts}      
\usepackage{nicefrac}    
\usepackage{microtype}
\usepackage{cleveref}
\usepackage{colortbl}
\usepackage{makecell}
\usepackage{dsfont}
\usepackage{array}
\usepackage{orcidlink}
\usepackage{slashed}
\usepackage{lipsum}

\newcommand{\vect}[1]{\boldsymbol{#1}}

\definecolor{Gray}{gray}{0.85}
\definecolor{LightCyan}{rgb}{0.88,1,1}

\newcolumntype{a}{>{\columncolor{Gray}}c}
\newcolumntype{b}{c}

\begin{document}

\title{Astrophysical constraints on color-superconducting phases in compact stars within the RG-consistent NJL model}
 
\author{Hosein Gholami~\orcidlink{0009-0003-3194-926X}}
\email{mohammadhossein.gholami@tu-darmstadt.de}
\affiliation{Technische Universit\"{a}t Darmstadt, Fachbereich Physik, Institut f\"{u}r Kernphysik, Theoriezentrum, Schlossgartenstr.~2, D-64289 Darmstadt, Germany}
\author{Ishfaq Ahmad Rather~\orcidlink{0000-0001-5930-7179}}
\email{rather@astro.uni-frankfurt.de}
\affiliation{Institut f\"{u}r Theoretische Physik, Goethe Universit\"{a}t, 
Max-von-Laue-Str.~1, D-60438 Frankfurt am Main, Germany}
\author{Marco~Hofmann~\orcidlink{0000-0002-4947-1693}}
\email{marco.hofmann@tu-darmstadt.de}
\affiliation{Technische Universit\"{a}t Darmstadt, Fachbereich Physik, Institut f\"{u}r Kernphysik, Theoriezentrum, Schlossgartenstr.~2, D-64289 Darmstadt, Germany}
\author{Michael Buballa~\orcidlink{0000-0003-3747-6865}}
\email{michael.buballa@tu-darmstadt.de}
\affiliation{Technische Universit\"{a}t Darmstadt, Fachbereich Physik, Institut f\"{u}r Kernphysik, Theoriezentrum, Schlossgartenstr.~2, D-64289 Darmstadt, Germany}
\affiliation{Helmholtz Forschungsakademie Hessen f\"{u}r FAIR (HFHF), 
	GSI Helmholtzzentrum f\"{u}r Schwerionenforschung,
	Campus Darmstadt,
	D-64289 Darmstadt,
	Germany}
\author{Jürgen Schaffner-Bielich~\orcidlink{0000-0002-0079-6841}}
\email{schaffner@astro.uni-frankfurt.de}
\affiliation{Institut f\"{u}r Theoretische Physik, Goethe Universit\"{a}t, 
Max-von-Laue-Str.~1, D-60438 Frankfurt am Main, Germany}

\begin{abstract}

We determine parameters of the renormalization group-consistent three-flavor color-superconducting Nambu–Jona-Lasinio (NJL) model that are suited to investigate possible compact-star configurations. 
Our goal is to provide quark-matter equations of state (EoS) that can be used for hadron-quark hybrid-star constructions. To that end, we mainly focus on the parameters of the quark-matter model.
By varying the vector and diquark coupling constants, we analyze their impact on the 
EoS, the speed of sound, the maximum diquark gap, and the mass-radius relation.
In almost all configurations, a stable color-flavor-locked (CFL) phase appears in the core of the maximum-mass configurations, typically spanning several kilometers in radius. In other cases, the star's two-flavor color-superconducting (2SC) branch of the EoS becomes unstable before reaching the CFL transition density. At neutron-star densities, the speed of sound squared reaches up to $c_s^2 \sim 0.6$ and the CFL gap up to $\Delta\sim250\,$\text{MeV}.
We argue that adding a hadronic EoS at lower densities by performing a Maxwell construction does not increase the maximum mass substantially. Thus we use the $2.0\,M_{\odot}$ constraint to constrain the NJL model parameters that are suited for the construction of hybrid-star EoS.
We construct three examples of the hybrid-star model, demonstrating that there is room for different color-superconducting compositions. The hybrid EoSs obtained in this way can have no 2SC matter or different ratios of 2SC and CFL quark matter in the core. We show that early hadron-quark transitions are possible that can modify the tidal deformability at 1.4\,$M_\odot$. We find that these EoSs are consistent with the imposed constraints from astrophysics and perturbative QCD. They allow for different hybrid-star scenarios with a hadronic EoS that is soft at low to intermediate densities ($\sim 1-3\, n_{\text{sat}}$).

\end{abstract}

\maketitle

\section{Introduction}

Understanding the theory of strong interactions under extreme conditions is a major goal of modern nuclear physics. Neutron stars (NS) provide a unique environment for studying strong-interacting matter at densities of several times nuclear saturation density. Their internal structure, stability, and observable properties, like mass and radius, are governed by the equation of state (EoS) of dense matter, which is in large parts still unknown.

Constraints on the EoS of strong interacting matter come from a variety of sources. 
The measurements of several pulsars with masses of around 2.0\,$M_\odot$ from radio observations of binaries \cite{Demorest:2010bx,Antoniadis:2013pzd,Fonseca:2021wxt} 
necessitates a strongly repulsive interaction so that the neutron star can withstand the strong attraction of gravity.
With the detection of gravitational waves (GWs) from a binary neutron star merger, GW170817 \cite{PhysRevLett.119.161101, LIGOScientific:2017ync, PhysRevLett.121.161101}, 
a new observational window opened for studying the properties of the interior of neutron stars. 
The constraint on the tidal deformability from GW170817 significantly constrained the EoS, ruling out models that predict a highly incompressible or stiff equation of state 
\cite{PhysRevC.81.015803, PhysRevC.55.540}. Further constraints on the EoS from the measurement of the mass and radii from X-ray observations of neutron stars 
by the NICER collaboration \cite{Riley:2021pdl,miller2021,Salmi:2024aum,dittmann2024precisemeasurementradiuspsr,Choudhury:2024xbk} 
provide further constraints on the properties of high-density matter in the interior of neutron stars.

In addition to astrophysical observations, constraints on the EoS in the density regime relevant for neutron stars come from calculations of perturbative QCD (pQCD) at high baryon chemical potentials \cite{Komoltsev:2021jzg,Somasundaram:2022ztm,Gorda:2023usm,Komoltsev:2023zor,Gorda:2022jvk,PhysRevLett.127.162003} and chiral effective field theory at low densities \cite{Hebeler:2013nza,Drischler:2016djf}, combined with the requirement that the equation of state must be causal \cite{Annala_2020,Altiparmak:2022bke,Semposki:2024vnp}.  Additionally, Bayesian inference methods have been employed to explore 
the properties of hybrid stars, i.e., stars that have a core of deconfined quark matter surrounded by hadronic matter,
by integrating hadronic and quark matter models, imposing both pQCD constraints and observational data from NICER and gravitational wave events \cite{albino2024hybridstarpropertiesnjl}.

All these advancements open the tantalizing perspective to probe the high-density behavior of QCD matter under the conditions prevailing in the core of neutron stars 
with an unprecedented distinctiveness.
At very high densities, QCD matter is expected to be in a color-superconducting (CSC) state \cite{Son:1998uk,Schafer:1999jg,Pisarski:1999tv}, in which quarks form Cooper pairs (``diquark condensates''). For a review on color superconductivity, see Ref. \cite{Alford:2007xm}. At high densities, the color-flavor locked (CFL)  phase \cite{Alford:1998mk}, in which all three colors (red, green, blue) and all three flavors (up, down, strange) are paired is the energetically favored phase~\cite{Schafer:1999fe,Shovkovy:1999mr,Alford:1998mk}. At lower densities, where pairing with strange quarks is not yet preferred, a two-flavor superconducting phase (2SC) phase, where only up and down quarks are paired, 
could be present \cite{Alford:1999pa} although this is still under debate \cite{Alford:2002kj,Yuan:2023dxl}. 

If the hadron-quark phase transition takes place at the densities reached in neutron stars, there will be hybrid stars.
The case for hybrid stars has been investigated extensively using different models and parametrizations for the hadronic part and the quark-matter part of the EoS (see e.g. \cite{Alford:2004pf,Alford:2006vz,Alford:2013aca,Annala_2020,Contrera:2022tqh,Annala:2023cwx,Christian:2023hez}). Meanwhile, the nature of the phase transition is unknown. It is either modeled as a first-order phase transition, e.g., via a Maxwell or Gibbs construction \cite{Glendenning:1992vb}, or as a smooth transition or crossover,
constructed by interpolating a low-density hadronic EoS and a high-density quark EoS (see e.g. Refs. \cite{Baym:2019iky,Huang:2022jiu}). To this date, hybrid stars and neutron stars without quark matter are both consistent with constraints from astrophysics and inferred constraints from chiral effective field theory at low densities and perturbative QCD at large densities \cite{PhysRevD.99.023009, Tews:2018kmu, Annala_2020,Annala:2023cwx,Rutherford:2024srk}.

In this work, we focus on the possibility that the quark core of a hybrid star is in a CSC state.
The densities of the outer and inner cores of neutron stars lie in the strong coupling regime of the strong interactions, where perturbative QCD does not converge.  
Moreover, lattice QCD, which is the most powerful non-perturbative method at vanishing baryon density, is plagued by the sign problem in the low-temperature high-density regime and therefore is not applicable either. Non-perturbative continuum approaches to QCD, such as Dyson-Schwinger Equations \cite{Fischer:2018sdj} or the Functional Renormalization Group (FRG) \cite{Fu:2019hdw}, are promising candidates to fill this gap. 
Dyson-Schwinger studies of various CSC phases have been performed in Refs.~\cite{Nickel:2006vf,Nickel:2006kc,Marhauser:2006hy,Nickel:2008ef,Muller:2013pya,Muller:2016fdr}.
The pressure and the speed of sound of two-flavor color superconductivity can be estimated with FRG calculations \cite{Braun:2021uua,Braun:2022jme,Braun:2022olp}, including the strong coupling to next-to-leading order at high densities \cite{Geissel:2024nmx}.
These approaches have not yet reached the degree of sophistication needed to predict the EoS of CSC quark matter under hybrid-star conditions.
In this situation, one has to rely on QCD-inspired models to get insights.
The three-flavor Nambu-Jona Lasinio (NJL) model allows to study the effect of dynamically generated quark masses due to chiral symmetry breaking on the CSC pairing structure. 

Hybrid stars with quark cores described within the three-flavor NJL model have been studied first in Ref.~\cite{PhysRevC.60.025801} without quark pairing and in Refs.~\cite{Baldo:2002ju} and \cite{Buballa:2003et} including CSC phases. Matching the quark-matter EoS to several hadronic EoSs it was found that in most cases the transition to quark matter renders the star gravitationally unstable so that either hybrid stars do not exist or collapse well below reaching the mass of 2\,$M_{\odot}$. 
This problem was solved in Ref.~\cite{Klahn:2006iw}, where a repulsive vector interaction was added to the NJL model. 
It was shown that the vector interaction stiffens the equation of state enough to reach hybrid stars with a mass of 2\,$M_{\odot}$ and more, however, no stable hybrid stars with a CFL core were found.

In Ref.~\cite{Pagliara:2007ph}, hybrid stars with stable quark matter cores containing 2SC, CFL, or 2SC and CFL matter were constructed with the NJL model for the quark phase and a relativistic mean-field model for the hadronic phase. Later work focused on confronting the equation of state of these hybrid models with astrophysical data in order to constrain the values of the diquark coupling and the vector coupling \cite{Bonanno:2011ch,Klahn:2013kga,Baym:2017whm,Alaverdyan:2020xnv,Alaverdyan:2022foz, PhysRevC.110.045802} (see Ref. \cite{Tanimoto:2019tsl} for studies with the two-flavor NJL model), showing that a stable, color-superconducting core is consistent with astrophysical data. In addition, the case for color-superconducting hybrid stars was investigated with a variety of different models for the quark phase, such as non-local NJL models \cite{Shahrbaf:2021cjz,Blaschke:2022egm,Ivanytskyi:2024zip} or parameterized quark matter models with strong perturbative corrections \cite{Zhang:2020jmb}. A recent work \cite{christian2024orderphasetransitionsquark} has explored first-order phase transitions to quark matter, highlighting their impact on the mass-radius relation and the potential observation of twin stars, and emphasizing that additional observational data such as tidal deformability measurements are crucial to confirm the presence of such transitions. The possibility of twin and triplet stars and for a third and forth family of compact stars due to sequential first-order phase transitions in a hybrid star with different quark matter phases was studied with the synthetic constant speed of sound parametrization \cite{Alford:2017qgh,Li:2019fqe,Li:2023zty}. In Ref.~\cite{Ranea-Sandoval:2017ort}, color-superconducting twin stars with an NJL-model for the quark matter EoS were constructed. In that work, a 2SC+s pairing, but no CFL pairing was taken into account. Unless a very stiff hadronic EoS was used, the authors did not find twin branches that reached $2\,M_{\odot}$.

A well-known shortcoming of the NJL model is the fact that it is non-renormalizable. As a consequence, the regulators that have to be introduced to render loop integrals finite cannot be removed by absorbing them in existing model parameters but have to be kept finite. Typically, a momentum cutoff of the order of 600~MeV is employed, which results from a fit to vacuum observables \cite{Rehberg:1995kh}. Unfortunately, this is not much larger than the quark chemical potential possibly encountered in the cores of hybrid stars. Indeed, severe cutoff artifacts are affecting the phase structure and the size of the CSC gaps in this regime \cite{Gholami:2024diy}.
To overcome these problems, a regularization based on the concept of renormalization group (RG) consistency \cite{Braun:2018svj} was formulated and applied for the three-flavor NJL model with CSC in Ref. \cite{Gholami:2024diy}.

In this work, we perform a systematic survey of the CSC quark-matter EoSs obtained within the three-flavor NJL model with RG-consistent regularization. 
To this end, we extend the model of Ref.~\cite{Gholami:2024diy} with a repulsive vector interaction and vary the value of the diquark and vector couplings.
The goal is to provide viable quark-matter EoSs that can be used for the construction of hybrid stars. 
For this, we will explore the properties of pure quark stars while leaving the actual hybrid construction for future work. 
The underlying rationale is that the maximum mass of a hybrid star is mainly determined by the high-density quark matter part of the equation of state. This means that, if the quark matter EoS does not reach a certain maximum mass, a hybrid construction with that EoS at large densities and a hadronic EoS at lower densities will hardly reach a higher mass.
In this way, we can constrain the NJL model parameters already without referring to a specific hadronic EoS by requiring that the pure quark star reaches a maximum mass of not much less than 2\,$M_\odot$. 
Another interesting result of this analysis which is independent of the hybrid construction is how the phase realized in the centers of the maximum mass stars, 2SC or CFL, depends on the parameters.
We also calculate the radii of the quark stars. Although the correspondence to the radii of hybrid stars is less clear, this can give us insights into the expected properties of the quark cores.

This work is organized as follows: In section Sec.~\ref{sec:model} we present the RG-consistent NJL model for CSC quark matter, in Sec.~\ref{sec:var} we investigate the effect of variations in the diquark coupling and the vector coupling on the diquark condensates, the EoS and the speed of sound. We also solve the Tolman-Oppenheimer-Volkoff (TOV) equation for our EoSs and the calculate mass-radius relations of quark stars and the radial profiles of the corresponding maximum-mass stars.
In Sec.~\ref{sec:parameterstudy} we constrain the 
vector and diquark coupling constants
by requiring configurations with a maximum mass of 2.0\,$M_{\odot}$ or more and provide further constraints from astrophysical observations that may become relevant depending on the density at which the hadron-quark phase transition takes place. At the end of Sec.~\ref{sec:parameterstudy}
we discuss the stability of the stars under 2SC-CFL phase transitions. In Sec.~\ref{sec:hybrid} we test the validity of the 2.0\,$M_{\odot}$ constraints for a hybrid star construction with a chosen hadronic EoS based on a relativistic mean field model serving as a proof-of-principle.  
Furthermore, we provide tabulated EoSs of the quark-matter model for three combinations of coupling constants representing cases with different mass-radius profiles. The resulting hybrid equations of state with the relativistic mean field model  
have different quark core compositions while fulfilling the present astrophysical constraints. Finally, we give a summary and draw a conclusion in Sec.~\ref{sec:conclusion}.


\section{The RG-consistent NJL model}
\label{sec:model}


In this work, we model quark matter within a Nambu-Jona Lasinio (NJL) model with a diquark interaction, allowing for the formation of color-superconducting condensates. We hereby follow the recent advancement for an RG-consistent description of the NJL model, see Ref. \cite{Gholami:2024diy},
which removes artifacts of the conventional regularization and allows for a consistent investigation of the phase structure at high chemical potentials.
The Lagrangian of our model is written as
\begin{equation}
\mathcal{L}=\mathcal{L}_0 + \mathcal{L}_{\bar q q}+\mathcal{L}_{qq}+\mathcal{L}_{\text{L}}.
\label{eq:Lagrangian}
\end{equation}
Here, 
\begin{equation}
\label{eq:L0}
\mathcal{L}_0 = \bar{\psi}(i\slashed{\partial}+\gamma^0\hat{\mu}-\hat{m})\psi 
\end{equation}
denotes the kinetic Lagrangian for the quark fields $\psi_\alpha^a$ with flavors $\alpha=u,d,s$ and colors $a=r,g,b$. 
The diagonal matrix in flavor space $\hat{m}=\text{diag}_f(m_u,m_d,m_s)$ contains the bare quark masses of the different flavors, 
and the chemical potential matrix
\begin{equation}\label{eq:mu_matrix}
    \hat{\mu}^{\alpha\beta}_{ab}=(\mu\delta^{\alpha\beta}+\mu_Q Q^{\alpha\beta})\delta_{ab}+[\mu_3 (\lambda_3)_{ab}+ \mu_8 (\lambda_8)_{ab}]\delta^{\alpha\beta}
\end{equation}
includes the electric charge operator $Q=\text{diag}_f(2/3,-1/3,-1/3)$ and the third and eighth Gell-Mann matrices in color space, $\lambda_3$ and $\lambda_8$.
The term 
\begin{eqnarray}
\mathcal{L}_{\bar q q} &=&
G_S\sum_{a=0}^8\left[(\bar{\psi}\tau_a\psi)^2 + (\bar{\psi} i \gamma_5 \tau_a \psi)^2\right]
\nonumber\\&&-K [\text{det}_{\text{f}}(\bar{\psi}(\mathds{1}+\gamma_5)\psi) + \text{det}_{\text{f}}(\bar{\psi}(\mathds{1}-\gamma_5)\psi)] \nonumber\\
&&-G_V\left(\bar\psi\gamma^\mu\psi\right)^2
\label{eq:Lagrangian_qbarq}
\end{eqnarray}
includes the $U(3)_L \times U(3)_R$ - symmetric scalar and pseudoscalar four-point interactions with the coupling constant $G_S$ and the $U_A(1)$ breaking six-point interaction with the coupling constant $K$ \cite{Kobayashi:1970ji,tHooft:1976rip}. 
The matrices $\tau_a$ denote the Gell-Mann matrices in flavor space for $a=1,..,8$, complemented by $\tau_0=\sqrt{2/3}\mathds{1}_f$. In addition to the terms present in Ref. \cite{Gholami:2024diy}, we include a repulsive four-point vector interaction
\begin{equation}
 \mathcal{L}_{V}=-G_V\left(\bar\psi\gamma^{\mu}\psi\right)^2
 \end{equation}
with the vector coupling $G_V$. The diquark interaction 
\begin{equation}
\label{eq:Lqq}
\mathcal{L}_{qq}= G_D\sum_{\gamma,c}(\bar{\psi}^a_\alpha i\gamma_5 \epsilon^{\alpha\beta\gamma} \epsilon_{abc}(\psi_C)^b_\beta)
    ((\bar{\psi}_C)^r_\rho i\gamma_5 \epsilon^{\rho\sigma\gamma} \epsilon_{rsc} \psi^s_\sigma)
\end{equation}
enables quark pairing in the spin-zero color and flavor anti-triplet channel in the Hartree approximation. The charge-conjugate spinors are defined as 
$\psi_C=C\bar{\psi}^T$ and $\bar{\psi}_C=\psi^TC$
with the charge conjugation operator $C=i\gamma^2\gamma^0$. Finally, we include kinetic terms for (free) electrons and muons
\begin{equation*}
    \mathcal{L}_L=\sum_{L=e,\mu} \bar{\psi}_L (i \slashed{\partial}-m_L)\psi_L
\end{equation*}
with the electron mass $m_e=0.511\,$MeV and the muon mass $m_\mu=105.66\,$MeV, respectively.

In the mean-field approximation, we consider only the chiral condensates
\begin{equation}
\phi_f=\langle \bar{\psi}_f \psi_f \rangle
\end{equation}
for flavors $f = u, d, s$, the three diquark condensates
\begin{equation}
\label{eq:DeltaA}
\Delta_A=-2G_D\langle \bar{\psi}^a_\alpha i\gamma_5 \epsilon^{\alpha\beta A} \epsilon_{abA}(\psi_C)^b_\beta \rangle
\end{equation}
with $A = 1,2,3$ and the quark number density
\begin{equation}
n=\langle\bar\psi\gamma^0\psi\rangle
\end{equation}
to be nonzero. We do not consider nonzero vector condensates of the form $\langle \bar{\psi} \gamma^i \psi \rangle$, $i=1,2,3$, which would break the rotational invariance of the system.

The chiral condensates give rise to the dynamical ``constituent'' quark masses
\begin{equation}
M_\alpha = m_\alpha -4G_S \phi_\alpha + 2K\phi_\beta \phi_\gamma,
\end{equation}
where $(\alpha,\beta,\gamma)$ is any permutation of $(u,d,s)$.
Similarly, the vector interaction effectively shifts the chemical potential term in the kinetic Lagrangian to 
\begin{equation}
\tilde{\mu}=\mu-2G_V n.
\end{equation}
The 2SC phase is characterized by pairing between the red and green up and down quarks only ($\Delta_1=\Delta_2=0$, $\Delta_3 \neq 0$) whereas in the CFL phase, quarks of all flavors and colors participate in the pairing ($\Delta_1,\Delta_2,\Delta_3\neq0$).
For notational convenience, we combine the chiral condensates and diquark condensates to a vector $\vect\chi=(\phi_u,\phi_d,\phi_s,\Delta_1,\Delta_2,\Delta_3)$ and the constituent quark masses to the vector $\vect M = (M_u,M_d,M_s)$. We also combine the independent chemical potentials of Eq.~\eqref{eq:mu_matrix} to the vector $\vect{\mu} =(\mu,\mu_Q,\mu_3,\mu_8)$.

The NJL model is most conveniently regularized with a sharp three-momentum cutoff $\Lambda'$. 
Due to the non-renormalizability of the model, the cutoff cannot be absorbed into renormalized quantities.
Instead, the results depend on the value of the cutoff. 
As mentioned in the Introduction and
highlighted in Ref.~\cite{Gholami:2024diy} the conventional cutoff regularization leads to unphysical artifacts as soon as the model is evaluated at scales of the order of the cutoff. 
To avoid these artifacts, a regularization based on the concept of renormalization-group consistency \cite{Braun:2018svj}, adapted for a three-flavor NJL model with color superconductivity was proposed in Ref.~\cite{Gholami:2024diy}. 

Renormalization group consistency is most straightforwardly formulated in the context of the functional renormalization group. 
It states, that the scale $\Lambda$ at which the flow equation is initialized, must be chosen large enough such that the effective action of the model does not depend on this scale in the infrared \cite{Braun:2018svj}. 
In the context of a non-renormalizable model in the mean-field approach, this large initial scale corresponds to a large momentum cutoff. For a detailed explanation of the significance of RG-consistency for mean-field models, we refer to Refs.~\cite{Braun:2018svj} and \cite{Gholami:2024diy}.
In short, the renormalization-group consistent regularization employs a large cutoff scale $\Lambda \gg \Lambda'$, while the divergent vacuum contribution is still regularized with the conventional cutoff $\Lambda'$. 
In the presence of diquark condensates, additional medium-dependent divergences arise, which must be balanced by suitable counterterms leading to physical results independent of the chosen large cutoff scale.

Following this procedure,
the mean-field effective potential per volume in the RG-consistent regularization can be written as
\begin{alignat}{1}\label{eq:Omega_eff}
\Omega_{\text{eff}}(\vect\mu,T&,\vect\chi,\tilde{\mu}) = \mathcal{V}(\vect\chi,\tilde{\mu}) 
\nonumber\\ 
- \frac{1}{2\pi^2} \bigg[& \int_0^{\Lambda} dp\, p^2 \mathcal{A}(\vect p;\vect{\mu},T,\vect\chi,\tilde{\mu}) 
- \int_{\Lambda'}^{\Lambda} dp\, p^2 \mathcal{A}_{\text{vac}}(\vect p;\vect\chi) \nonumber\\
&- \int_{\Lambda'}^{\Lambda} dp\, p^2 \sum \frac{1}{2}\mu_{\alpha a,\beta b}^2\nonumber\\
&\times  
\left(\frac{\partial^2}{\partial \mu_{\alpha a,\beta b}^2} \mathcal{A}(\vect p;\vect\mu,0,\vect\chi)\right) \bigg|_{\vect\mu=\tilde{\mu}=\textbf{M}=0;\Delta_{\alpha a,\beta b}\neq 0} \bigg] \nonumber\\[1mm]
+\, \Omega_L(&\mu_e,T).
\end{alignat}

 The first term on the right-hand side, 
\begin{eqnarray}
    \label{eq:V_def}
    \mathcal{V}(\vect\chi,\tilde{\mu})&=&2G_S(\phi_u^2+\phi_d^2+\phi_s^2)-4K\phi_u\phi_d\phi_s
    \nonumber\\&&+\frac{1}{4G_D}\sum_{A=1}^3\vert\Delta_A\vert^2 -\frac{(\mu - \tilde{\mu})^2}{4G_V},
\end{eqnarray}
is the potential for the condensates. 
The function
\begin{equation}
\mathcal{A}(\vect p;\vect{\mu},T,\vect\chi,\tilde{\mu})=
\frac{1}{2}T\sum\limits_n \text{tr}\ln\frac{S^{-1}(i\omega_n,\vect{p};\vect{\mu},\vect\chi,\tilde{\mu})}{T}
\label{eq:Adef}
\end{equation}
depends on the inverse Nambu-Gorkov propagator $S^{-1}$ which in turn depends on the chemical potentials and condensates, and $A_\text{vac}$ denotes the same quantity in a vacuum, i.e., at $T=\vect\mu=0$: $\mathcal{A}_\text{vac}(\vect p;\vect\chi)=\mathcal{A}(\vect p;\vect{\mu}=0,T=0,\vect\chi,\tilde{\mu}=0)$. The trace in Eq. \eqref{eq:Adef} is evaluated over Nambu-Gorkov, flavor, color, and Dirac space, and the sum runs over fermionic Matsubara frequencies $\omega_n=(2n+1)\pi T$.

The three momentum integrals in the parentheses of Eq.~\eqref{eq:Omega_eff} correspond to the kinetic term integrated up to the large cutoff $\Lambda$, the renormalization group flow contribution in the vacuum between $\Lambda$ and $\Lambda'$, and a counterterm, respectively.
The renormalization group contribution and the counterterm ensure that the model is RG-consistent. For $\Lambda=\Lambda'$, both terms vanish and 
$\Omega_{\text{eff}}$ reduces to the effective potential in the conventional cutoff regularization.
The first term in the parentheses in Eq.~\eqref{eq:Omega_eff} includes contributions that scale as $\sim\sum \mu_{\alpha a,\beta b}^2\Delta_{\alpha a,\beta b}^2 \ln(\Lambda)$, where $\Delta_{\alpha a,\beta b}$ denotes the gap for pairing quarks of flavor $\alpha$ and color $a$ with flavor $\beta$ and color $b$, and 
$\mu_{\alpha a,\beta b}=(\mu_{\alpha a}+\mu_{\beta b})/2$ is the average chemical potential of the paired quarks. These logarithmic divergences arise from nonzero diquark condensates and are balanced by the counterterm. 
The counterterm includes second derivatives with respect to the average chemical potentials. The derivatives are evaluated in vacuum ($T=\vect\mu=\tilde{\mu}=0$), and $\Delta_{\alpha a,\beta b}\neq0$ denotes that all \textit{other} diquark gaps that are not associated with the chemical potential $\mu_{\alpha a,\beta b}$ are set to zero. 
Furthermore, all quark masses are set to zero in the counterterm. This corresponds to the \textit{massless} renormalization scheme of Ref. \cite{Gholami:2024diy}. For further details about the effective potential in the renormalization group-consistent regularization, we refer to Ref. \cite{Gholami:2024diy}. 

The leptonic contribution to the effective potential is modeled as a free relativistic gas
\begin{eqnarray} &&\Omega_L(\mu_e,T)=\nonumber\\
&&-2T\sum_{l=e,\mu}\int \frac{d^3 p}{(2\pi^3)}\left(
    \ln(1+e^{-\frac{E_l-\mu_l}{T}})+ \ln(1+e^{-\frac{E_l+\mu_l}{T}})
    \right),\nonumber\\
\end{eqnarray}
with the energy $E_l=\sqrt{\vect{p}^2+m_l^2}$.
In this work, we assume that the system is in equilibrium with respect to the weak interactions and that neutrinos can leave the system freely. 
Then, the electron and muon chemical potentials are given by the (negative) charge chemical potential $\mu_e=\mu_{\mu}=-\mu_Q$.

The physical values for the condensates are found as the stationary points of the effective potential
\begin{align}\label{eq:gapeqs}
    \frac{\partial \Omega_\text{eff}}{\partial \phi_f}\bigg|_{\phi_f=\bar{\phi}_f}=\frac{\partial \Omega_\text{eff}}{\partial \Delta_A}\bigg|_{\Delta_A=\bar{\Delta}_A}=\frac{\partial \Omega_\text{eff}}{\partial \tilde{\mu}}\bigg|_{\tilde{\mu}=\bar{\tilde{\mu}}}=0.
\end{align}
The last equation sets $\tilde{\mu}$ to be thermodynamic consistent with the definition of the number density 
$n=-\frac{\partial \Omega_{\text{eff}}}{\partial \mu}$, see also Ref. \cite{BUBALLA2005205}. 
The charge and color chemical potentials are fixed by requiring local electric and color neutrality
\begin{equation}\label{eq:neutrality}
    \frac{\partial \Omega_\text{eff}}{\partial \mu_Q}\bigg|_{\mu_Q=\bar{\mu}_Q}=\frac{\partial \Omega_\text{eff}}{\partial \mu_3}\bigg|_{\mu_3=\bar{\mu}_3}=\frac{\partial \Omega_\text{eff}}{\partial \mu_8}\bigg|_{\mu_8=\bar{\mu}_8}=0.
\end{equation}
Quantities with a bar denote the physical values of the condensates and chemical potentials. 
If for given $\mu$ and $T$, the gap equations \eqref{eq:gapeqs} and the neutrality constraints, Eq.~\eqref{eq:neutrality}, have more than one solution, the physical solution is the one with the lowest effective potential. Plugging this solution into the effective potential \eqref{eq:Omega_eff} gives the grand canonical potential per volume 
\begin{alignat}{1}
&\Omega(\mu,T)=\nonumber\\
&\Omega_{\text{eff}}(\mu,\mu_Q=\bar{\mu}_8,\mu_3=\bar{\mu}_3,\mu_8=  \bar{\mu}_8,T,\vect\chi= \bar{\vect\chi},\tilde{\mu}=\bar{\tilde{\mu}}).
\end{alignat}
The pressure is then obtained as
\begin{eqnarray}
    P&=&-(\Omega-\Omega_0),\label{eqvacbag}
\end{eqnarray}
where we fix the pressure in the vacuum to be zero by subtracting $\Omega_0=\Omega(\mu=T=0)$. As the NJL model lacks the description of a confined phase at low densities, it is valid to fix the pressure in the vacuum to a different value by subtracting a different "bag pressure", and setting the pressure to zero in the vacuum must be interpreted as a special choice \cite{Bonanno:2011ch}. We discuss the implications of changing the bag parameter on some of our results in Sec.~\ref{sec:cflstability}.

Other thermodynamic quantities which can be calculated from the grand potential per volume are the quark number density $n=-\partial \Omega / \partial \mu \vert_T$, the entropy density $s=\partial \Omega/ \partial T \vert_{\mu}$, the energy density $\epsilon=-P+\mu n + sT$ and the speed of sound squared 
\begin{align}
    c_s^2=\frac{\partial P}{\partial \epsilon} \Big\vert_{\frac{s}{n}}.
\end{align}

For numerical calculations, we set $\Lambda'=602.3\,$MeV, $G_S\Lambda'^2=1.835$, $K\Lambda'^5=12.36$ and the bare quark masses $m_{u,d}=5.5$\,MeV and $m_s=140.7$\,MeV. These parameters have been fitted to the pseudoscalar meson octet in vacuum \cite{Rehberg:1995kh}. The ratios of the vector coupling $\eta_V=G_V/G_S$ and the diquark coupling $\eta_D=G_D/G_S$ to the scalar coupling are treated as free parameters.
Within the RG-consistent regularization, the model can be evaluated at arbitrary high scales, provided that $\Lambda$ is chosen sufficiently large. For this work, we choose $\Lambda=10 \Lambda'=6023$\,MeV, which we find to be sufficient to study the densities reached in neutron stars. 

\section{Comprehensive parameter study for NJL quark matter} \label{sec:var}

For comprehensive studies of the properties of the RG-consistent NJL model, the diquark coupling and the vector coupling
are varied between $\eta_D$ = 0.90 and $\eta_D$ = 1.80, and $\eta_V$ = 0.00 and $\eta_V$ = 1.50, respectively. In this section, the impact of changing one parameter while keeping the other parameter constant on the diquark gaps, the equation of state, the speed of sound, and the mass-radius profiles are investigated.

\begin{figure*}
		\begin{minipage}[t]{0.32\textwidth}		 		
  \includegraphics[width=\textwidth]{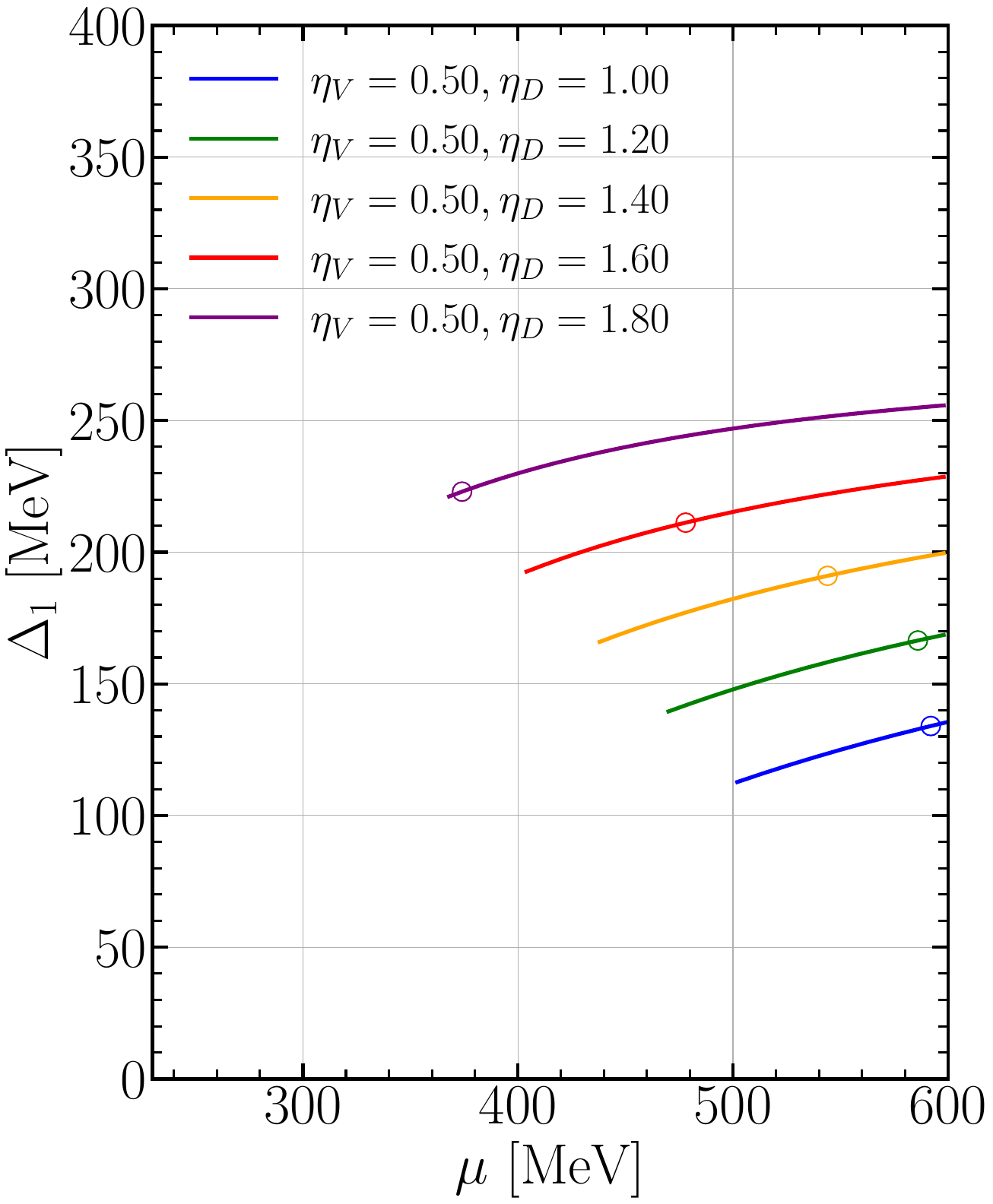}
			 	\end{minipage}
		 		\begin{minipage}[t]{0.32\textwidth}
			 		\includegraphics[width=\textwidth]{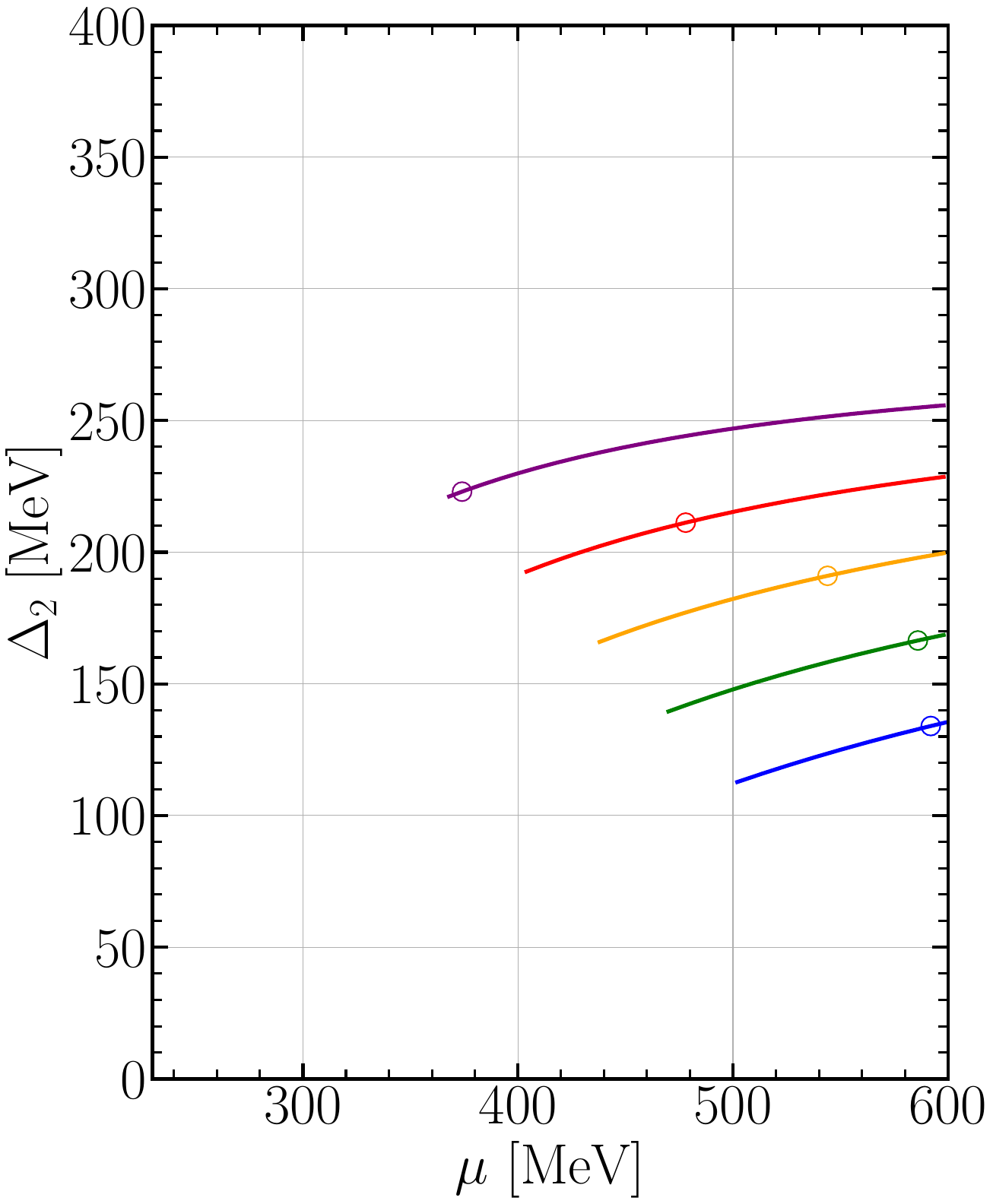}
			 	\end{minipage}
		 	\begin{minipage}[t]{0.32\textwidth}
		\includegraphics[width=\textwidth]{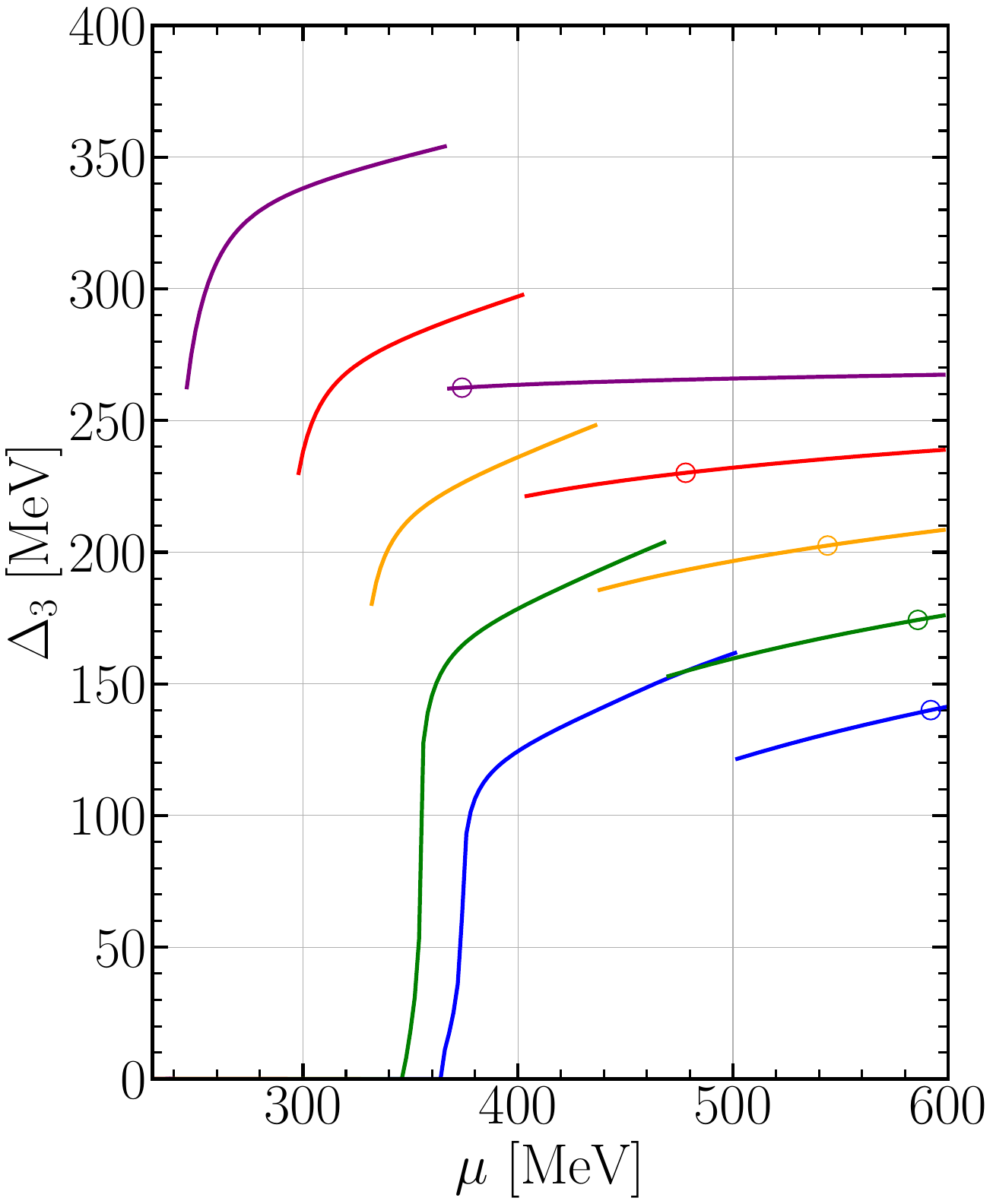}
			 	\end{minipage}
			 			\caption{Diquark gaps $\Delta_1$ (left), $\Delta_2$ (middle) and $\Delta_3$ (right) for constant vector coupling $\eta_V$ = 0.50 and  different diquark couplings $\eta_D$. 
        For each $\eta_V$-$\eta_D$ coupling combination, the open circle marks the value reached in the center of the maximum-mass solution of the TOV equation.
       }
		\label{figgaps1}	 	
\end{figure*}	

\begin{figure*}
		\begin{minipage}[t]{0.32\textwidth}		 		
  \includegraphics[width=\textwidth]{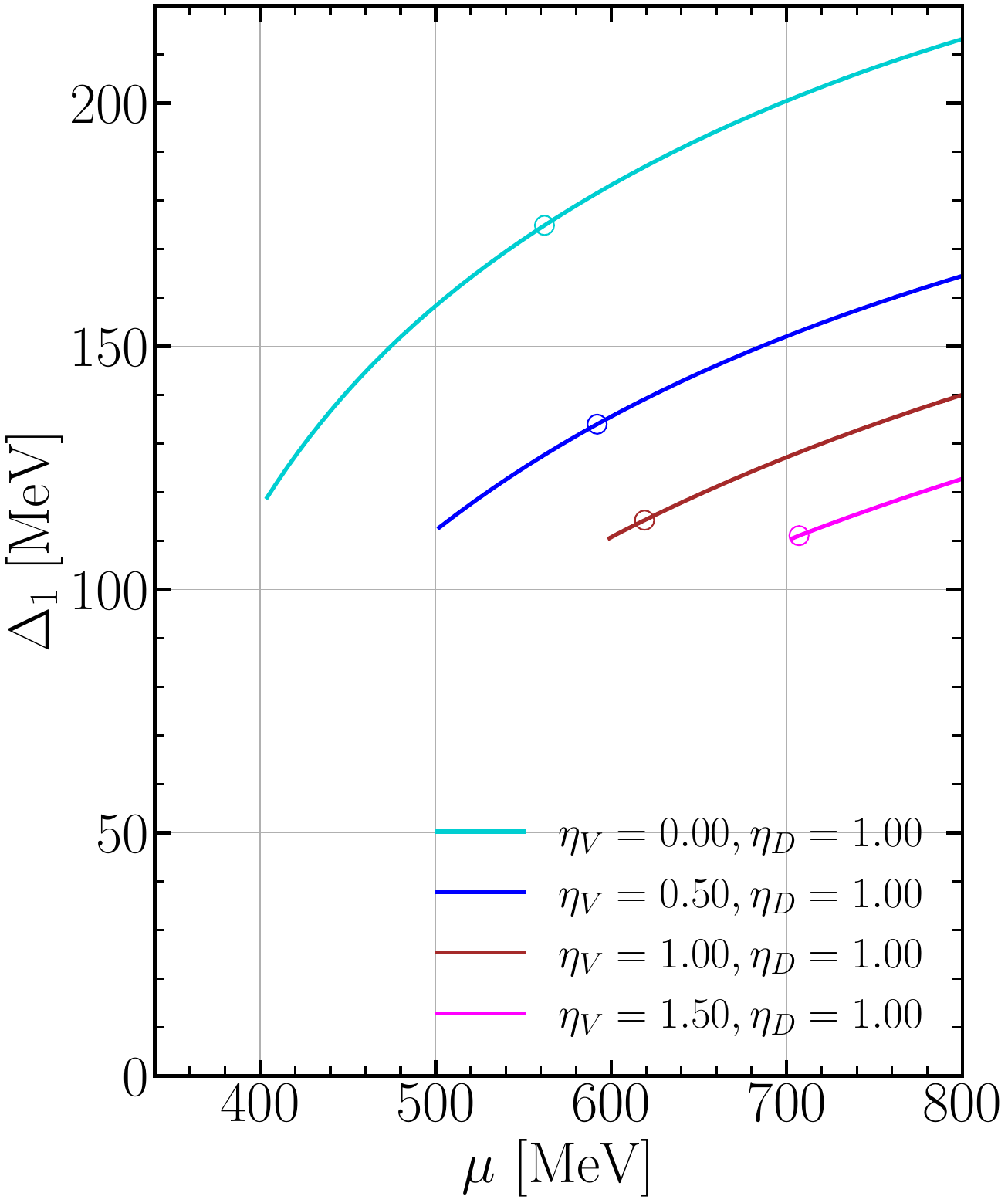}
			 	\end{minipage}
		 		\begin{minipage}[t]{0.32\textwidth}
			 		\includegraphics[width=\textwidth]{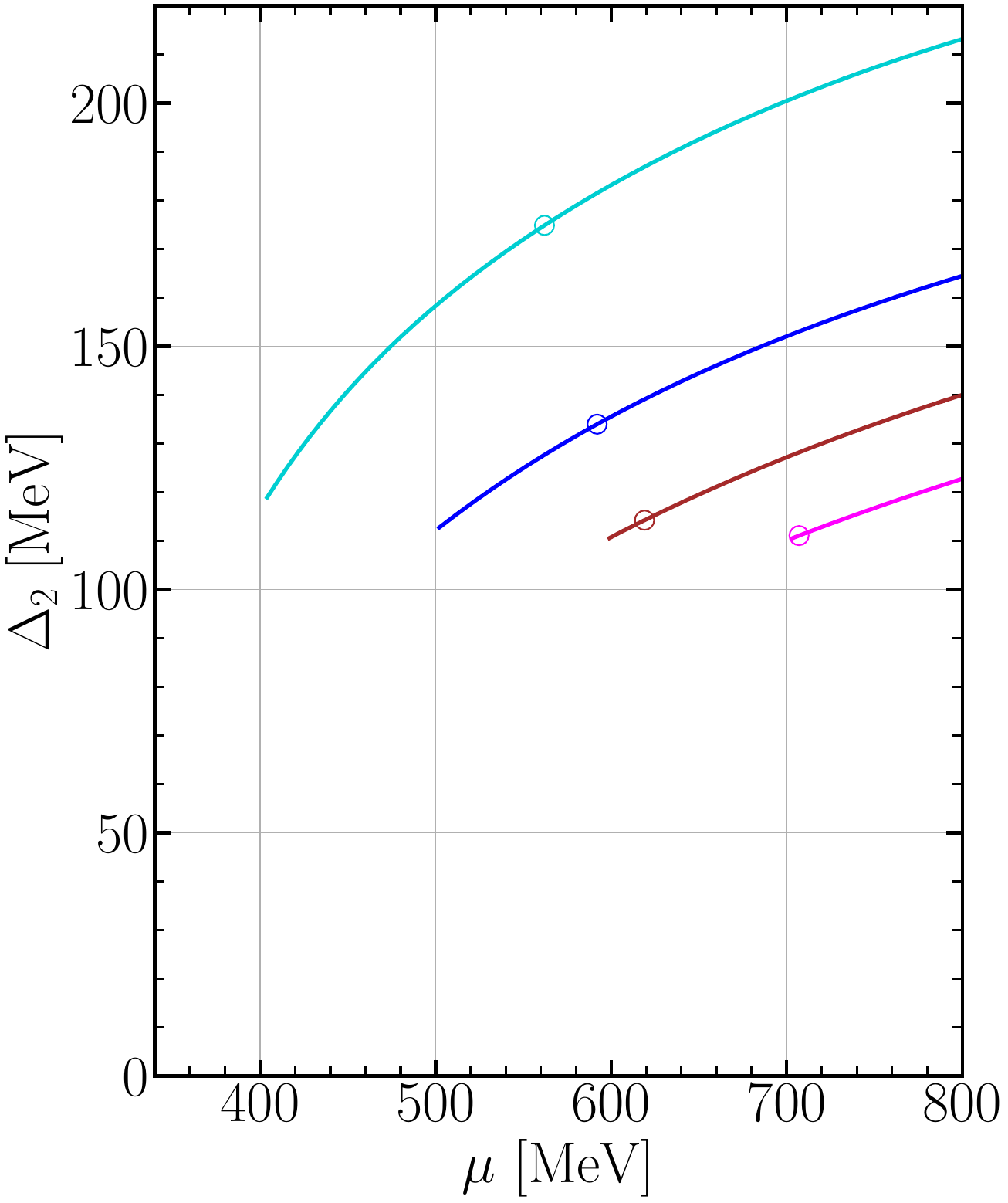}
			 	\end{minipage}
		 	\begin{minipage}[t]{0.32\textwidth}
		\includegraphics[width=\textwidth]{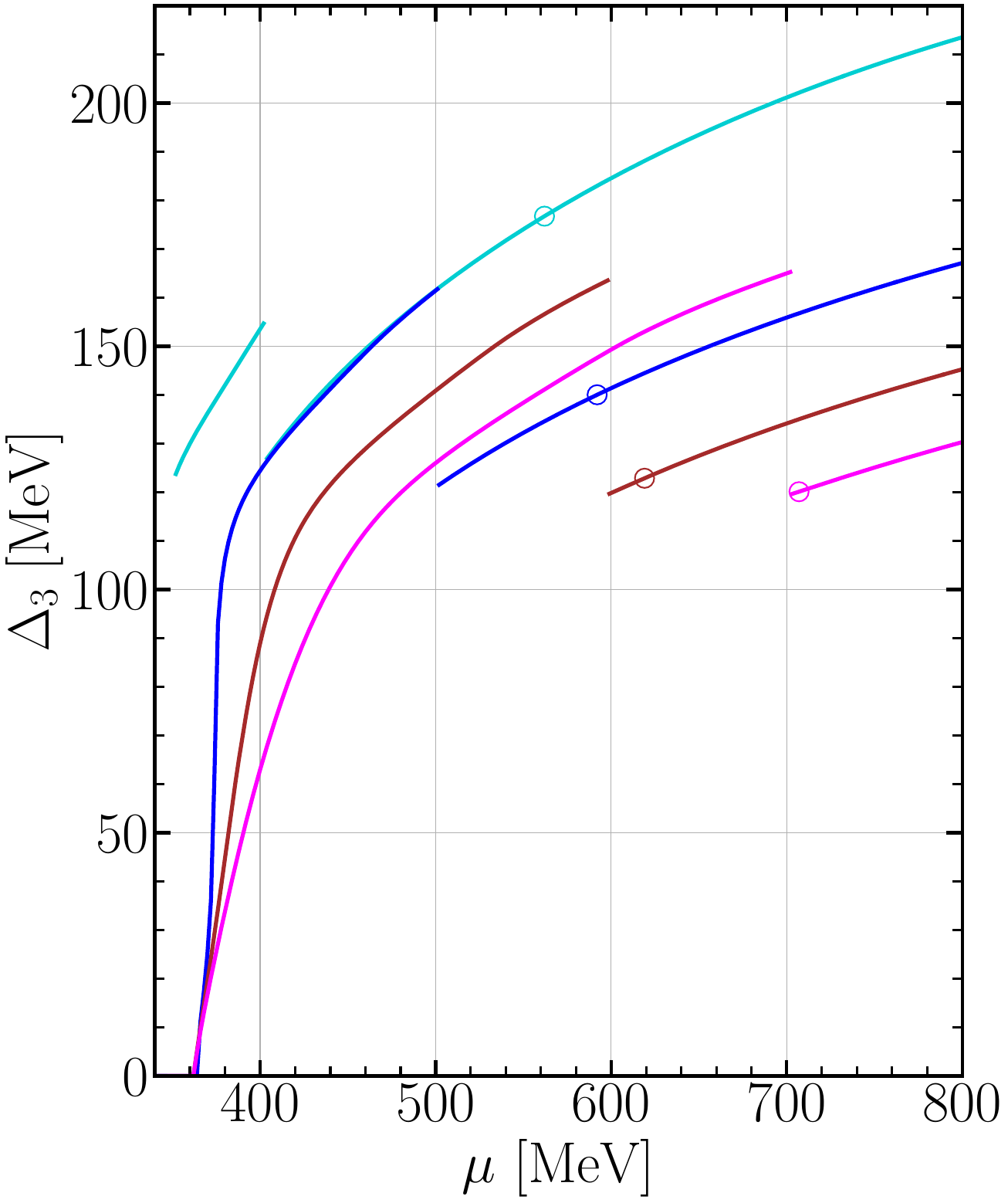}
			 	\end{minipage}
			 			\caption{Same as Figure \ref{figgaps1}, but for a constant diquark coupling $\eta_D$ = 1.00 and different vector couplings $\eta_V$.}
		\label{figgaps2}	 	
     \end{figure*}

\subsection{Diquark gaps}

As highlighted in Refs.~\cite{Braun:2018svj,Gholami:2024diy}, using conventional cutoff regularization in effective models becomes problematic near the cutoff energy scales due to unphysical artifacts that arise from the finite cutoff. This is particularly evident in the conventional regularization of the NJL model with three flavors and color superconductivity, where, above a certain value of the chemical potential,
the diquark gaps decrease with increasing $\mu$ and eventually even vanish, leading to unphysical results.
In contrast, the RG-consistent model overcomes these limitations and enables calculations at arbitrarily high chemical potentials without encountering such cutoff artifacts. 

The solutions to the gap equations and neutrality conditions vary significantly for each combination of the diquark coupling $\eta_D$ and the vector coupling $\eta_V$.
Figure \ref{figgaps1} displays the diquark gaps as a function of the quark chemical potential $\mu$ for various choices of the diquark coupling between $\eta_D$ = 1.00 and $\eta_D$ = 1.80 and constant vector coupling $\eta_V$ = 0.50. The open circles indicate the values of the diquark gaps reached in the centers of the maximum-mass solution of the TOV equation, see Sec.~\ref{sec:mr-relations} for more details. We ensured the validity of our results by extending our analysis to high values of chemical potential, maintaining RG consistency.

At low chemical potential, the matter is in the chiral-symmetry broken phase and all diquark condensates are zero. At zero temperature, this phase is identical to the vacuum. At intermediate chemical potential, the matter is in a 2SC phase, in which only $\Delta_3$ is nonzero, followed by a first-order phase transition to the CFL phase where all diquark gaps are nonzero. 
With increasing values of the diquark coupling, the phase transition from the chiral-symmetry broken phase to the 2SC phase and from the 2SC phase to the CFL phase move to lower chemical potential, and the value of the diquark gaps in both the 2SC and the CFL phase increase significantly.

In Figure \ref{figgaps2}, the vector coupling is varied between $\eta_V$ = 0.00 and $\eta_V$ = 1.50 while the diquark coupling is kept constant at $\eta_D$ = 1.00. With increasing vector coupling $\eta_V$, the 2SC to CFL phase transition shifts to a larger chemical potential, whereas the transition from the chiral-symmetry broken phase to the 2SC phase is only slightly delayed. However, the chemical potential at which $\Delta_3$ reaches a certain value shifts to higher values as $\eta_V$ increases. This behavior arises because the vector interaction causes the pressure to build up more gradually. 
In fact, as $\eta_V$ increases, the phase transition from the chiral-symmetry broken phase to the 2SC phase changes from a first-order transition to a smooth transition. 

The chemical potential at which the maximum mass is reached and the star turns unstable (circles in Figures \ref{figgaps1} and \ref{figgaps2}) moves to lower chemical potential with increasing $\eta_D$ and to higher chemical potential with increasing $\eta_V$. Note that for large values of $\eta_V$ the chemical potentials reached in the maximum-mass stars are higher than the vacuum cut-off $\mu=\Lambda'$. With conventional cutoff regularization, describing these points without running into severe cutoff artifacts would be impossible. Therefore, the use of the RG consistent regularization, where no cutoff artifacts occur, is crucial for our analysis.
As will be seen in Sec.~\ref{sec:constraints}, for extremely large $\eta_V$, the maximum mass occurs in the 2SC phase.

  \begin{figure*}
		\begin{minipage}[t]{0.49\textwidth}		 		
  \includegraphics[width=\textwidth]{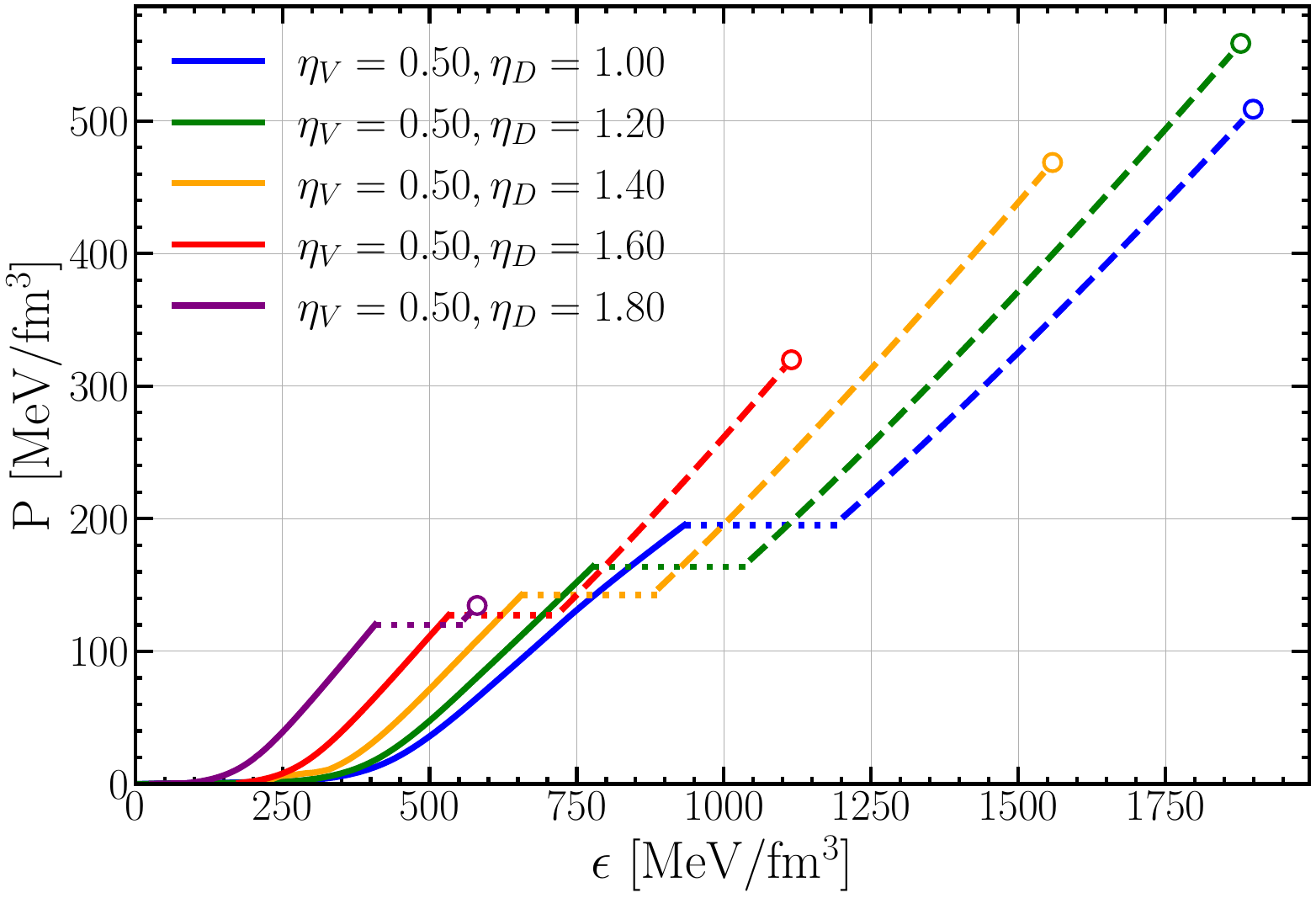}
			 	\end{minipage}
		 		\begin{minipage}[t]{0.49\textwidth}
			 		\includegraphics[width=\textwidth]{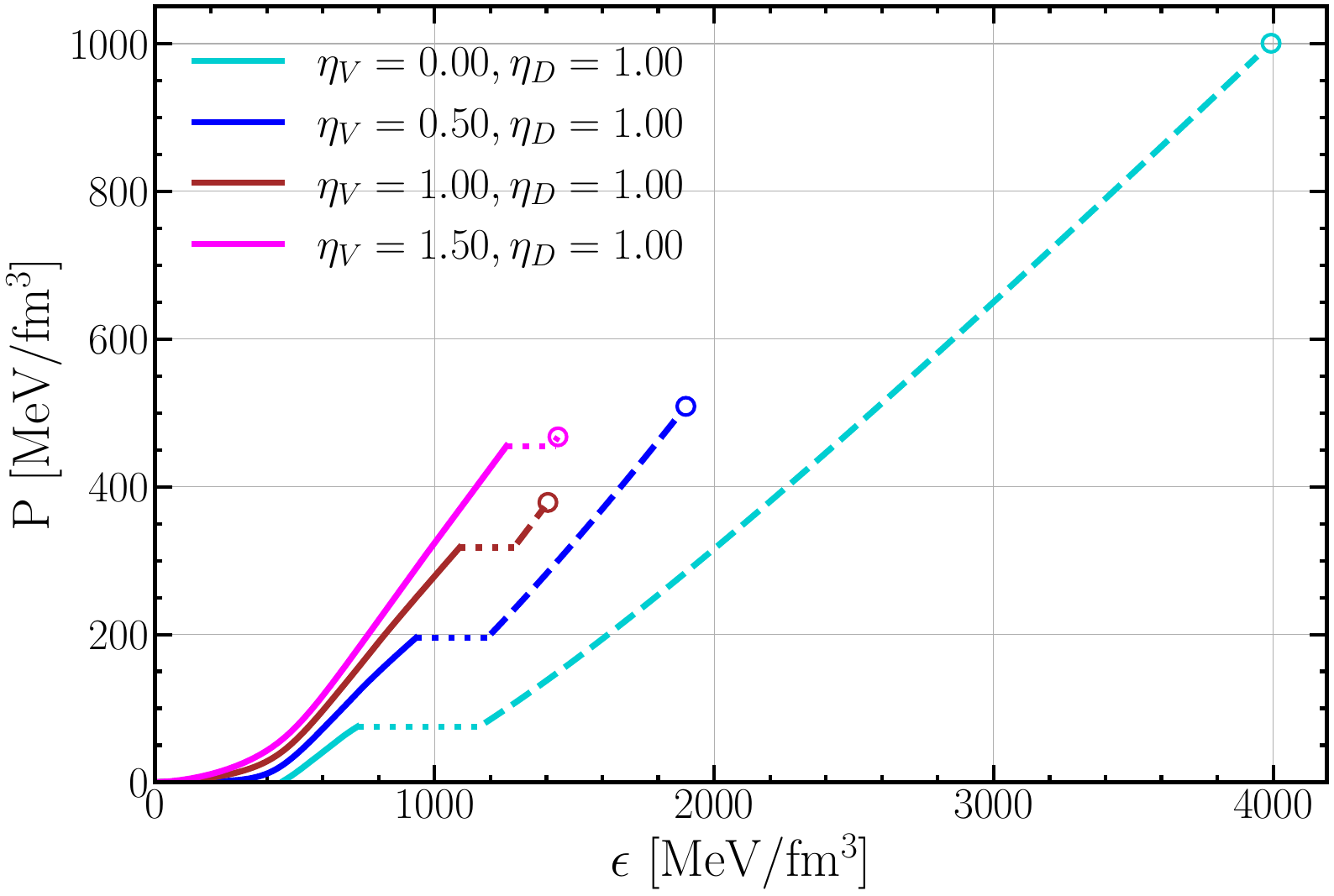}
			 	\end{minipage}
			     \caption{Equation of state in the form of pressure versus energy density for some combinations of $\eta_V$-$\eta_D$ 
           coupling constants. Left plot: EoS at a fixed value of the vector coupling strength $\eta_V$ = 0.50. 
       Right plot: EoS at a fixed value of the diquark coupling strength $\eta_D$ = 1.00. 
       Solid lines correspond to the 2SC phase, while dashed lines correspond to the CFL phase. 
       The dotted lines represent the phase transition from the 2SC to the CFL phase.
       For each $\eta_V$-$\eta_D$ coupling combination, the open circle marks the value reached in the center of the maximum-mass solution of the TOV equation.
       }
		\label{figeos}	 	
     \end{figure*}
     
\subsection{Equation of State}
   
Figure \ref{figeos} displays the pressure $P$ versus energy density $\epsilon$ for fixed $\eta_V$ = 0.50 (left panel) and at fixed 
$\eta_D$ = 1.00 (right panel).  The solid lines in each curve correspond to the 2SC phase while dashed lines correspond to the CFL phase. The dotted lines connecting the solid and dashed lines represent the transition from the 2SC to the CFL phase. The open circle at the end of each curve marks the values reached in the center of the maximum-mass solution of the TOV equation for this $\eta_V$-$\eta_D$ coupling combination. 
Working within the grand canonical ensemble, we computed the model up to quark chemical potentials at which the mass-radius configuration turns unstable. This corresponds to different maximal values of energy density and number density depending on the coupling parameters.

 For the EoS at fixed $\eta_V$, increasing the diquark coupling from $\eta_D$ = 1.00 to $\eta_D$ = 1.80 shifts the EoS to higher pressure values. This is because higher $\eta_D$ coupling leads to a stronger Cooper pairing and larger gaps, which in turn leads to higher pressure.
 The phase transition from the 2SC to the CFL phase occurs at lower density and lower pressure values for higher values of $\eta_D$, 
and the jump in energy density becomes smaller, i.e.\ the latent heat $\Delta \epsilon$, decreases.  For low values of $\eta_D$, the energy density range of the 2SC phase is large and decreases with increasing values of the diquark coupling. For higher values $\eta_D\sim$ 1.80, the star turns unstable almost immediately after the onset of the phase transition to the CFL phase at its center. 

For the case of a fixed diquark parameter  $\eta_D$ = 1.00 (right plot), we find that with zero vector coupling, $\eta_V$ = 0.00, the EoS reaches a vanishing pressure at a non-vanishing energy density. 
When the vector coupling value is increased, the slope of the equation of state (EoS) increases, suggesting a stiffer EoS with an expanded density range for the 2SC phase, albeit with reduced latent heat. In contrast, the energy-density range of CFL quark matter up to the point of the maximum mass decreases with increasing vector coupling, becoming almost zero at $\eta_V$ = 1.50. The central pressure at the maximum mass configuration decreases, then increases again with increasing the vector coupling, contrary to the increasing diquark coupling (left plot) where it increases first and then decreases again.

\begin{figure*}
		\begin{minipage}[t]{0.49\textwidth}		 		
  \includegraphics[width=\textwidth]{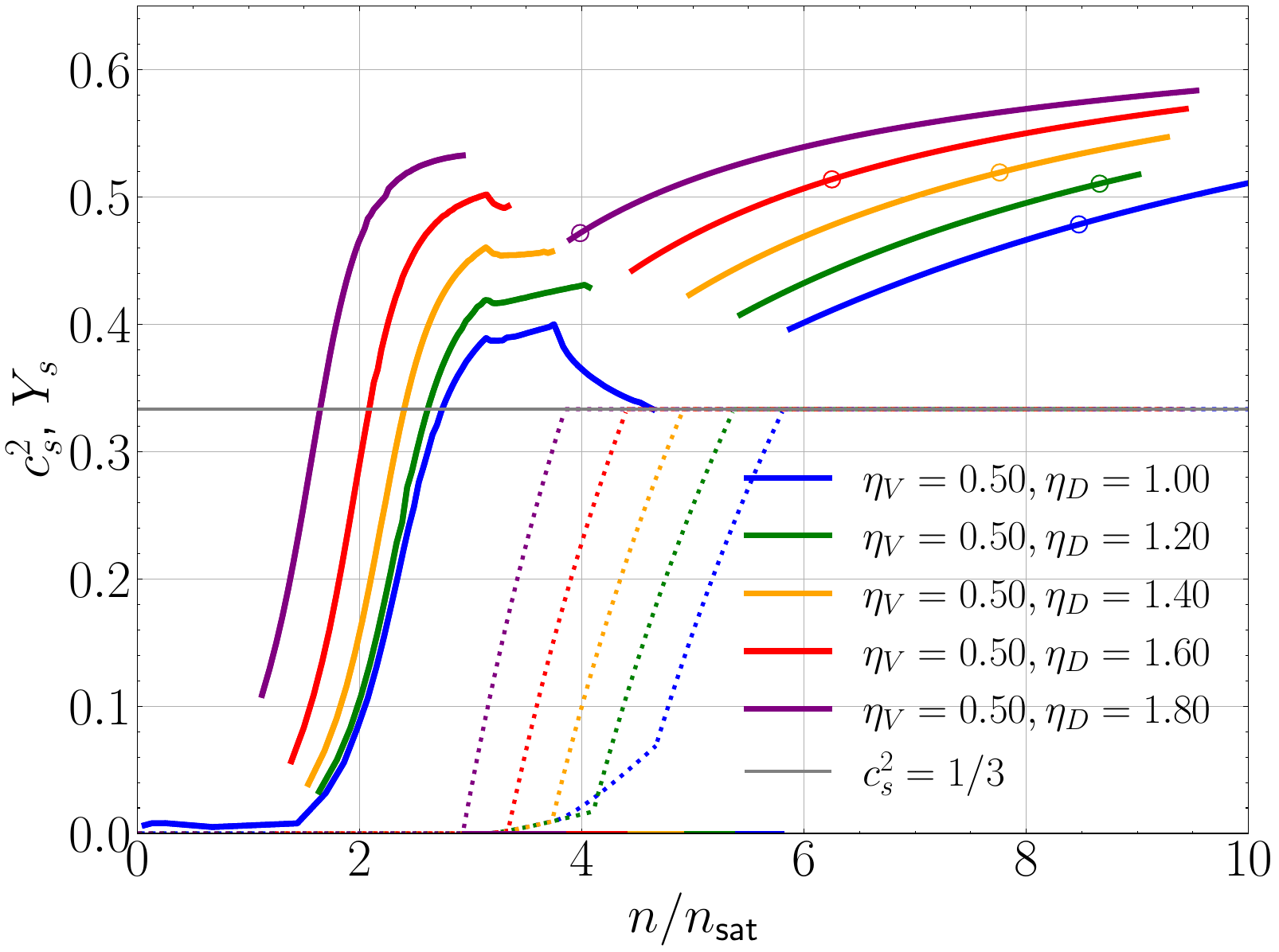}
			 	\end{minipage}
		 		\begin{minipage}[t]{0.49\textwidth}
			 		\includegraphics[width=\textwidth]{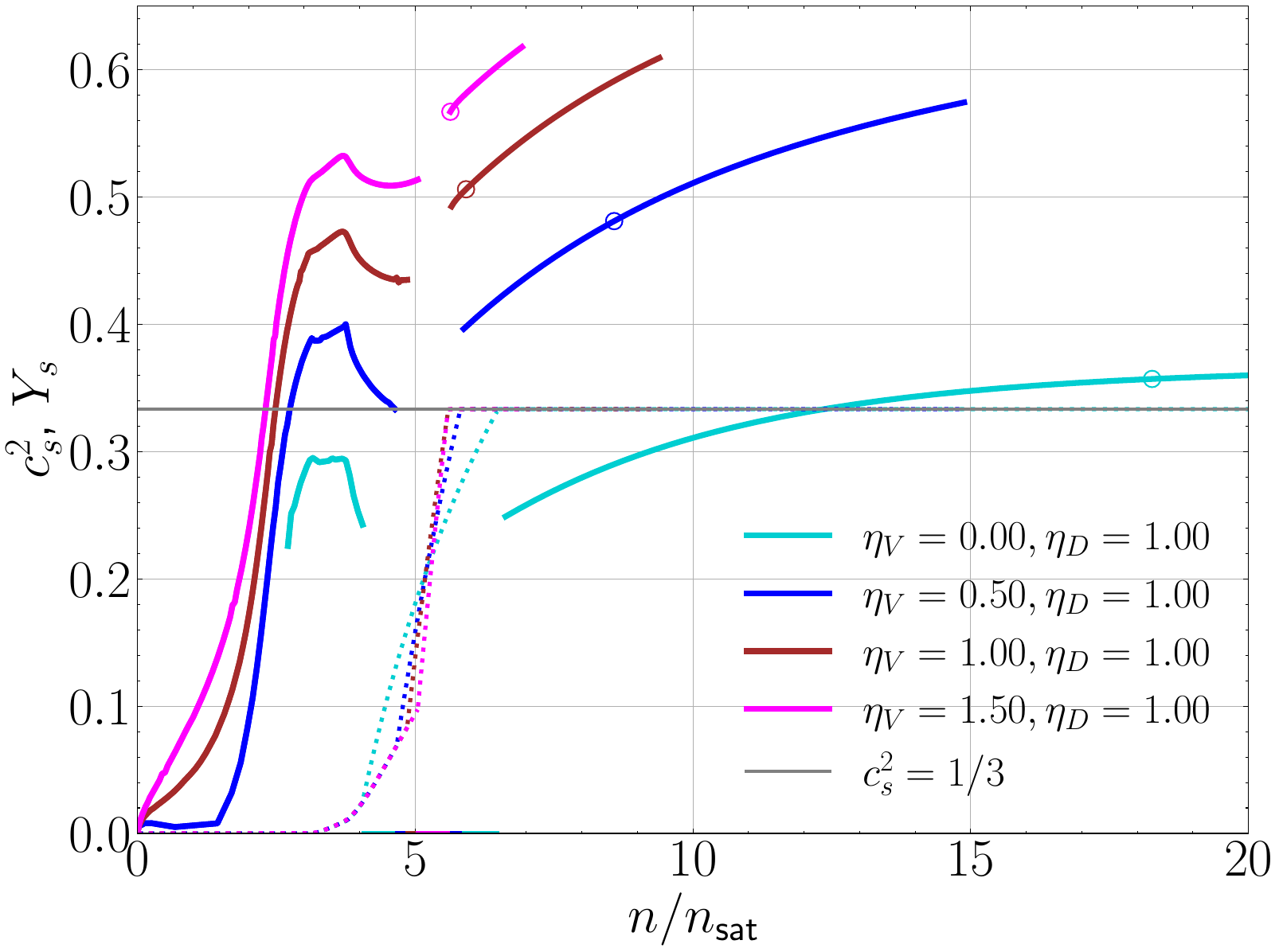}
			 	\end{minipage}
			 			\caption{Speed of sound squared $c_s^2$ (solid) and strangeness fraction $Y_s$ (dotted) plotted over the baryon density in units of the saturation density $n_\text{sat}$.
       Left plot: $\eta_D$ varies between $\eta_D$ = 1.00 and 1.80 with constant $\eta_V$ = 0.50. 
       Right plot: $\eta_V$ varies between $\eta_V$ = 0.00 and 1.50 with constant $\eta_D$ = 1.00. 
        The thin gray line corresponds to the conformal limit of $c_s^2$ = 1/3. The $\eta_V$-$\eta_D$ coupling combinations where $c_s^2$ starts from a finite value correspond to the self-bound configurations, whereas the ones where $c_s^2$ starts from zero correspond to gravitationally bound configurations. For each $\eta_V$-$\eta_D$ coupling combination, the open circle marks the value reached in the center of the maximum-mass solution of the TOV equation. }
		\label{figcs}	 	
\end{figure*}
	
\subsection{Speed of sound}

The RG-consistent NJL model can be evaluated at large chemical potential and energy scales to study the behavior of the speed of sound at higher densities without cutoff artifacts. If the vector interaction is set to zero ($\eta_V$ = 0.00) as in Ref. \cite{Gholami:2024diy}, the speed of sound approaches the conformal limit of $\frac{1}{3}$ at high densities, consistent with expectations for a relativistic gas of massless particles. However, for a nonzero $\eta_V$, $c_s^2$ increases towards \(1\) at high densities, approaching the causal limit where the speed of sound equals the speed of light. This feature of approaching $c_s^2=1$ with a vector interaction is generic for an interaction-dominated equation of state and is well-known as
the Zeldovich equation of state \cite{Zeldovich:1961sbr}.
It is clear that the NJL model, even with RG-consistent regularization, cannot be trusted in the realm of perturbative QCD. As we will see, for densities reached in the centers of compact stars, which is the focus of our investigations, the speed of sound stays well below the causal limit.

Figure~\ref{figcs} shows the behavior of the speed of sound squared for different combinations of the coupling parameters $\eta_D$ and $\eta_V$, plotted as a function of the ratio of the baryon number density $n=(n_u+n_d+n_s)/3$ to the nuclear saturation density $n_\text{sat} = 0.16$ fm$^{-3}$. 
Note that the density ranges plotted are different in both plots shown. 

A stiff EoS corresponds to a high speed of sound, as the pressure responds strongly to changes in energy density. Both plots demonstrate that increasing either the diquark coupling $\eta_D$ or the vector coupling $\eta_V$ results in a higher speed of sound for a given density, thus corresponding to a stiffer equation of state (EoS) in both the 2SC and CFL phases. A higher diquark coupling enhances the pairing interactions, leading to larger diquark gaps and increased pressure at a given density. Similarly, a higher vector coupling introduces additional repulsion between quarks, stiffening the EoS. 

The speed of sound increases with density at low to moderate densities reaching values above the conformal limit of $c_s^2=1/3$ for all cases.
Characteristic features in the speed of sound associated with the phase transition can be seen. The transition to the CFL phase, when present, is indicated by a discontinuity in the speed of sound curve as it is a first-order phase transition. In the 2SC-CFL mixed phase, the pressure is constant (Maxwell construction) and the speed of sound is zero. For low values of the diquark coupling constant, the speed of sound decreases in the 2SC phase before the CFL phase transition happens. This is accompanied by a nonzero strangeness fraction $Y_s=n_s/(n_u+n_d+n_s)$, depicted by the dotted line in Figure  \ref{figcs}. This is because the ungapped strange quarks in the 2SC phase, which become abundant after the partial chiral symmetry restoration in the strange-quark sector, soften the equation of state before the phase transition. Strange quarks appear for the first time at a density of $n\sim 3-4 n_\text{sat}$. This onset density decreases with increasing diquark coupling.

In the mixed phase, the strangeness fraction is computed from the volume fractions $x_{\text{2SC}}$ and $x_\text{CFL}$. The volume fractions are computed from $x_\text{2SC}\cdot n_{\text{2SC}}+x_\text{CFL}\cdot n_{\text{CFL}}=n$ with the last density point in the pure 2SC phase $n_{\text{2SC}}$ and the onset density of the CFL phase $n_{\text{CFL}}$, respectively, and $x_{\text{2SC}}+x_\text{CFL}=1$.

In the CFL phase, the speed of sound squared increases again, reaching values of close to $c_s^2\sim0.6$ for high vector and diquark coupling constants in the center of maximum mass configurations. Due to the symmetric pairing structure in the CFL phase, all quarks come in the same abundance, and $Y_s=1/3$.

     \begin{figure*}
		\begin{minipage}[t]{0.47\textwidth}		 		
  \includegraphics[width=\textwidth]{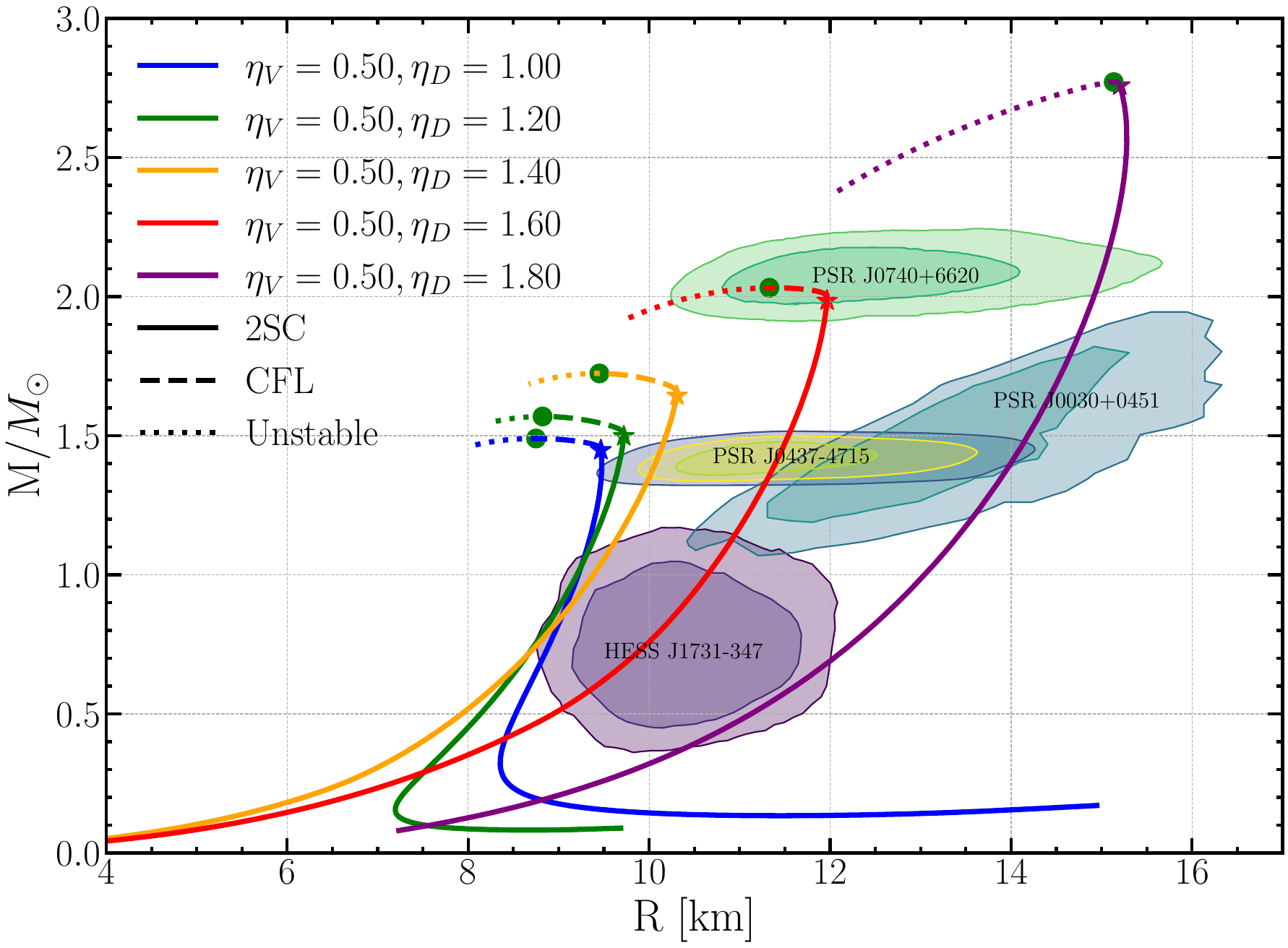}
			 	\end{minipage}
		 		\begin{minipage}[t]{0.49\textwidth}
			 		\includegraphics[width=\textwidth]{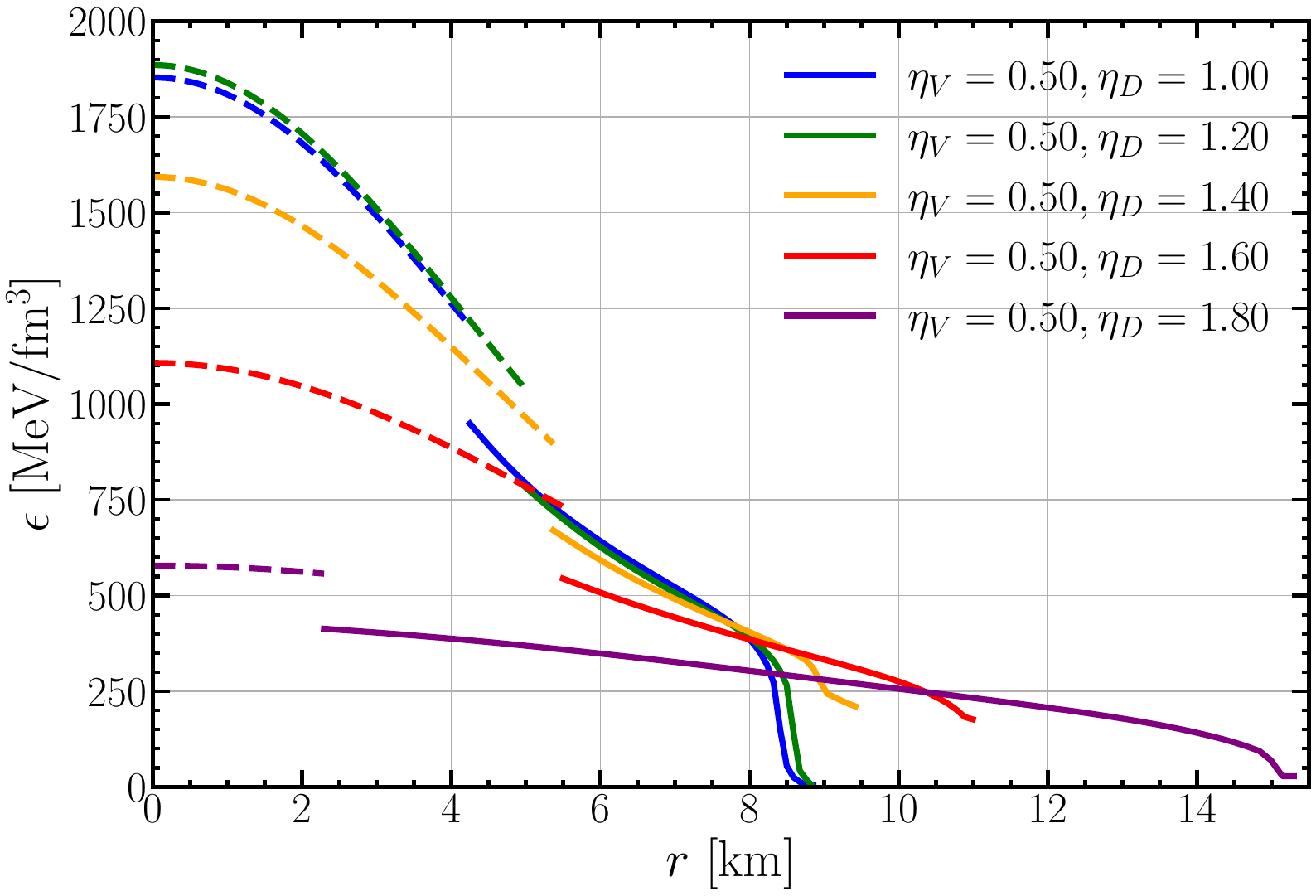}
			 	\end{minipage}
			 			\caption{Left: Mass-radius relation for different diquark coupling constants $\eta_D$ at a fixed vector coupling $\eta_V$ = 0.50. 
       Solid lines correspond to the quark star configurations with a 2SC phase only, while dashed lines correspond to configurations with a CFL phase in the core. 
       The dotted lines correspond to the unstable configurations. The star symbol marks the onset of the CFL phase in the core, 
       and the solid green symbol represents the maximum mass configuration. The various shaded areas are 
       credibility regions for mass and radius inferred from the analysis of PSR J0740+6620, PSR J0030+0451, PSR J0437-4715, and HESS J1731-347 \cite{miller2021,2021ApJ...918L..27R, Miller_2019a, Riley_2019, Choudhury:2024xbk, Doroshenko2022}.
       Right: Radial profiles of the energy density as a function of the radial coordinate for the maximum mass configurations of the same choices of the coupling constants as shown in the left plot. Solid (dashed) lines indicate matter in the 2SC (CFL) phase. Configurations where the energy density reaches zero smoothly at the surface correspond to gravitationally bound stars, whereas configurations with a non-zero surface energy density are self-bound stars.}
		\label{figmr}	 	
     \end{figure*}	

     \begin{figure*}
		\begin{minipage}[t]{0.47\textwidth}		 		
  \includegraphics[width=\textwidth]{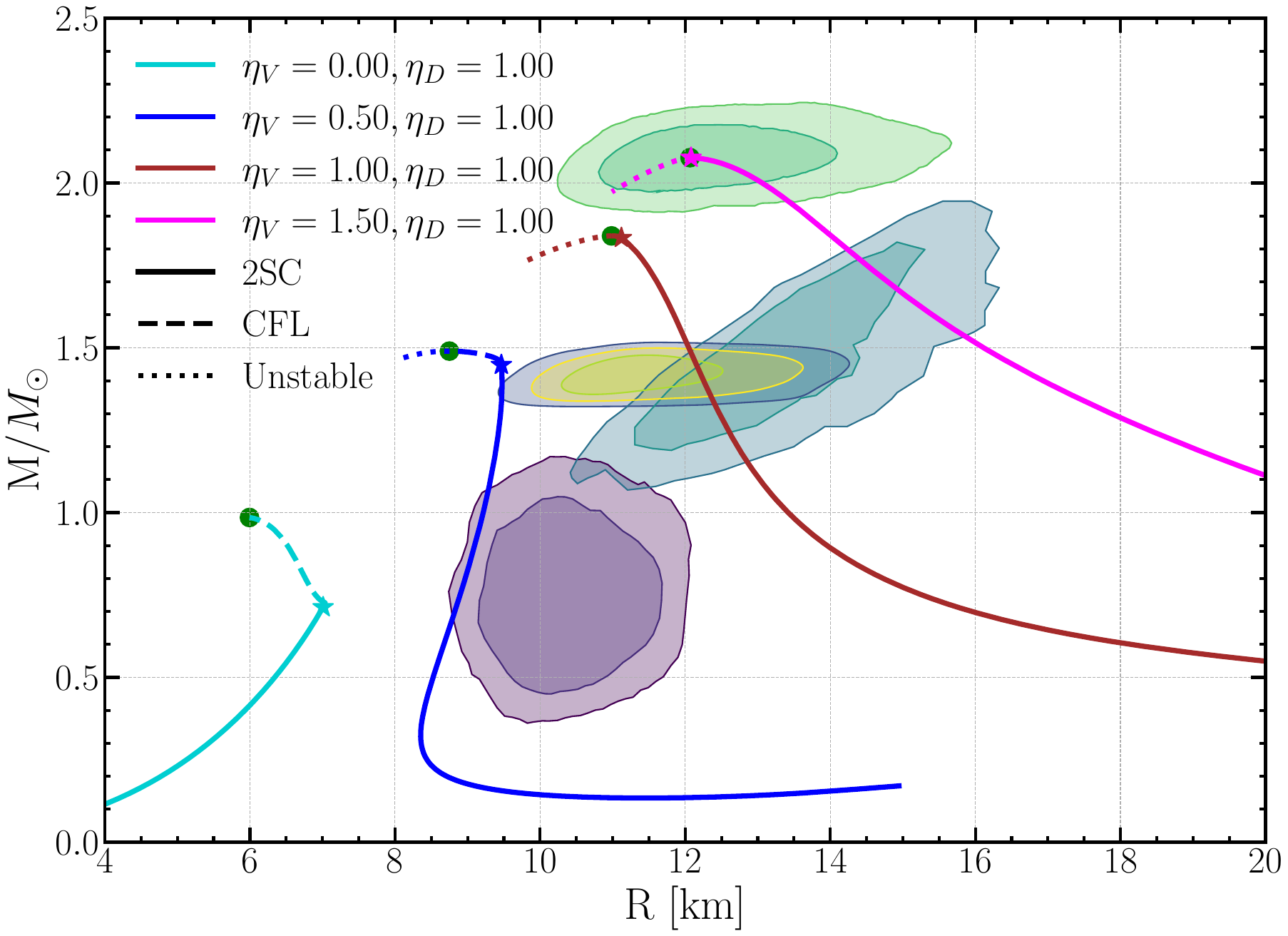}
			 	\end{minipage}
		 		\begin{minipage}[t]{0.49\textwidth}
			 		\includegraphics[width=\textwidth]{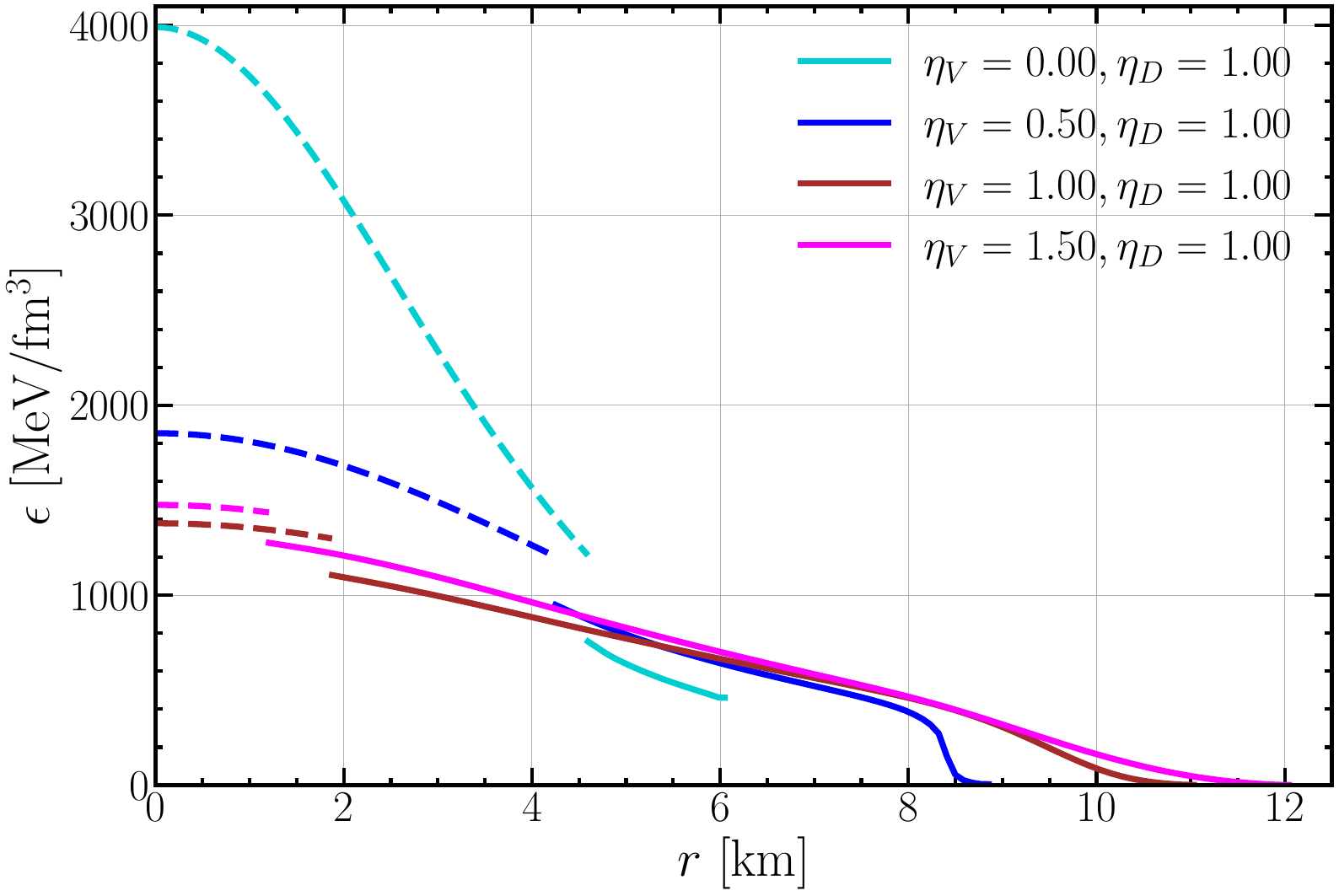}
			 	\end{minipage}
			 			\caption{ Same as Figure \ref{figmr} for different vector coupling constants $\eta_V$ and a fixed diquark coupling constant of $\eta_D$ = 1.00. }
		\label{radial}	 	
     \end{figure*}	

\subsection{Mass-Radius Relations and Radial Profiles of the Maximum-Mass Configurations}\label{sec:mr-relations}

Solving the Tolman-Oppenheimer-Volkoff (TOV) equation \cite{PhysRev.55.374, PhysRev.55.364}, we calculate the mass and radius for the above-defined $\eta_V$-$\eta_D$ coupling combinations.
Figure \ref{figmr} shows the mass-radius (MR) profiles (depicted on the left) of the EoS defined above and the radial profiles of the corresponding maximum-mass configurations in terms of the energy density as a function of radius (depicted on the right), evaluated at a constant vector coupling parameter of $\eta_V$ = 0.50.  Different line styles correspond to the 2SC or CFL phases as described in the figure captions. The star symbol on each curve in the left plot separates pure 2SC stars from stars with a CFL core. The green dot represents the last stable point, marking the maximum mass obtained, while the subsequent dotted curve represents the unstable branch. To guide the eye, we show various astrophysics constraints, described in the caption. To what extent the credibility intervals by NICER and the lower bound of 2$\,M_{\odot}$ for the maximum mass can be used to constrain the parameters of the model is the subject of Sec.~\ref{sec:constraints}.

With the fixed vector coupling of $\eta_V$ = 0.50 and $\eta_D$  = 1.00, the MR relation resembles a gravitationally bound star with a maximum mass of 1.49\,$M_{\odot}$ and a corresponding radius of 8.74 km. For the case of ($\eta_V$, $\eta_D$)  = (0.50, 1.20), the maximum mass and the corresponding radius change to a value of 1.57\,$M_{\odot}$ and 8.82 km, respectively. Increasing the diquark coupling to $\eta_D$ = 1.40 alters the MR relation from gravitationally bound to a self-bound profile because the EoS reaches a vanishing pressure at a non-vanishing energy density (see also the radial plot). With increasing the diquark coupling to $\eta_D=1.60$, the maximum mass reaches 2.02\,$M_{\odot}$ with a radius of approximately 11.4 km at 1.4\,$M_{\odot}$ and 11.96 km at 2.0\,$M_{\odot}$. The choice $\eta_D=1.80$ results in a self-bound star with a maximum mass of 2.77\,$M_{\odot}$ and a radius of 14\,km at 1.4\,$M_{\odot}$.

The configurations where the energy density at the star's surface reduces to zero are indicative of gravitationally bound stars so that the radius increases for lower masses. Cases with a non-zero surface energy density are denoted as self-bound stars, where the radius decreases for arbitrarily lowering the mass.  With increasing diquark coupling, the CFL core expands in volume but then contracts at even higher values, becoming confined to a smaller core region with lower energy densities.  This non-monotonic behavior with increasing diquark coupling can also be seen in the central energy density of the core.

For the maximum-mass configurations, the 2SC part of the core reaches higher energy densities at lower values of $\eta_D$. The radial extension of the 2SC core changes non-monotonically with $\eta_D$; it is up to 5~km for $\eta_D$ = 1.00, then decreases to a value of 4~km for $\eta_D$ = 1.40 before increasing again to 13~km for $\eta_D$ = 1.80.
As the density increases beyond the phase transition point, the matter in the core is in the CFL phase, resulting in more compact radial configurations. For $\eta_D$ = 1.00, the CFL phase covers a radial extent of 4~km, which decreases to 2~km for $\eta_D$ = 1.80. 

Figure \ref{radial} shows again the MR-profiles (left) and the radial profiles of the energy density (right), but this time for a constant diquark coupling parameter of $\eta_D$ = 1.00 and varying vector coupling constant $\eta_V$. The MR relation (left plot) at $\eta_V$ = 0.00 shows a self-bound star with a maximum mass below 1.0\,$M_{\odot}$ and a core in the CFL phase. Increasing the vector coupling to $\eta_V$ = 0.50 results in a gravitationally bound star. For higher $\eta_V$ values of 1.00 and 1.50, the maximum masses reach values of 1.87 and 2.08\,$M_{\odot}$, respectively. While the configuration ($\eta_V$, $\eta_D$) = (1.50, 1.00) meets the 2.0\,$M_{\odot}$ limit, the radius at 1.4\,$M_{\odot}$ is 16.9 km, outside the NICER measurements' constraint region from PSR J0437-4715. This means that only a hybrid construction with matching at high densities to the CSC EoS could satisfy this constraint. 
As can be seen from the radial profiles on the right, there is a small CFL core present for the maximum-mass configurations of the $\eta_V$ = 1.00 and 1.50 cases, although it is difficult to see this in the mass-radius relations on the left.

The central energy density in the CFL phase is significantly larger for $\eta_V$ = 0.00 compared to nonzero vector coupling. The radial extension of the CFL phase in this configuration is 5~km, whereas the 2SC phase has a smaller radial size of 1.5~km. Given that the mass-radius relation is the one for a self-bound star, the energy density is non-zero at the surface. 
Increasing the vector coupling constant changes the MR relation from self-bound to gravitationally bound, visible in the radial profiles with a vanishing energy density at the surface. Specifically, the CFL phase core contracts to about 1~km for $\eta_V$ = 1.50, while the 2SC phase mantle extends to  $r\sim12~km$. 

From the radial profiles, we see that in all cases shown here the maximum-mass configurations have a CFL core, which decreases in size as the diquark coupling is increased for a fixed vector coupling or the vector coupling is increased for a fixed diquark coupling.
     
\section{Constraints on NJL parameters from Astrophysics and Stability of the 2SC-CFL phase transition}
\label{sec:parameterstudy}


As pointed out in the introduction, in this work we restrict ourselves to a comprehensive study of quark-matter EoSs, focusing on the CSC phases. However, as already found in earlier works with the NJL model \cite{BUBALLA2005205, Buballa:1998pr}, none of our EoSs fulfills the requirements for the existence of absolutely stable strange-quark matter \cite{Witten1984, Bodmer1971, Farhi:1984qu}, and, hence, strange stars do not exist in our model. This means quark matter can only be present in the cores of hybrid stars, surrounded by hadronic matter.
A full investigation of this case involves the inclusion of a low-density hadronic EoS and the calculation of the hybrid star EoS.
However, the hadronic EoS for neutron stars beyond saturation density is not known, so the resulting mass-radius relation for hybrid stars would depend on the chosen low-density hadronic model.

In the following, we discuss how the parameter space of the RG-consistent NJL model can be pre-constrained without specifying the hadronic EoS. This is useful for finding good parameters for future studies of hybrid stars with the RG-consistent NJL model.

\subsection{Astrophysical constraints for diquark and vector coupling constants}\label{sec:constraints}

\begin{figure*}
\begin{minipage}[t]{0.49\textwidth}
			 		\includegraphics[width=\textwidth]{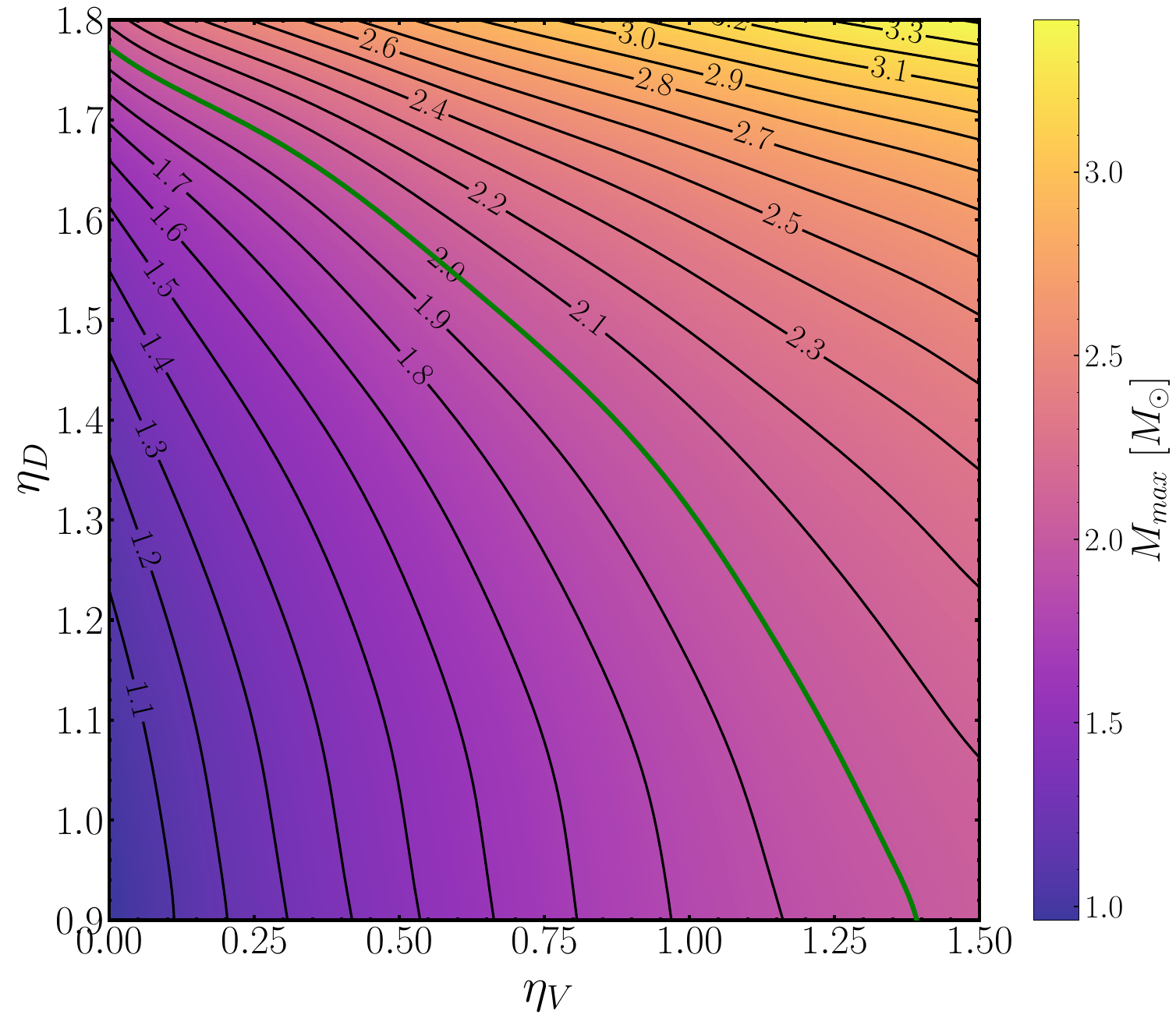}\\
			 	\end{minipage}
     \begin{minipage}[t]{0.49\textwidth}
			 		\includegraphics[width=\textwidth]{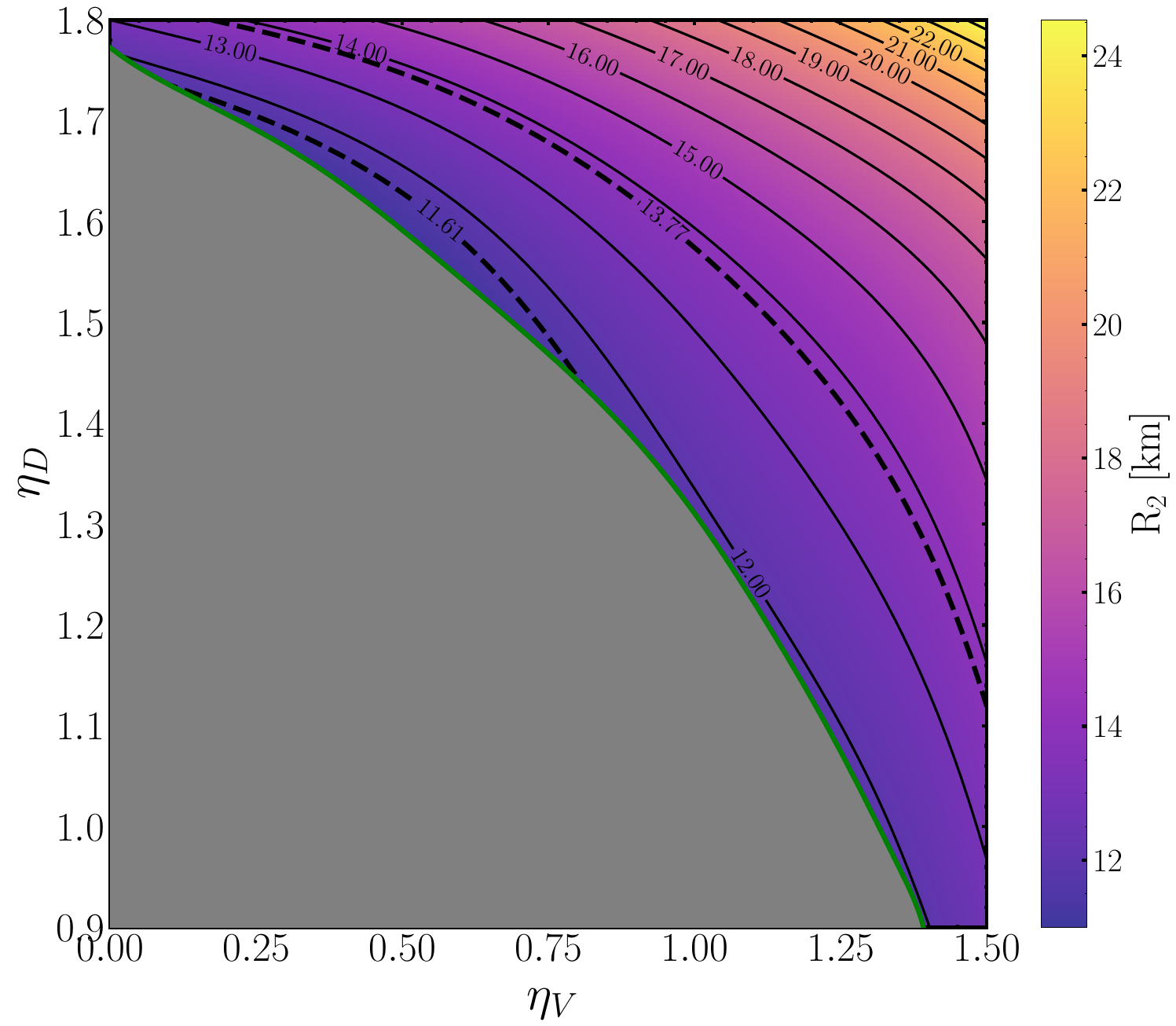}\\
			 	\end{minipage}
	\caption{Left: Maximum mass density plot for the equation of state with $\eta_V\in[0.00,1.50]$ and $\eta_D\in[0.90,1.80]$  and contours of constant mass. The green line shows the contour of EoSs with a maximum mass of 2.0\,$M_{\odot}$. Right: Density plot and contours for the radius at 2 solar masses, $R_{2.0}$. In the gray-shaded region, the maximum mass of the EoS is smaller than 2.0\,$M_{\odot}$. The dashed contours represent the 68\% confidence intervals from the NICER measurements of PSR J0740+6620 \cite{Salmi:2024aum, dittmann2024precisemeasurementradiuspsr}. 
}\label{figMR2}
\end{figure*}

In this section, we put initial constraints on the two-dimensional parameter space spanned by the vector coupling $\eta_V$ and the diquark coupling $\eta_D$. To this end,
we evaluate the equation of state of the RG-consistent NJL model for the whole parameter space between $\eta_V$=0.00 and $\eta_V$ = 1.50, and $\eta_D$ = 0.90 and $\eta_D$ = 1.80. The main constraint on the coupling parameters comes from the fact that the equation of state of dense matter must be able to describe neutron stars of $2\,M_{\odot}$, according to the mass measurements of the massive pulsars J1614-2230 ($M=1.97\pm 0.04~M_\odot$) \cite{Demorest:2010bx}, PSR J0348+0432 ($M=2.01\pm 0.04~M_\odot$) \cite{Antoniadis:2013pzd}, and PSR J0740+6620 ($M=2.08\pm 0.07~M_\odot$) \cite{Fonseca:2021wxt}. This suggests that the EoS must remain sufficiently stiff at high densities.

Measurements of X-rays from the pulsar PSR J0740+6620 by the Neutron Star Interior Composition
Explorer (NICER) and X-ray Multi-Mirror (XMM) constrain the radius to $R = 12.39^{+1.30}_{-0.98}$ km \cite{Riley:2021pdl} and $R = 13.7^{+2.6}_{-1.5}$ km \cite{miller2021}. Recently, these results were updated to $R = 12.49^{+1.28}_{-0.88}$ \cite{Salmi:2024aum}, and  $R = 12.76^{+1.49}_{-1.02}$ \cite{dittmann2024precisemeasurementradiuspsr}.  The radius measurements for the PSR J0030+0451 at 1.4\,$M_{\odot}$ are, $R = 13.02^{+1.24}_{-1.06}$\,km by \citet{Miller_2019a} and $R = 12.71^{+1.14}_{-1.19}$\,km by \citet{Riley_2019}. These two limits also include the recent measurement by \citet{Vinciguerra:2023qxq}. The mass-radius measurement of PSR J0437-4715 to $R = 11.36^{+0.95}_{-0.63}$ km for a mass of $M = 1.418 \pm 0.037~M_{\odot}$ by \citet{Choudhury:2024xbk} favors softer dense matter EoS. In addition, the measurements of the object
HESS J1731-347 with very low mass $M = 0.77^{+0.20}_{-0.17}~M_{\odot}$ and radius $R = 10.4^{+0.86}_{-0.78}$ km \cite{Doroshenko2022} further challenge existing models for the equation of state (see e.g. Refs. \cite{Brodie:2023pjw,Tewari:2024qit,Mariani:2024gqi}).
In addition to the mass and radius constraints from several measurements, there is the dimensionless tidal deformability measurement at 1.4\,$M_{\odot}$ to $\Lambda$ = 190$^{+390}_{-120}$ \cite{PhysRevLett.121.161101} from the GW170817 measurement \cite{PhysRevLett.119.161101, LIGOScientific:2017ync, PhysRevLett.121.161101}.

To ensure smoothness and continuity in our results, particularly when generating contour plots and M-R curves, we interpolated between the computed data points. We then fitted analytical functions to our quantities of interest, such as the maximum mass and radius at a given mass. This approach enabled us to produce smooth contours across the parameter space.

Figure \ref{figMR2} shows the maximum mass, $M_{\text{max}}$ (left), and the radius at 2 solar masses, $R_{2.0}$ (right), as a color density plot in the $\eta_V-\eta_D$-plane. Darker (lighter) colors indicate lower (higher) values. Contours of the same maximum mass and $R_{2.0}$ are also shown, respectively.

 All coupling parameters above the 2.0\,$M_{\odot}$ contour (green in Figure \ref{figMR2} left) give a quark matter EoS with a maximum mass greater than 2.0\,$M_{\odot}$. Thus, raising the maximum mass can be achieved either by increasing the diquark coupling, the vector coupling, or both parameters. Even for zero vector coupling, we still obtain  2.0\,$M_{\odot}$ stars at a high diquark coupling constant of $\eta_D$ = 1.80. Higher values of $\eta_V$ and $\eta_D$ produce stars with a maximum mass greater than 3.0\,$M_{\odot}$.

 We use the 2.0\,$M_{\odot}$ constraint to pre-select parameter combinations in the $\eta_D-\eta_V$ plane for the study of hybrid compact stars with color-superconducting cores. In general, the maximum mass of the hybrid star EoS depends not only on the quark matter EoS but also on the hadronic EoS and the nature of the transition. However, as we are going to show in Sec.~\ref{sec:hybrid}, the maximum mass of the hybrid EoS is not larger than the maximum mass of the quark EoS, as long as the hadron-quark transition happens at sufficiently low densities. As will be seen in Sec.~\ref{sec:hybrid}, even for high-density hadron-quark transitions, the rise in the maximum mass by adding a hadronic envelope to the quark core is expected to be small. Thus we can interpret the green line as a strict lower constraint on the coupling parameters if the hadron-quark transition happens at low densities, whereas in addition, points slightly below the green line can be used to construct massive hybrid stars with a high-density hadron-quark transition.
 
The right plot in Figure \ref{figMR2} represents the change in the radius at 2.0\,$M_{\odot}$  when the $\eta_V$ - $\eta_D$ coupling parameter set is varied across the studied parameter space. The shaded gray area represents the parameter sets where the quark matter EoS does not reach two solar masses, so there are no data points available for that region. The two dashed lines with a value of 11.61 km and 13.77 km represent the 1\,$\sigma$ NICER constraints for PSR J0740+6620, $R = 12.49^{+1.28}_{-0.88}$ km by \citet{Salmi:2024aum} which also includes the recent measurement of $R = 12.76^{+1.49}_{-1.02}$ km by \citet{dittmann2024precisemeasurementradiuspsr} at 2.0\,$M_{\odot}$.

With these constraints, only specific parameter combinations of high $\eta_V$-low $\eta_D$ or low $\eta_V$-high $\eta_D$ allow for quark star configurations that can support a maximum mass greater than 2.0\,$M_{\odot}$ while also satisfying the radius constraints from NICER observations.
Typically, the radius at 2.0\,$M_{\odot}$ is larger in a hybrid equation of state compared to the value of the EoS with only quark matter, provided that the hadron-quark phase transition happens before the hybrid EoS reaches two solar masses, such that all these stars already contain a quark core. Thus the contour of 13.77\,km in the right plot of Figure~\ref{figMR2} provides an upper bound for $\eta_D$ and $\eta_V$ for hybrid EoSs that contain a quark matter core already at $2\,M_{\odot}$, or earlier. Together with the $2\,M_{\odot}$ constraint, this leads to a band of allowed parameters in the $\eta_V-\eta_D$ plane for these kinds of EoSs.

\begin{figure}
\centering
	\includegraphics[scale=0.34]{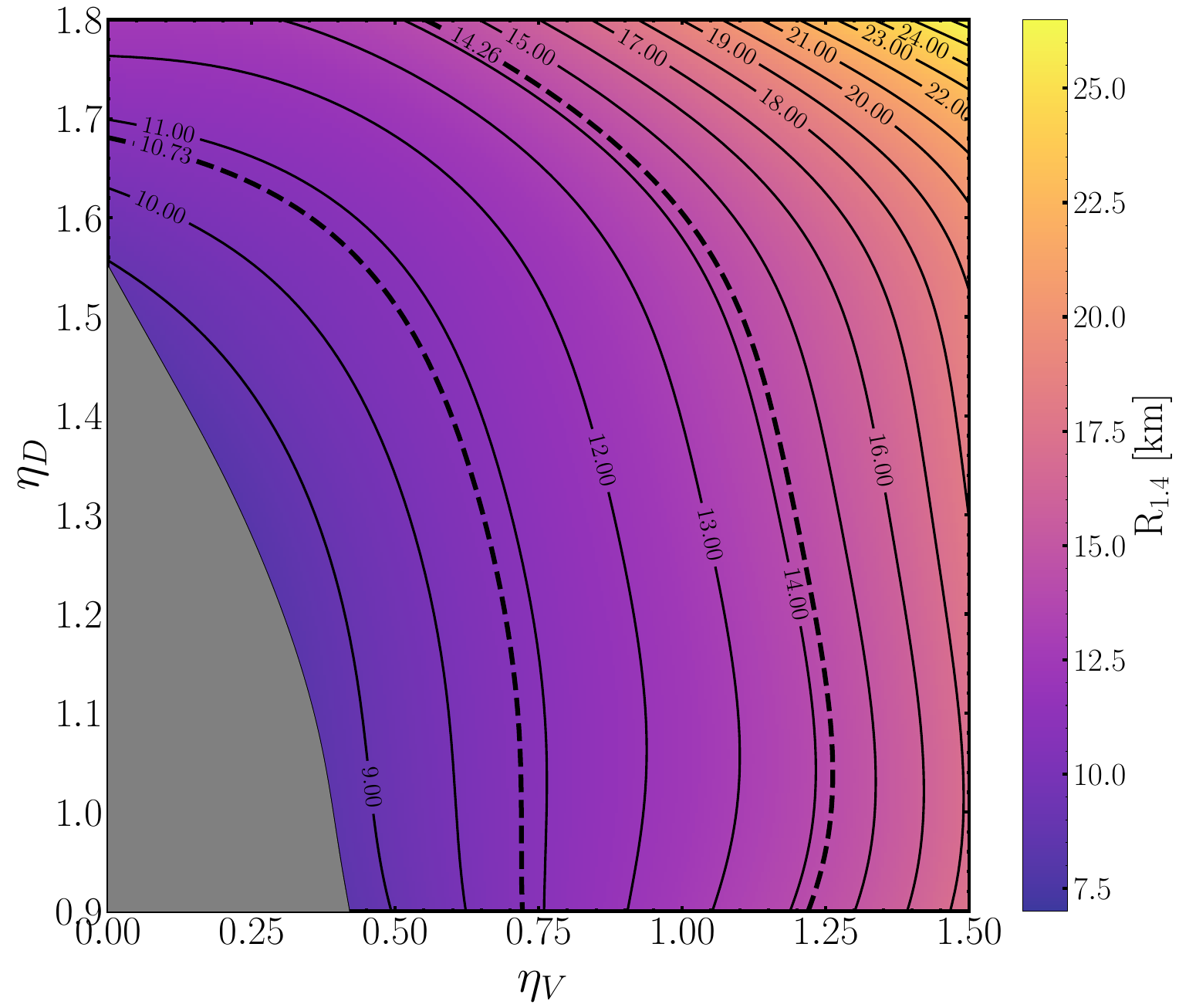}
	\caption{Density plot for the radius at 1.4\,$M_{\odot}$ for $\eta_V\in[0.00,1.50]$ and $\eta_D\in[0.90,1.80]$ with contours of constant radii. The dashed lines represent the 68\% confidence interval of the NICER constraints for PSR J0030+0451 \cite{Miller_2019a, Riley_2019}, and PSR J0437-4715 \cite{Choudhury:2024xbk} at 1.4\,$M_{\odot}$. The gray-shaded region represents the $\eta_V$-$\eta_D$ parameter combinations where the maximum mass is less than 1.4\,$M_{\odot}$. }
	\label{figRL14} 
\end{figure}

Figure~\ref{figRL14} shows the density plot with contours for the radius at 1.4\,$M_{\odot}$. The dashed line with a value of 10.73\,km represents the lower limit of the 68\% confidence interval on the radius at 1.4\,$M_{\odot}$ for PSR J0437-4715, $R = 11.36^{+0.95}_{-0.63}\,$ km by \citet{Choudhury:2024xbk} whereas the line with a value of 14.26 represents the upper limit from NICER measurements, $R = 13.02^{+1.24}_{-1.06}$\,km from PSR J0030+0451 by \citet{Miller_2019a}. These two limits also include the NICER constraints from other measurements for 1.4\,$M_{\odot}$ as well \cite{Riley_2019, Vinciguerra:2023qxq}. 
The gray-shaded region represents the $\eta_V$-$\eta_D$ parameter combinations where the maximum mass lies below 1.4\,$M_{\odot}$. Again, we argue that for the same coupling parameters in a hybrid star construction, the radius of the hybrid EoS at 1.4$\,M_{\odot}$ would be larger than the $R_{1.4}$ of the pure quark matter EoS, provided that the hadron-quark transition takes place already in stars with $1.4\,M_{\odot}$. In this way, the $R_{1.4}=14\,$km contour in Figure~\ref{figRL14} is an upper bound to the $\eta_D-\eta_V$-values for hybrid EoSs with a low-density hadron-quark transition (below $1.4\,M_{\odot}$) and thus for hybrid EoSs with a large quark core.

\begin{figure*}
\centering
\begin{minipage}[t]{0.49\textwidth}
			 		\includegraphics[width=\textwidth]{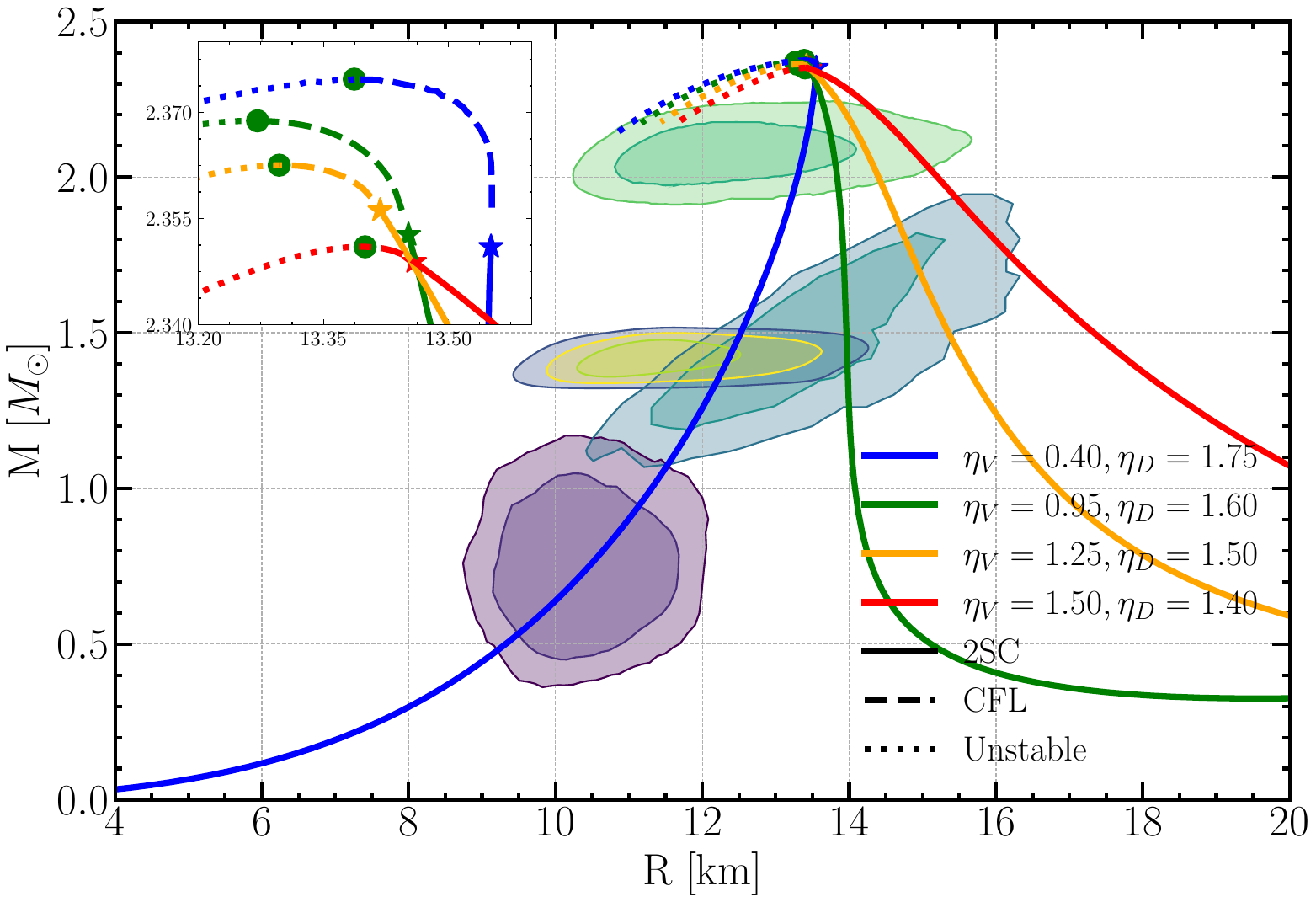}\\
			 	\end{minipage}
     \begin{minipage}[t]{0.49\textwidth}
			 		\includegraphics[width=\textwidth]{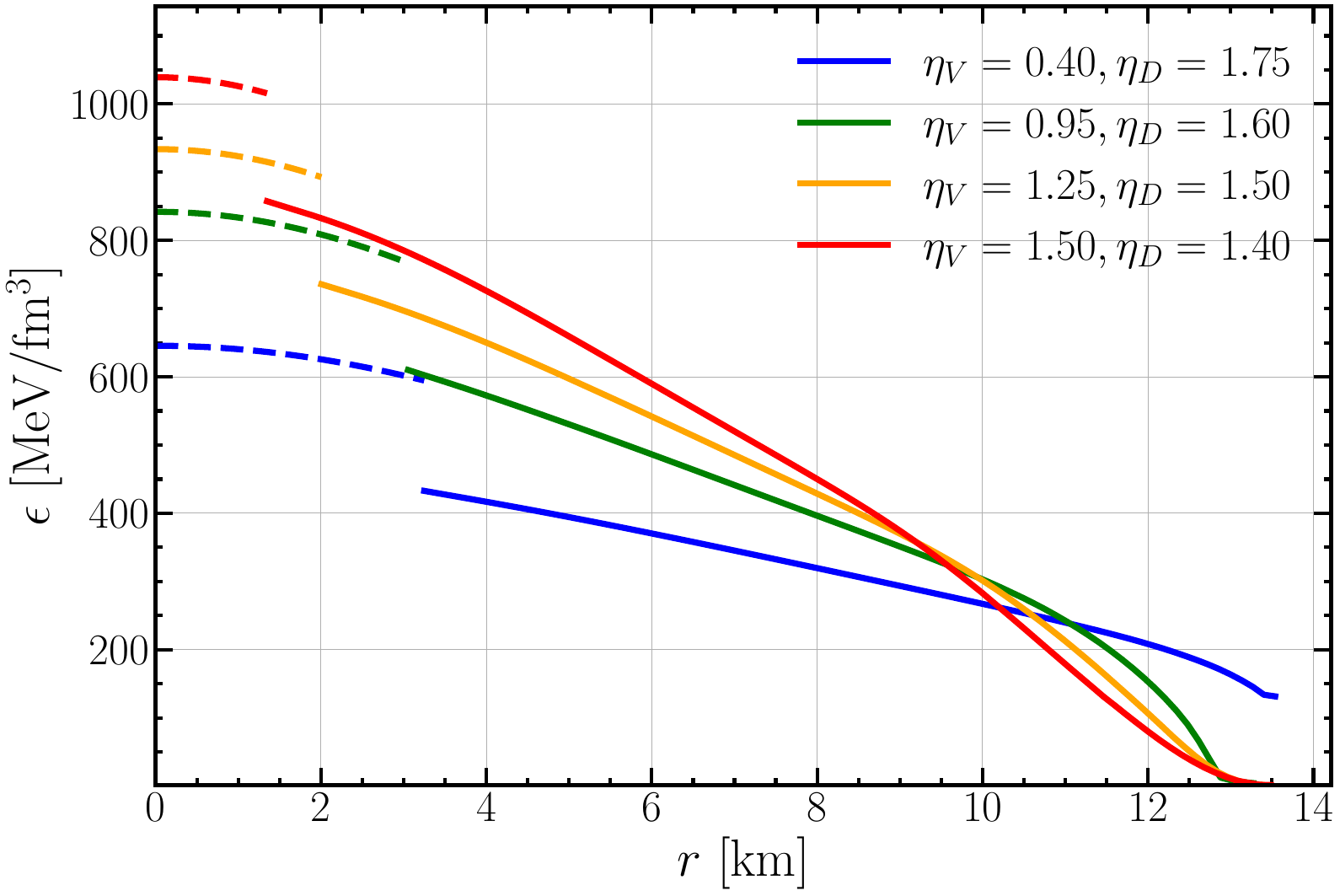}\\
			 	\end{minipage}	\caption{Mass-radius relation (left) and energy-density radial profiles of the corresponding maximum-mass stars (right) for combinations of $\eta_D$-$\eta_V$ parameters with an approximately equal maximum mass of 2.37\,$M_{\odot}$. Solid lines correspond to the 2SC phase whereas the dashed lines represent the CFL phase. The same astrophysical constraints as in Figure \ref{figmr} are depicted in addition.}
	\label{figConfig} 
\end{figure*}

In Figure~\ref{figConfig} we analyze quark stars for EoSs with several different combinations of the $\eta_V$-$\eta_D$ coupling parameters that have approximately the same maximum mass of $M_{\text{max}}\sim2.37\,M_{\odot}$. The mass-radius profiles are shown in the left plot. The various shaded regions are the same credibility regions for mass and radius as in Figure \ref{figmr}. The energy density radial profiles of the maximum-mass stars for the same $\eta_V$-$\eta_D$ coupling combinations are pictured in the right plot.

The inset in the left plot shows the zoomed-in region of the mass-radius relation around the maximum mass.  The solid line in each MR curve represents the quark-star configurations which are entirely in the 2SC phase, with the star symbol marking the end of these pure 2SC phase configurations. The region from the star symbol to the solid dot (dashed line) marks the configurations with a CFL core, surrounded by a 2SC mantle. The green dot corresponds to the last stable point in the curve and hence represents the maximum mass. The dotted line after the solid green dot represents the unstable region.

For a low vector coupling of $\eta_V=0.40$ and a high diquark coupling of $\eta_D=1.75$ the extension of the CFL core is more than $3\,$km. With increasing the vector coupling and decreasing the diquark coupling, the radial extension of the CFL core shrinks below $2\,$km.

At the same time, the mass-radius relations shift from being self-bound to gravitationally
bound configurations. This is also visible in the radial profiles, where, as mentioned earlier, gravitationally bound configurations correspond to EoSs for which the energy density goes to zero at the surface, whereas for self-bound configurations, the energy density stays non-zero.

In the mass-radius plot, the radius at 2.0\,$M_{\odot}$ increases by about 2$\,$km when increasing the vector coupling and decreasing the diquark coupling parameter in this way. The radius at 1.4\,$M_{\odot}$ is much more affected by a change of the coupling constants, changing from around 12.5 km for ($\eta_V$, $\eta_D$) = (0.40, 1.75) to about 17.5 km for ($\eta_V$, $\eta_D$) = (1.50, 1.40).

Notably, the maximum mass for these coupling combinations is closer by less than 0.01\,$M_{\odot}$, whereas the maximum energy density, the radial extension of the CFL phase, and the origin of stability (self-bound vs. gravitationally bound configurations) vary between the parameters.

\begin{figure}
\centering
	\includegraphics[scale=0.35]{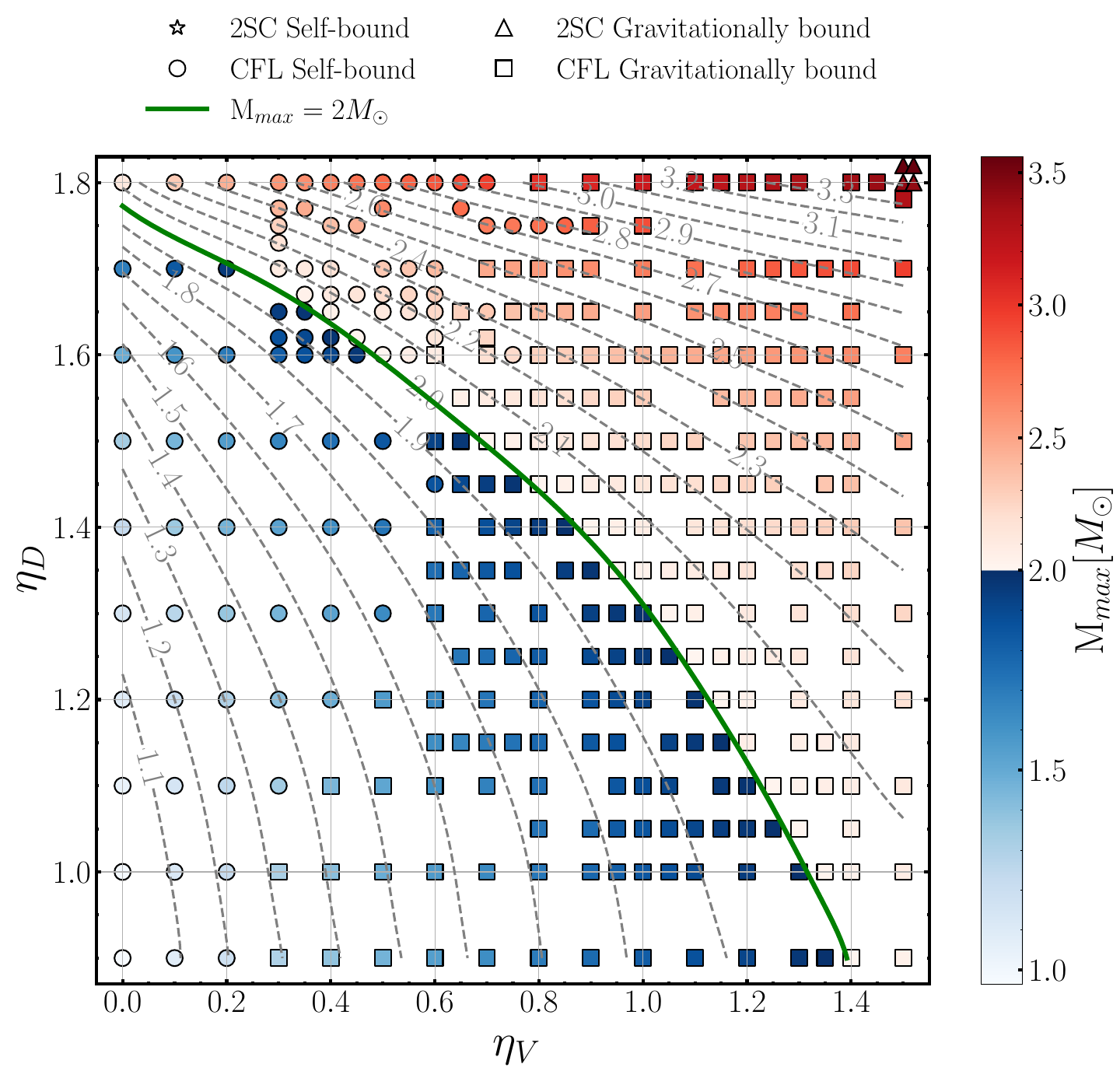}
	\caption{Properties of quark stars across the entire $\eta_V$-$\eta_D$ parameter space studied. 
 The color bar indicates the maximum mass reached at each specific point in parameter space. Gray dashed lines are the contour lines of maximum mass. EoSs with couplings above or right of the green contour reach a maximum mass of at least 2.0\,$M_{\odot}$.
 The circle (square) symbols represent self-bound (gravitationally bound) mass-radius relations with a stable CFL core in the maximum mass configuration. 
 The upward triangle symbols represent gravitationally bound mass-radius relations with a 2SC core in the maximum mass configuration. We do not find self-bound mass-radius relations with a 2SC core in the maximum mass configuration.
 }

	\label{figRegion} 
\end{figure}

Figure \ref{figRegion} shows the characteristic features of compact star configurations, the maximum mass, the type of mass-radius relation, and the composition for the whole $\eta_V$-$\eta_D$ parameter space studied.
The maximum masses are delineated as a contour plot, added with a color bar for the individual choices of the coupling constants. 
The gray dashed lines correspond to the contour lines of maximum mass with values from 1.1 to 3.4\,$M_{\odot}$. Equations of state with couplings above or right of the green contour reach a maximum mass of at least 2.0\,$M_{\odot}$. Different symbols represent different MR configurations obtained throughout the whole parameter space. The circle (square) symbols represent self-bound (gravitationally bound) mass-radius relations with a stable CFL core in the maximum mass configuration. 

For low values of the vector coupling, including $\eta_V$ = 0.00, and the whole diquark coupling range, we find self-bound stars with CFL cores in the maximum-mass configurations.

For a fixed value of $\eta_V$, the amount of CFL present in the maximum mass configuration decreases with increasing $\eta_D$. For $\eta_D$ = 1.80 and $\eta_V\lesssim 1.50$ the maximum-mass configuration still contains a small CFL core, only for $\eta_D$ = 1.80 and $\eta_V\lesssim 1.50$ we find that the maximum-mass configuration is pure 2SC.
With increasing the $\eta_V$ parameter, the MR relation changes from self-bound to gravitationally bound. This is because the vector interaction is repulsive and thus weakens the binding.
For intermediate values of the vector coupling, we therefore observe that the maximum-mass stars still have CFL cores but are gravitationally bound.

For large vector coupling, maximum mass configurations are possible with small CFL cores, but for very high vector and high diquark coupling, we only obtain pure 2SC maximum mass configurations. The coupling combination ($\eta_V$, $\eta_D$) = (1.50, 1.80) represents a 2SC gravitationally bound configuration, meaning that the maximum-mass star entirely consists of quark matter in the 2SC phase. That is, before a phase transition to the CFL phase occurs in the center, the star becomes unstable. This can be explained by the fact that, as we saw in Figure \ref{figeos}, with higher diquark values, the maximum energy density reached for the stable configurations gets lower. We also observed that with higher vector coupling values, the maximum configuration densities get closer to the CFL phase transition point (see Figure~\ref{radial}). Having both effects simultaneously can lead to the last stable star being a pure 2SC configuration.
We do not find any self-bound equations of state that have the maximum mass in the 2SC phase stars within our explored parameter range.

For all the parameters we scanned, either the maximum mass configurations become unstable with a 2SC core, without reaching the CFL phase transition point, or the configuration is stable with a CFL core. More precisely, we find that the transition from 2SC to CFL does not make the star unstable. This will be analyzed in more details in Sec.~\ref{sec:cflstability}.

Figure \ref{figRegion} allows selecting coupling parameters to study equations of state with a desired maximum mass and composition. As we argued in the discussion of Figure \ref{figMR2} and are going to show in more detail in Sec.~\ref{sec:hybrid}, the maximum masses in Figure \ref{figRegion} already provide a good estimate for the maximum mass in a hybrid construction. The core composition of the resulting hybrid stars depends further on whether the hadron-quark phase transition is a hadron-2SC transition or a hadron-CFL transition.

\begin{figure*}
\centering
\begin{minipage}[t]{0.49\textwidth}
			 		\includegraphics[width=\textwidth]{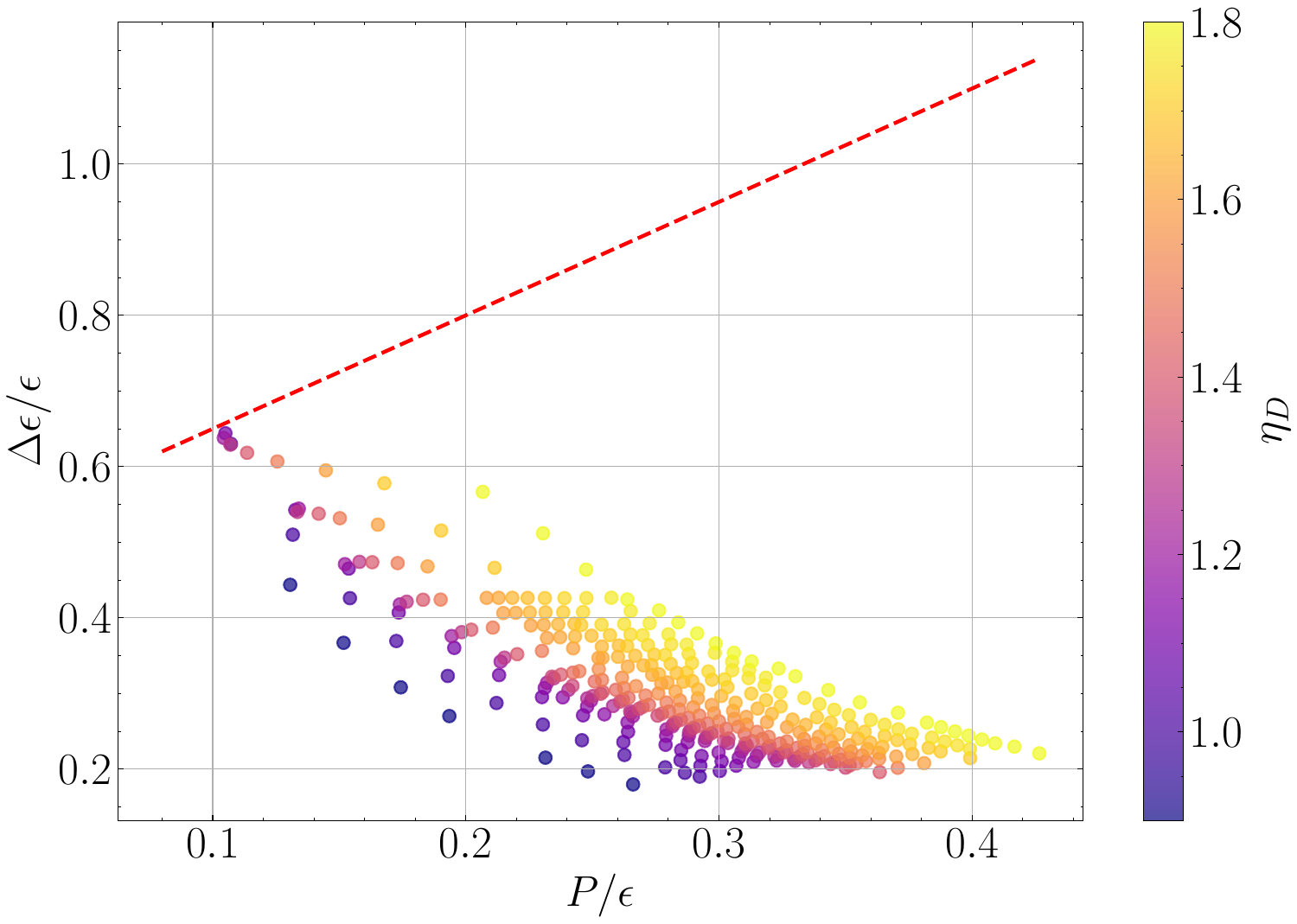}
			 	\end{minipage}
\begin{minipage}[t]{0.49\textwidth}
			 		\includegraphics[width=\textwidth]{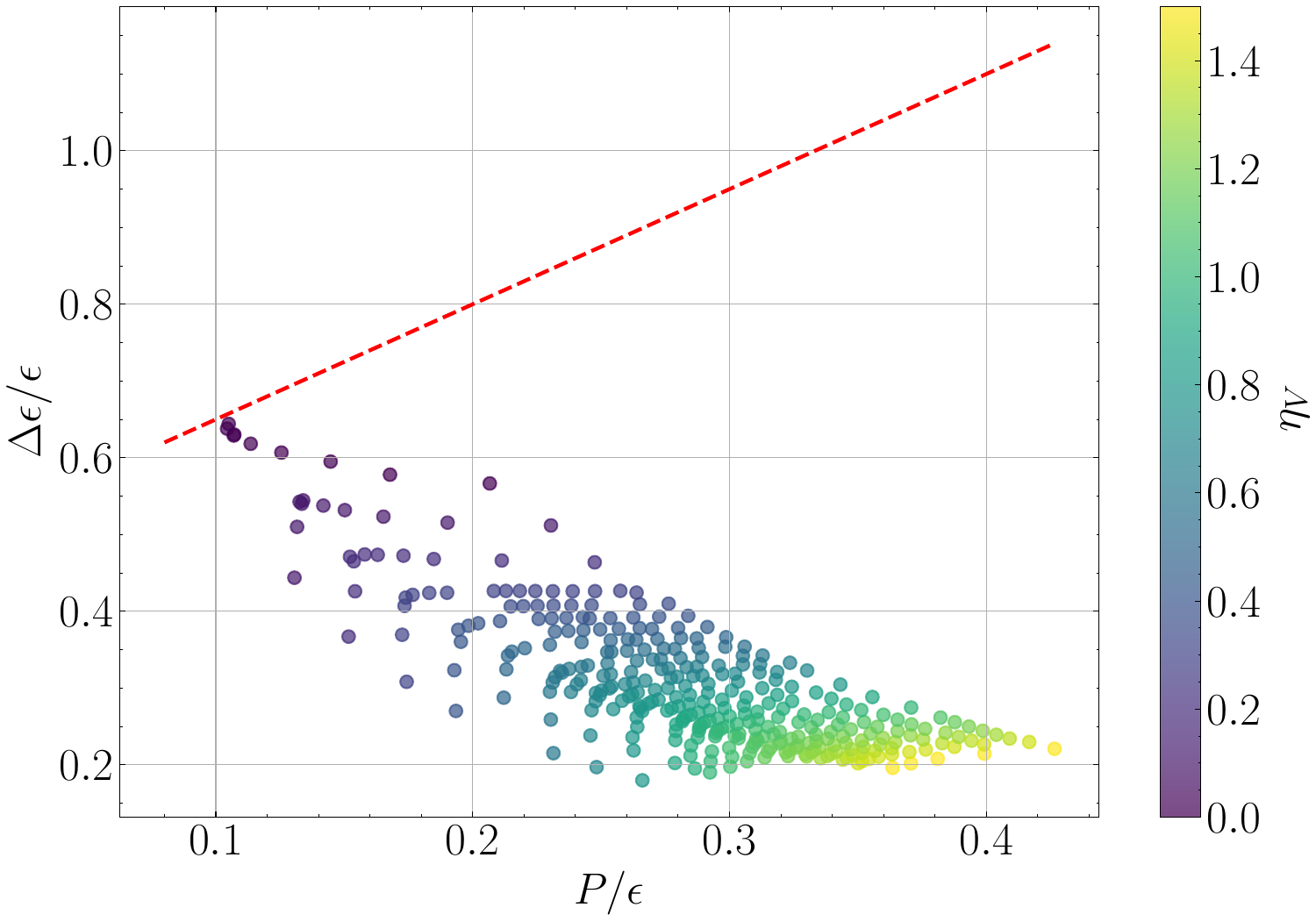}
			 	\end{minipage}
	\caption{Relation between the relative change in energy density $\Delta \epsilon / \epsilon$ and the pressure-to-energy density ratio ($ P/\epsilon$) across 
the 2SC$\to$CFL phase transition for
 all equations of states examined in this study that have such a phase transition in the stable branch. The red dashed line represents the Seidov limit, Eq.~\eqref{eq:Seidov}. 
 Left plot: color shading according to different diquark coupling constants $\eta_D$. Right plot: color shading by varying the vector coupling constant $\eta_V$.}
	\label{figSeidov} 
\end{figure*}

\subsection{Stability of the 
star under 2SC-CFL phase transitions
}\label{sec:cflstability}

In Figure~\ref{figRegion}, the triangle points associated with gravitationally bound 2SC-type equations of state indicate scenarios where the stellar configurations with 2SC core become unstable before the central pressure reaches the CFL transition point.
All other points represent cases where the transition from the 2SC to the CFL phase can be realized in the stable branch. This implies that the 2SC$\to$CFL phase transition does not render the star unstable.

Figure~\ref{figSeidov} further supports this observation by showing the relationship between the relative change in energy density ($\Delta \epsilon / \epsilon$) and the pressure-to-energy density ratio ($P / \epsilon$) 
at the first-order 2SC$\to$CFL phase transition for all EoSs exhibiting such a phase transition in the stable branch.
Here, both in $\Delta \epsilon / \epsilon$ and $P / \epsilon$, $\epsilon$ refers to the energy density immediately before the phase transition, i.e., on the 2SC-side of the EoS.
The red dashed line in Figure~\ref{figSeidov} represents the Seidov limit, defined by: 
\begin{equation}
\label{eq:Seidov}
\frac{\Delta \epsilon}{\epsilon} = \frac{3}{2} + \frac{1}{2} \left( \frac{P}{\epsilon} \right),
\end{equation}
as given in Refs.~\cite{1971SvA....15..347S, Kampfer:1981zmq}. This limit serves as a criterion for gravitational stability during a first-order phase transition in a neutron star: if the fractional increase in energy density $\left( \Delta \epsilon / \epsilon \right)$ exceeds this limit, the star cannot maintain hydrostatic equilibrium and becomes gravitationally unstable right after the phase transition. We find that none of the cases studied here is lying above the Seidov limit,
thereby leading to continuously stable configurations along the mass-radius curve at the phase transition point.  The most critical ones lying close to the Seidov limit are the ones with a low vector coupling constant. The higher the vector coupling constant the lower the jump in the relative energy density and the higher the critical pressure-to-energy density ratio leading to configurations further away from the Seidov limit. The dependence on the diquark coupling constant
is more complicated: higher values of the diquark coupling constant increase both the relative jump in the energy density and the critical pressure-to-energy density ratio such that
at low diquark coupling values, increasing the diquark coupling value leads to configurations that get closer to the Seidov limit without exceeding it and at higher diquark coupling values, the configurations diverge from the limit.

An important factor affecting these results would be the introduction of an additional bag constant. Although the NJL model generates a non-zero bag pressure dynamically by chiral-symmetry breaking in a vacuum, which in Eq.~\eqref{eqvacbag} we have accounted for by setting the pressure of the non-trivial vacuum equal to zero, it has been argued that other effects, like confinement, could modify this value \cite{Bonanno:2011ch}. This holds particularly for hybrid-star EoSs, where there is no reason to set the vacuum pressure of the quark EoS equal to zero. 
We could thus go away from this special choice by introducing an additional bag constant $B$. 

In the following, we briefly discuss the impact of an additional bag constant on the stability. With $B\neq 0$, the pressure and energy density change as: 
\begin{equation*}
P \rightarrow P - B, \quad \epsilon \rightarrow \epsilon + B. 
\end{equation*}
This means that the pressure $P$ and energy density $\epsilon$ change due to the bag constant, affecting the ratios $P / \epsilon$ and $\Delta \epsilon / \epsilon$. Specifically, assuming $B>0$, the pressure-to-energy density ratio $P / \epsilon$ decreases because $P$ decreases while $\epsilon$ increases, moving the points leftward on the horizontal axis of Figure~\ref{figSeidov}. Simultaneously, the relative change in energy density $\Delta \epsilon / \epsilon$ decreases because, while $\Delta \epsilon$ remains constant, $\epsilon$ increases by $B$, shifting points downward on the vertical axis.

We can analyze these changes quantitatively by expanding the expressions in powers of $B/\epsilon$: 
\begin{eqnarray} 
\frac{P}{\epsilon} &\rightarrow& \frac{P}{\epsilon} - \frac{B}{\epsilon} 
-\frac{PB}{\epsilon^2}
+ \mathcal{O}\left( \left( \frac{B}{\epsilon} \right)^2 \right), \\ \frac{\Delta \epsilon}{\epsilon} &\rightarrow& \frac{\Delta \epsilon}{\epsilon} - \frac{B}{\epsilon} \cdot \frac{\Delta \epsilon}{\epsilon} + \mathcal{O}\left( \left( \frac{B}{\epsilon} \right)^2 \right).
\end{eqnarray} 
As we can see, both ratios decrease by terms proportional to $B/\epsilon$. However, the change in $P / \epsilon$ is a direct subtraction of $B / \epsilon$, whereas the change in $\Delta \epsilon / \epsilon$ is proportional to $(B / \epsilon) \cdot (\Delta \epsilon / \epsilon)$. 
Since $\Delta \epsilon / \epsilon$ is less than one in our cases studied near the Seidov limit, 
the fractional change in $\Delta \epsilon / \epsilon$ is smaller than that in $P / \epsilon$.
As a result, the points on the Seidov plot shift more leftward than downward, potentially moving above the Seidov limit line and leading to unstable configurations. This shift implies that even EoSs that were marginally stable without the bag pressure may become unstable when a positive bag pressure is introduced. Conversely, having a negative value for the bag pressure would increase both $P / \epsilon$ and $\Delta \epsilon / \epsilon$, shifting points more rightward than upward, thereby making the star even more stable under the phase transition to the CFL phase.
\begin{figure*}
\centering
\begin{minipage}[t]{0.49\textwidth}
			 		\includegraphics[width=\textwidth]{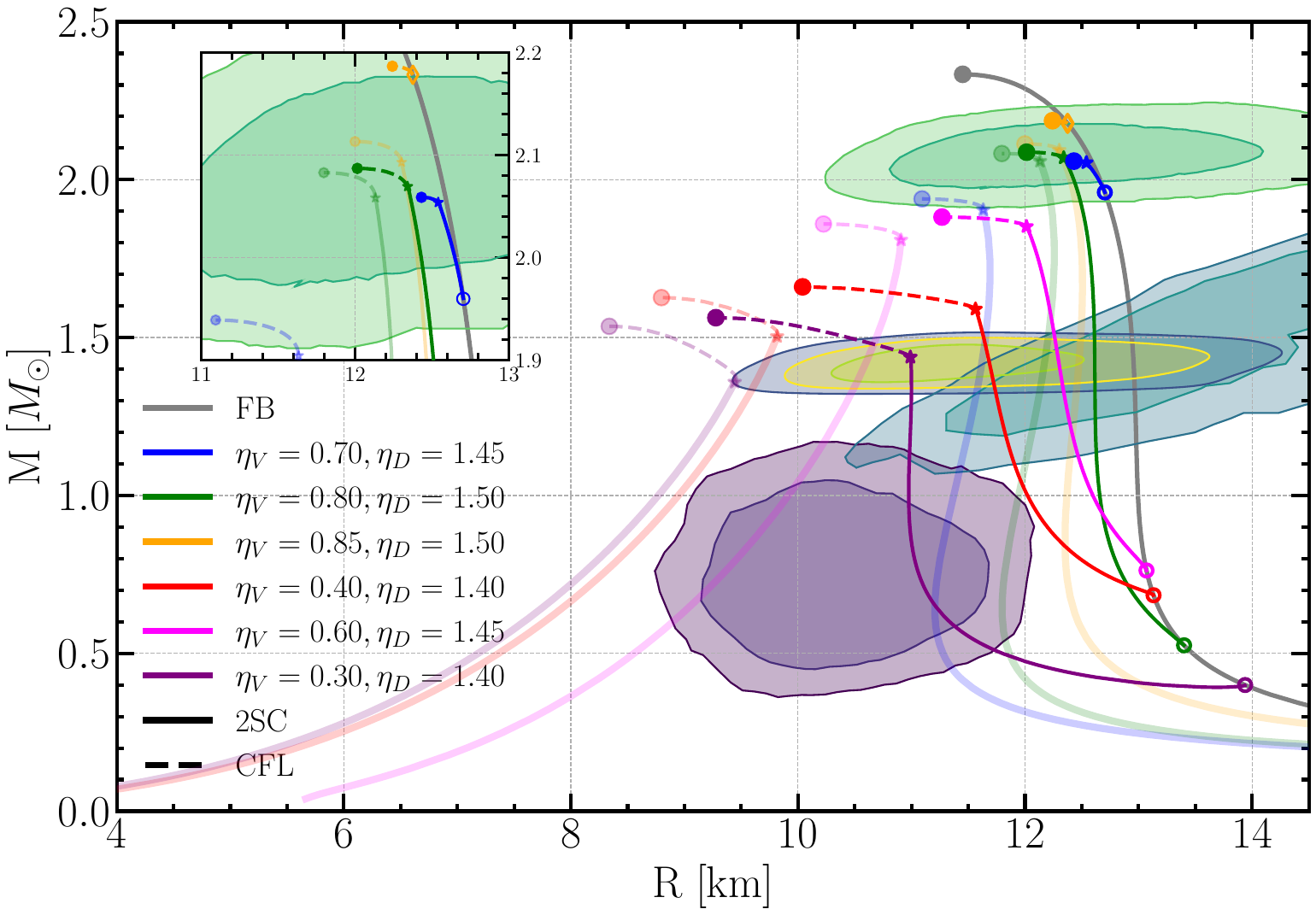}
			 	\end{minipage}
\begin{minipage}[t]{0.49\textwidth}
			 		\includegraphics[width=\textwidth]{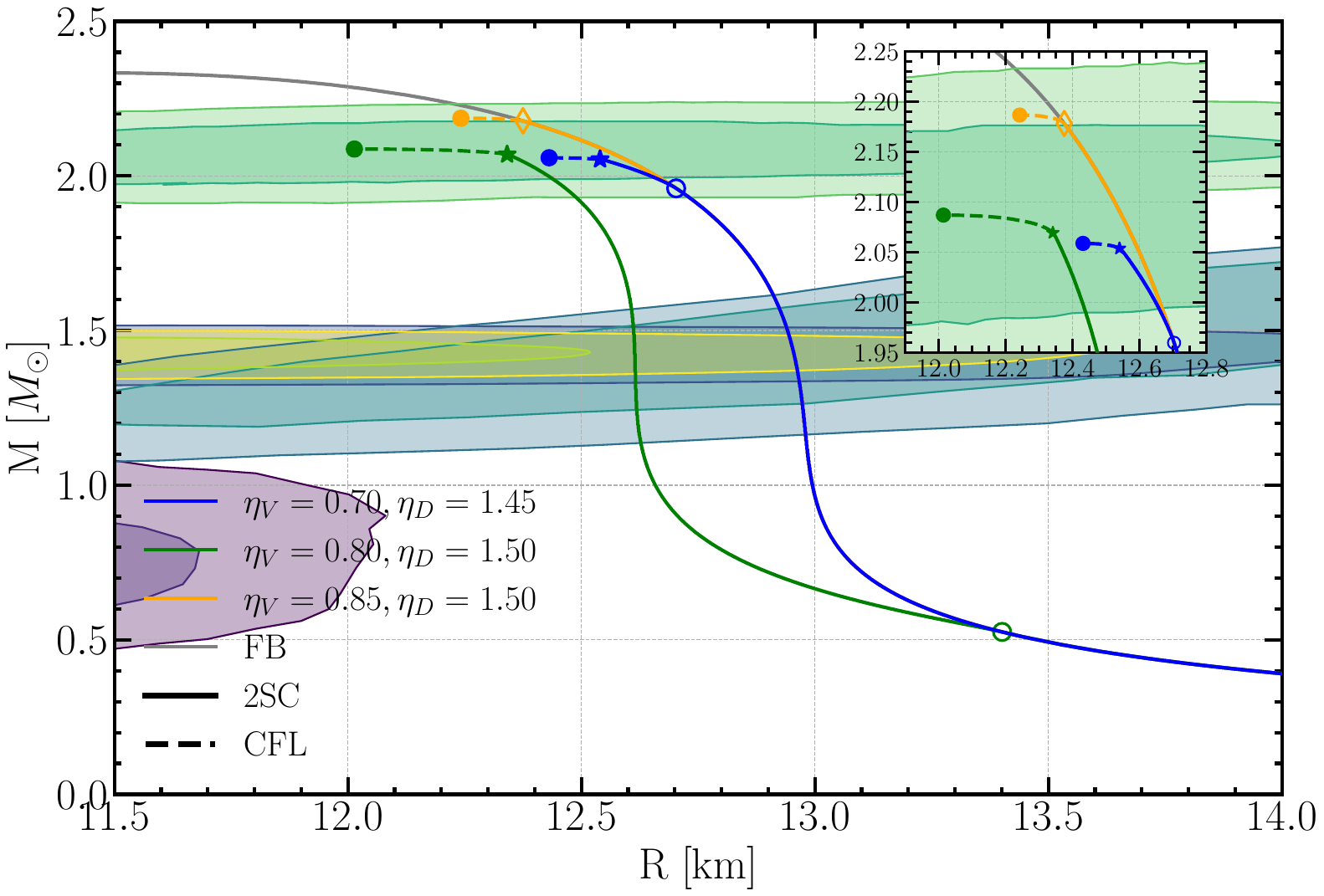}
			 	\end{minipage}
	\caption{Left: Mass-radius relation for the hybrid stars with FB hadronic EoS (in gray) and different $\eta_V$-$\eta_D$ couplings from the NJL model. The pure quark star MR relations are also shown for comparison (light-shaded). Right: Mass-radius relation for hybrid stars, including the hadronic EoS (gray), with a maximum mass more than 2\,$M_{\odot}$.  For the case of hybrid stars, the phase transition from hadronic matter to the 2SC phase is marked by an empty circle, whereas the transition to the CFL phase directly is marked by an empty rhombus. Along the solid line from the empty circle to the star symbol lie configurations with the 2SC phase in the center. Along the dashed line lie configurations with the CFL phase in the center. The filled circle indicates the last stable point. 
 The insets show the zoomed version of the hybrid stars at their maximum mass. The same astrophysical constraints as in Figure \ref{figmr} are shown.}
	\label{fig:hyb} 
\end{figure*}

As a final remark, we note that this analysis applies to the 2SC-CFL transition in hybrid stars for as well. If the 2SC phase is present after matching to a hadronic EoS, the values of pressure and energy density remain the same values across the 2SC$\to$CFL phase transition. Therefore, this analysis remains valid for hybrid constructions, provided that the matching occurs in the 2SC phase and the 2SC phase remains stable until the transition to the CFL phase.
For completeness, we note that this analysis does not exclude the possibility of twin stars. There could be cases where a gravitational instability is reached at a
central density that is slightly higher than the critical density of the phase transition, but at higher densities, the EoS becomes stable again (case B in the classification of \cite{Alford:2013aca}). However, we do not find such cases in the parameter range studied.
In a hybrid star, a first-order hadron-quark phase transition could in principle lead to configurations exceeding the Seidov limit. A thorough investigation of the stability of the hadron-quark phase transition and the possibility of twin star configurations is required and will be addressed in future work.

\section{Hybrid construction}
\label{sec:hybrid}

\begin{table*}[t]
    \centering
    \begin{minipage}{0.47\linewidth}
        \centering
        \begin{tabular}{|c|c|c|c|}
            \hline
            & RG-NJL1 & RG-NJL2 & RG-NJL3\\
            \hline
            ($\eta_V$,$\eta_D$) & (0.70,1.45) & (0.80,1.50) & (0.85,1.50) \\
            \hline
            $M_{\text{max}}$ [$M_{\odot}$] & 1.94 & 2.08 & 2.11 \\
            \hline
            $R_{\text{max}}$ [km] & 11.1 & 11.8 & 12.0 \\
            \hline
            $R_{2.0}$ [km] & - & 10.4 & 10.3 \\
            \hline
            $R_{1.4}$ [km] & 11.6 & 12.2 & 12.5 \\
            \hline
            $\Lambda_{1.4}$ & 430 & 432 & 477 \\
            \hline
            ${c_s^{2}}_c^*$ & 0.52 & 0.53 & 0.53 \\
            \hline
            ${c_s^{2}}^*_{\text{max, 2SC}}$ & 0.50 & 0.52 & 0.53 \\
            \hline
            $P_c^*$ [MeV/fm$^3$] & 321 & 272 & 266 \\
            \hline
            $n_c^*$ [$n_\text{sat}$] & 6.03 & 5.27 & 5.28 \\
            \hline
            $\Delta r_{2SC}^*$ [km] & 4.60 & 3.80 & 3.85 \\
            \hline
            $\Delta r_{CFL}^*$ [km] & 6.49 & 8.00 & 8.15 \\
            \hline
            $M_{2SC}^*$ [$M_{\odot}$] & 1.59 & 1.89 & 1.92 \\
            \hline
            $M_{CFL}^*$ [$M_{\odot}$] & 0.35 & 0.19 & 0.19 \\
            \hline
            $\Delta\epsilon$ [MeV/fm$^3$] & 203 & 188 & 183 \\
            \hline
            $P_T$ [MeV/fm$^3$] & 171 & 181 & 184 \\
            \hline
            $n_T$ [$n_{\text{sat}}$] & 3.8-4.8 & 3.6-4.6 & 3.6-4.5 \\
            \hline
            ${\Delta_{1,2}^*}_{c}$ [MeV] & 185 & 186 & 185 \\
            \hline
            ${\Delta_{3}^*}_{c}$ [MeV] & 202 & 207 & 206 \\
            \hline
        \end{tabular}
         \caption{Properties of pure quark stars at different ($\eta_V$,$\eta_D$) combinations: maximum mass ($M_{\text{max}}$), radius at maximum mass ($R_{\text{max}}$), at 2.0\,$M_{\odot}$ ($R_{2.0}$), and at 1.4\,$M_{\odot}$ ($R_{1.4}$). Dimensionless tidal deformability at 1.4\,$M_{\odot}$ ($\Lambda_{1.4}$), central speed of sound squared ($c_{sc}^{2*}$), maximum speed of sound squared in the 2SC phase (${c_s^{2}}^*_{\text{max, 2SC}}$), central pressure ($P_c^*$,) central number density ($n_c^*$), radial extension of the 2SC phase ($\Delta r_{2SC}^*$) and the CFL phase ($\Delta r_{CFL}^*$), mass of the 2SC ($M_{2SC}^*$) and the CFL phase ($M_{CFL}^*$), the latent heat $\Delta\epsilon$, 2SC to CFL transition pressure, $P_T$ and transition density range, $n_T$, CSC gap values, ${\Delta_{1,2,3}^*}_{c}$. The starred symbols correspond to the values at the maximum mass configuration. 
         }

\label{tab:eos-CSC}
\end{minipage}\hfill
\begin{minipage}{0.47\linewidth}
    \centering
    \begin{tabular}{|c|c|c|c|a|}
         \hline
       FB + & RG-NJL1 & RG-NJL2 & RG-NJL3 & FB\\
        \hline
        ($\eta_V$,$\eta_D$) & (0.70,1.45) & (0.80,1.50) & (0.85,1.50) & - \\
        \hline
        $M_{\text{max}}$ [$M_{\odot}$] & 2.06 & 2.09 & 2.19 & 2.33 \\
        \hline
        $R_{\text{max}}$ [km] & 12.4 & 12.0 & 12.2 & 11.4 \\
        \hline
        $R_{2.0}$ [km] & 11.2 & 12.4 & 12.7 & 12.7 \\
        \hline
        $R_{1.4}$  [km] & 13.0 & 12.6 & 13.0 & 13.0 \\
        \hline
        $\Lambda_{1.4}$ & 570 & 472 & 570 & 570 \\
        \hline
        ${c_s^2}^{*}_c$ & 0.48 & 0.52 & 0.53 & 0.77 \\
        \hline
        ${c_s^{2}}^*_{\text{max, 2SC}}$ & 0.50 & 0.52 & - &-\\
            \hline
            ${c_s^{2}}^*_{\text{max, had}}$ & 0.58 & 0.13 & 0.66 & 0.77 \\
            \hline
        $P_c^*$ [MeV/fm$^{3}$] & 202 & 266 & 267 & 578 \\
        \hline
            $n_c^*$ [n$_{sat}$] & 5.0 & 5.2 & 5.3 & 5.8\\
        \hline
        
        $\Delta r_{2SC}^*$ [km] & 1.92 & 6.61 & - & - \\
        \hline
        $\Delta r_{CFL}^*$ [km] & 2.38 & 3.85 & 2.67 & - \\
        \hline
        $\Delta r_{\text{had}}^*$ [km] & 8.13 & 1.55 & 9.57 & 11.5 \\
        \hline
        $M_{2SC}^*$ [$M_{\odot}$] & 0.15 & 1.67 & - & - \\
        \hline
        $M_{CFL}^*$ [$M_{\odot}$] & 0.05 & 0.19 & 0.67 & - \\
        \hline
        $M_{\text{had}}^*$ [$M_{\odot}$] & 1.86 & 0.23 & 1.52 & 2.33 \\
        \hline
        $\Delta\epsilon_{\text{had}}$ [MeV/fm$^{3}$] & 32.5 & 22.0 & 204 & - \\
        \hline
        $P_{T,\text{had}}$ [MeV/fm$^{3}$] & 135 & 9.17 & 225 & - \\
        \hline
        $n_{T,\text{had}}$ [$n_{\text{sat}}$] & 3.3-3.5 & 1.4-1.6 & 4.0-4.8 & - \\
        \hline
        ${\Delta_{1,2}^*}_{c}$ [MeV] & 175 & 186 & 185 & - \\
        \hline
        ${\Delta_{3}^*}_{c}$ [MeV] & 196 & 207 & 206 & - \\
        \hline
    \end{tabular}
        \caption{Same as Table \ref{tab:eos-CSC}, but for hybrid stars with properties of a hadronic phase using the FB EoS included: maximum speed of sound squared (${c_s^{2}}^*_{\text{max, had}}$), radial extension ($\Delta r_{\text{had}}^*$), and mass ($M_{\text{had}}^*$). The latent heat $\Delta\epsilon_{\text{had}}$, the pressure $P_{T,\text{had}}$, and the density range $n_{T,\text{had}}$ of the hadron-quark transition are also tabulated. The corresponding quantities for the transition from the 2SC to the CFL phase are given in Table~\ref{tab:eos-CSC} and do not change for the hybrid star configurations. The properties of a pure hadronic star are given in the right column.}
    \label{tab:eos-hyb}
    \end{minipage}
\end{table*}

The preliminary constraints put on the $\eta_D-\eta_V$ parameter space of the RG-consistent NJL model in Sec.~\ref{sec:constraints} are useful to study hybrid stars. To illustrate this, we construct hybrid equations of state with a hadronic EoS at the low-density part from the literature that is consistent with current nuclear and astrophysics constraints, followed by a phase transition to the CSC phases. For the hadronic equation of state, we use the relativistic mean-field (RMF) model \cite{Serot:1984ey, Glendenning:1997wn, PhysRevLett.86.5647, Walecka:1974qa, Mueller:1996pm, Serot:1997xg, Gambhir:1990uyn} which is widely used to describe the interior structure of neutron stars based
on fits to the nuclear properties. We follow the work of \citet{PhysRevC.98.065804}, where the parameter values such as effective mass, slope parameter, and symmetry energy are varied in compliance with the predictions from chiral effective field theory \cite{PhysRevC.94.054307}. For the current study, we use $m^*/m$ = 0.70 for the effective nucleon mass, $L$ = 60 MeV for the slope parameter and $J$ = 32 MeV for the symmetry energy at saturation density $n_{\text{sat}}$ = 0.16 fm$^{-3}$. In coordination with the authors of Ref. \cite{PhysRevC.98.065804}, we abbreviate this hadronic EoS with FB (Frankfurt-Barcelona) in our work.

The phase transition from the hadronic EoS to the color-superconducting quark matter EoS is modeled via a Maxwell construction. This means, the critical chemical potential $\mu_c$ at which the transition takes place is given by the intersection of the hadronic and quark matter $P-\mu$ curves, such that the pressure of quark matter $P_{\text{QM}}$ is higher than the pressure of the hadronic EoS $P_\text{HM}$ beyond $\mu_c$: 
\begin{equation}
    P_\text{QM}(\mu_c)=P_\text{HM}(\mu_c) \text{ and } P_\text{QM}(\mu) \gtrless P_\text{HM}(\mu) \text{ for } \mu \gtrless \mu_c.
\end{equation}
By construction, this leads to a first-order phase transition with a nonzero latent heat. However, depending on the used EoSs, the latent heat can get arbitrarily small in principle.

Figure \ref{fig:hyb} shows hybrid mass-radius profiles for different parameters of the NJL model. The mass-radius relation of the RMF model is displayed in gray. In the left plot, the pure quark matter mass-radius curves are shown in light-shaded color for comparison.  At the point of the phase transition to quark matter, the hybrid profiles deviate from the RMF model curve. The phase transition is either to a 2SC phase (empty circle) or directly to a CFL phase (empty rhombus). The point where the center of the star changes from a 2SC to a CFL core is marked by a star and the last stable point, corresponding to the maximum mass, is marked as a filled circle. If the phase transition to quark matter happens at low densities, the maximum-mass hybrid star contains a substantial portion of the 2SC phase, and the maximum mass in the hybrid configuration and the maximum mass of the pure quark matter profile are very similar while the radius is shifted to larger values for the hybrid construction. If the phase transition happens at a higher density, the hybrid star contains little or no matter in the 2SC phase, and the maximum mass is increased by up to $\sim 0.15\,  M_\odot$. 

This means that the 2.0\,$M_{\odot}$ constraint depicted by the green line in Figure \ref{figRegion} correctly constrains the quark model parameters for a hybrid construction if the hybrid star has a hadronic-matter to 2SC interface and the transition happens at lower densities. For stars without or only a small amount of 2SC phase, two solar masses in the hybrid configuration can be achieved even with parameters that lie close, but below the green line in 
Figure~\ref{figRegion}.

The observed features in the mass-radius relation are an interplay of two effects: The 2SC phase contains ungapped blue and strange quarks, which soften the EoS (see the dip in the speed of sound squared in Figure \ref{figcs}). On the other hand, the hadron-quark transition, as well as the 2SC$\to$CFL transition (if present in the hybrid star), are both accompanied by a latent heat that reduces the integrated (radial) mass and shifts the mass-radius curves to lower radii after the phase transition.

For future studies of color-superconducting phases in hybrid stars, we provide tabulated EoSs for three parameter sets of the RG-consistent NJL model (RG-NJL1, RG-NJL2, and RG-NJL3). 
The three EoSs are chosen representatives for different astrophysical scenarios. All three hybrid EoSs (FB+RG-NJL1, FB+RG-NJL2, and FB+RG-NJL3) are consistent with current astrophysics constraints. The mass-radius relations for pure quark matter and the hybrid star are shown in Figure \ref{fig:hyb} (left). In the right plot, the mass-radius relations for the three representative hybrid EoSs are plotted again for better visibility. Various properties of the representative quark matter EoSs, such as the maximum mass $M_{\text{max}}$, the radius at the maximum mass $R_{\text{max}}$, and many more are tabulated in Table~\ref{tab:eos-CSC}
. The respective data for the hybrid EoSs is shown in Table~\ref{tab:eos-hyb}. This table contains, in addition, the maximum speed of sound squared (${c_s^{2}}^*_{\text{max, had}}$), radial extension $r_\text{had}^\ast$, and mass $M_\text{had}^\ast$ of the hadronic phase in the maximum mass configuration, as well as the latent heat $\Delta\epsilon_\text{had}$, pressure $P_{T,\text{had}}$ and density range $n_{T,\text{had}}$ of the hadron-quark phase transition.

We choose a soft (RG-NJL1, $\eta_V$ = 0.70, $\eta_D$ = 1.45), intermediate (RG-NJL2, $\eta_V$ = 0.80, $\eta_D$ = 1.50) and stiff (RG-NJL3, $\eta_V$ = 0.85, $\eta_D$ = 1.50) equation of state that lead to different interior compositions in the hybrid construction. The FB+RG-NJL1 EoS leads to a hadron-quark phase transition at 3.3\,$n_\text{sat}$. For the maximum-mass configuration, the quark core consists of both a 2SC and a CFL phase. For the FB+RG-NJL2 EoS, the hadron-quark phase transition happens already at 1.4 $n_{\text{sat}}$, resulting in a large 2SC phase around the CFL core in the maximum-mass star.  For the stiff FB+RG-NJL3 EoS, on the contrary, the hadron-quark phase transition happens at a higher density of 4.0\,$n_{\text{sat}}$ and the hybrid star contains no 2SC phase. However, the CFL core of the maximum-mass star is of similar size as for RG-NJL1. The three pure quark EoSs have different maximum masses and radii. In the hybrid star construction, the maximum mass and the radius of the maximum-mass configuration lie closer to each other, with both quantities shifted to larger values compared to the pure quark matter EoS ($M_\text{max}$ between 2.06 and 2.19\,$M_{\odot}$ and $R_{M_{\text{max}}}$ between 12.01 and 12.43\,km). Thus we find hybrid stars with very similar maximum masses but different quark core compositions. Hybrid stars with the hadron-quark phase transition happening at low densities (RG-NJL2) have smaller radii $R_{1.4}$ and lower dimensionless tidal deformabilities $\Lambda_{1.4}$ at 1.4 solar masses. Furthermore, the mass of the hadronic matter around the quark core can be very small in this case ($M_{\text{had}^\ast}$ = 0.23\,$M_{\odot}$ for RG-NJL2). 
For the hybrid stars with RG-NJL1 or RG-NJL3 EoS, the phase transition happens at high density, such that $R_{1.4}$ and $\Lambda_{1.4}$ are given by the corresponding values of the FB model. 

For the examples shown here, the latent heat associated with the phase transition from low-density hadronic matter to quark matter is relatively small for the RG-NJL1 and RG-NJL2 models ($\sim$ 20–30\,MeV). 
For these two cases, the latent heat for the 2SC to CFL transition ($\sim$ 180–200\,MeV) is an order of magnitude higher than the initial transition. For the RG-NJL3 model, the latent heat for the hadronic to quark matter phase transition is substantially higher, ($\sim$ 200\,MeV) due to a direct transition from hadronic matter to the CFL phase without an intermediate 2SC phase.

The speed of sound squared is tabulated in Table~\ref{tab:eos-CSC} and Table~\ref{tab:eos-hyb} both for the center ($c_{sc}^{2\ast}$) and the maximum value reached in the 2SC phase ($c_{s_{\text{max, 2SC}}}^{2\ast}$) of the maximum mass configuration. As all three hybrid EoSs have a CFL core in the maximum mass configuration and the speed of sound in our model increases monotonically in the CFL phase (see Figure~\ref{figcs}), $c_{sc}^{2\ast}$ is also the maximum value in the CFL phase of that configuration. However, the speed of sound in the CFL phase can start lower than its maximum value in the 2SC phase (compare $c_{sc}^{2\ast}$ with $c_{s_{\text{max, 2SC}}}^{2\ast}$ for FB+RG-NJL1 in Table~\ref{tab:eos-hyb} and Figure~\ref{figcs}). In the maximum-mass configuration, we find the maximum speed of sound squared reached in the 2SC and CFL phase to be very similar and about 0.5. This can give insights into which values to use in the synthetic constant speed of sound parametrization for 2SC and CFL quark matter \cite{Alford:2017qgh,Li:2019fqe,Li:2023zty}. The maximum speed of sound in the quark phase can be higher (FB+RG-NJL2) or lower (FB+RG-NJL1) than in the hadronic phase of the star.

We also checked our model against the criteria defined in Ref~\cite{Komoltsev:2021jzg} (see also Refs.~\cite{,Gorda:2022jvk,PhysRevLett.127.162003}) to check its consistency with pQCD calculations at high densities. These criteria involve comparing the pressure difference between our model's EoS and pQCD predictions at a high chemical potential, ensuring this difference falls within bounds determined by causality and thermodynamic stability. A key factor in these bounds is the limiting speed of sound squared, $c^2_{s,\text{lim}}$, which influences the stiffness of the EoS permitted by the criteria. We adopted $c^2_{s,\text{lim}}=0.5$ in our tests, imposing a stricter constraint compared to the most conservative choice of $c^2_{s,\text{lim}}=1$. Our three examples in Table\ref{tab:eos-hyb} satisfy these pQCD consistency conditions up to quark chemical potentials of approximately 680~MeV, corresponding to densities up to $8.8\,n_{\text{sat}}$. Beyond this chemical potential, our model no longer aligns with pQCD predictions under these stricter constraints, suggesting it may not be reliable at higher densities. With the most conservative choice of $c^2_{s,\text{lim}}=1$, these values change to chemical potentials up to $\sim$710 MeV and correspodning densities of $\sim$9.5 $n_{\text{sat}}$.

Ref.~\cite{PhysRevLett.132.262701} examines upper bounds on the CFL diquark gap by combining astrophysical observations with theoretical constraints from chiral effective field theory and pQCD. The study establishes ‘‘reasonable’’ ($c^2_{s,\text{lim}}=0.5$) and ‘‘conservative’’ ($c^2_{s,\text{lim}}=1$) bounds on the gap value at high baryon chemical potentials.
In our calculations, the CFL gaps are not identical; $\Delta_1$ and $\Delta_2$ reach up to approximately 190~MeV, while $\Delta_3$ reaches up to about 210~MeV, as shown in Table \ref{tab:eos-hyb}. To make a meaningful comparison, we take the root-mean-square (rms) average of these gaps, resulting in an rms CFL gap value of 197~MeV. This rms CFL gap is achieved at baryon chemical potentials $\mu_B = 3 \mu$ around 1500~MeV, with the speed of sound squared slightly exceeding 0.5 in the center of the maximum mass configurations.
For a direct comparison to the results of Ref.~\cite{PhysRevLett.132.262701}, we should extend our calculations up to $\mu_B = 2600$~MeV. However as mentioned before, at these chemical potential regions our EoS does not satisfy the pQCD constraints in these three examples and hence our model is not reliable there. Our last reliable chemical potential is $\mu_B\sim 2000$~MeV, with $c_s^2\sim0.62$ and rms CFL gap value of $\sim215$~MeV, which would most probably put us slightly above the ‘‘reasonable’’ bounds and into the ‘‘conservative’’ bounds discussed in that study. 

Matching these three quark EoSs to the FB model is a special choice. However, we checked that this choice of parameters also allows for hybrid stars with other hadronic equations of state commonly used in the literature such as DD2 \cite{Hempel:2009mc,Typel:2009sy} and the QMC-RMF model \cite{Alford:2022bpp,Alford:2023rgp}. We find that in order to achieve hybrid stars with two solar masses and more
with the three parameter sets tabulated in Table~\ref{tab:eos-hyb}, typically hadronic EoSs that are soft at low to intermeditate densities ($\sim 1-3\, n_{\text{sat}}$) and stiff at high densities must be chosen. For other hadronic EoSs, different parameters for the RG-consistent NJL model are a more suitable choice. Our results change further if we vary the bag pressure and thus effectively the transition density of the hadron-quark phase transition. We leave a comprehensive study of possible hybrid star configurations with the RG-consistent NJL model for future work.

\section{Summary and Conclusion}
\label{sec:conclusion}

In this work, we investigated a range of astrophysical properties of compact stars in the context of color superconductivity (CSC) phases using the Nambu--Jona-Lasinio (NJL) model with a Renormalization Group (RG)-consistent treatment. Specifically, we examined how these properties change in response to variations in the vector interaction coupling ($\eta_V$) and the diquark coupling ($\eta_D$) parameters. In order to calculate the properties of the compact star's cores, the NJL model had to be evaluated at chemical potentials that are close to or even larger than the cutoff $\Lambda'=602.3\,$MeV that is typically used in conventional cutoff-regularization. Thanks to the RG-consistent treatment, we could evaluate the model at these chemical potential without cutoff artifacts.

We performed a comprehensive study of the parameter dependence of the equation of state, the color superconducting gaps, the appearance of the 2SC and CFL phases, and the speed of sound.
Our analysis demonstrates that adjusting the diquark and vector couplings significantly influences the stiffness of the EoS and, 
consequently, the mass-radius relationships of the resulting compact stars. 
The larger the diquark or vector coupling constant the stiffer the equation of state and the larger the maximum mass. 
The transition from the 2SC to the CFL phase can happen at densities of several times saturation density. The transition density to the CFL phase is lowered for higher diquark coupling constants but shows small changes for different vector coupling constants. 
The speed of sound squared reaches values above the conformal value $c_s^2=1/3$ for all values of the vector coupling constant.
We find values of up to $c_s^2\sim 0.6$ in the 2SC and in the CFL phase for high vector and diquark coupling constants at high densities reached in the center of the maximum-mass configuration. This is lower than the value of $c_s^2\sim 0.7-1$ typically used for the high-density quark branch in the constant speed of sound model with two successive quark phases \cite{Alford:2017qgh,Li:2019fqe,Li:2023zty}.

The implications for the properties of compact stars with a transition from hadron to quark matter have been investigated in a comprehensive analysis, too. We examined the parameter constraints such as to keep the discussion as general as possible without specifying a hadronic EoS. 
Possible corrections by adding a hadronic equation of state are discussed and exemplified for a specific choice. 
We find that maximum masses of 2.0\,$M_\odot$ and more can be reached for large diquark or high vector coupling constants for the pure quark star configuration
which is rather unspoiled for hybrid star configurations.
The maximum mass constraint alone gives strong limits on the values of the vector and diquark coupling strength.
The radius of pure quark stars stays below the limit from the NICER observation of the heavy pulsar PSR J0740+6620.  We find that the CFL phase can appear
such that it produces a stable configuration in the mass-radius relation within the RG-consistent NJL approach used in this work.
For the maximum-mass configurations, the size of the CFL core can be large, at the order of several kilometers. In almost all cases
studied, the CFL phase appears in the maximum-mass configuration.
We show that the radius of the quark core if present in a light compact star with a mass of e.g.\ 1.4\,$M_\odot$ 
can also be sizable. The cores of these stars are always in the 2SC phase. This amount of 2SC core is substantial enough to reduce the dimensionless tidal deformability at 1.4\,$M_{\odot}$. In a hybrid-star calculation, we find that the quark core radius and mass change only slightly compared to the pure quark star case
which allows us to use the results of the NJL model calculations alone as a reasonable measure of the size of the quark core in a hybrid star 
independent of the chosen hadronic equation of state. 

For the mass-radius results of the pure quark stars, 
we observe that the stellar configurations change from self-bound states, similar to strange stars, 
to configurations that resemble gravitationally bound neutron stars as the parameters are varied \cite{Hanauske:2001nc}. We do not find absolutely stable configurations, i.e., quark matter with a lower energy per baryon number than nuclear matter,
in the whole
parameter range studied. We generally find that the phase transition from the 2SC phase to the CFL phase at the center of the star does not cause instability of the star.
Only for the highest values of the vector coupling and diquark coupling studied,
the star becomes unstable before the phase transition to the CFL phase occurs in the center.
The stability of the star under the 2SC to CFL phase transition is not altered by matching to a hadronic equation of state at lower densities.

We demonstrated explicitly for a chosen low-density equation of state that it is possible 
to construct hybrid stars and that the corresponding mass-radius relations are 
compatible with the astrophysical constraints from pulsar mass measurements, GW170817, and NICER observations.

We provide several $\eta_V$-$\eta_D$ coupling combinations that serve as a starting point for studying hybrid stars with color-superconducting cores. 
Depending on the hadronic equation of state, we expect to find configurations of quark matter cores with only 2SC, only CFL, or 2SC+CFL composition. 
We find these three examples remain consistent with perturbative QCD (pQCD) predictions up to quark chemical potentials of approximately 680\,MeV (710\,MeV with a more conservative speed of sound limit), corresponding to densities up to about $9\,n_{\text{sat}}$. The calculated CFL gap is within acceptable ranges established by combined astrophysical and theoretical constraints, in alignment with the conservative bounds from Ref. \cite{PhysRevLett.132.262701}. The selected set of equations of state for pure quark matter will be made available online for its use in the research 
community to construct hybrid stars with a chosen low-density hadronic equation of state.
For direct future astrophysical applications, the hybrid equations of state
presented in our work will be used to study, e.g., the formation of color-superconducting phases in neutron-star mergers.

In conclusion, our results highlight the impact that variations in the vector and diquark interaction parameters have on the properties of dense quark matter and the astrophysical observables of compact stars. This enables us to set constraints on the existence of CSC phases in neutron stars and emphasize the possible 
importance of incorporating CSC phases into neutron-star models to meet astrophysical constraints and to explore these phases in astrophysical scenarios.

\acknowledgements
The authors acknowledge Aleksi Kurkela for his valuable suggestions and comments.
We thank Jan-Erik Christian for providing us with a table of the RMF hadronic EoS used for the hybrid star calculations. 
H.G. and M.H. thank Martin Steil for his support in numerical calculations.
M.H. is supported by the F\&E program of GSI Helmholtzzentrum für Schwerionenforschung Darmstadt.
I.A.R. acknowledges support from the Alexander von Humboldt Foundation. 
The authors acknowledge support from the Deutsche Forschungsgemeinschaft (DFG, German Research Foundation) 
through the CRC-TR211 'Strong-interaction matter under extreme conditions' project number 315477589 – TRR 211. 

\twocolumngrid

\bibliography{references.bib}

\begin{thebibliography}{118}%
\makeatletter
\providecommand \@ifxundefined [1]{%
 \@ifx{#1\undefined}
}%
\providecommand \@ifnum [1]{%
 \ifnum #1\expandafter \@firstoftwo
 \else \expandafter \@secondoftwo
 \fi
}%
\providecommand \@ifx [1]{%
 \ifx #1\expandafter \@firstoftwo
 \else \expandafter \@secondoftwo
 \fi
}%
\providecommand \natexlab [1]{#1}%
\providecommand \enquote  [1]{``#1''}%
\providecommand \bibnamefont  [1]{#1}%
\providecommand \bibfnamefont [1]{#1}%
\providecommand \citenamefont [1]{#1}%
\providecommand \href@noop [0]{\@secondoftwo}%
\providecommand \href [0]{\begingroup \@sanitize@url \@href}%
\providecommand \@href[1]{\@@startlink{#1}\@@href}%
\providecommand \@@href[1]{\endgroup#1\@@endlink}%
\providecommand \@sanitize@url [0]{\catcode `\\12\catcode `\$12\catcode `\&12\catcode `\#12\catcode `\^12\catcode `\_12\catcode `\%12\relax}%
\providecommand \@@startlink[1]{}%
\providecommand \@@endlink[0]{}%
\providecommand \url  [0]{\begingroup\@sanitize@url \@url }%
\providecommand \@url [1]{\endgroup\@href {#1}{\urlprefix }}%
\providecommand \urlprefix  [0]{URL }%
\providecommand \Eprint [0]{\href }%
\providecommand \doibase [0]{http://dx.doi.org/}%
\providecommand \selectlanguage [0]{\@gobble}%
\providecommand \bibinfo  [0]{\@secondoftwo}%
\providecommand \bibfield  [0]{\@secondoftwo}%
\providecommand \translation [1]{[#1]}%
\providecommand \BibitemOpen [0]{}%
\providecommand \bibitemStop [0]{}%
\providecommand \bibitemNoStop [0]{.\EOS\space}%
\providecommand \EOS [0]{\spacefactor3000\relax}%
\providecommand \BibitemShut  [1]{\csname bibitem#1\endcsname}%
\let\auto@bib@innerbib\@empty
\bibitem [{\citenamefont {Demorest}\ \emph {et~al.}(2010)\citenamefont {Demorest}, \citenamefont {Pennucci}, \citenamefont {Ransom}, \citenamefont {Roberts},\ and\ \citenamefont {Hessels}}]{Demorest:2010bx}%
  \BibitemOpen
  \bibfield  {author} {\bibinfo {author} {\bibfnamefont {P.}~\bibnamefont {Demorest}}, \bibinfo {author} {\bibfnamefont {T.}~\bibnamefont {Pennucci}}, \bibinfo {author} {\bibfnamefont {S.}~\bibnamefont {Ransom}}, \bibinfo {author} {\bibfnamefont {M.}~\bibnamefont {Roberts}}, \ and\ \bibinfo {author} {\bibfnamefont {J.}~\bibnamefont {Hessels}},\ }\href {\doibase 10.1038/nature09466} {\bibfield  {journal} {\bibinfo  {journal} {Nature}\ }\textbf {\bibinfo {volume} {467}},\ \bibinfo {pages} {1081} (\bibinfo {year} {2010})}\BibitemShut {NoStop}%
\bibitem [{\citenamefont {Antoniadis}\ \emph {et~al.}(2013)\citenamefont {Antoniadis} \emph {et~al.}}]{Antoniadis:2013pzd}%
  \BibitemOpen
  \bibfield  {author} {\bibinfo {author} {\bibfnamefont {J.}~\bibnamefont {Antoniadis}} \emph {et~al.},\ }\href {\doibase 10.1126/science.1233232} {\bibfield  {journal} {\bibinfo  {journal} {Science}\ }\textbf {\bibinfo {volume} {340}},\ \bibinfo {pages} {6131} (\bibinfo {year} {2013})}\BibitemShut {NoStop}%
\bibitem [{\citenamefont {Fonseca}\ \emph {et~al.}(2021)\citenamefont {Fonseca} \emph {et~al.}}]{Fonseca:2021wxt}%
  \BibitemOpen
  \bibfield  {author} {\bibinfo {author} {\bibfnamefont {E.}~\bibnamefont {Fonseca}} \emph {et~al.},\ }\href {\doibase 10.3847/2041-8213/ac03b8} {\bibfield  {journal} {\bibinfo  {journal} {Astrophys. J. Lett.}\ }\textbf {\bibinfo {volume} {915}},\ \bibinfo {pages} {L12} (\bibinfo {year} {2021})},\ \Eprint {http://arxiv.org/abs/2104.00880} {arXiv:2104.00880 [astro-ph.HE]} \BibitemShut {NoStop}%
\bibitem [{\citenamefont {Abbott}\ and\ \citenamefont {Abbott~{\textit{et al.}}}(2017)}]{PhysRevLett.119.161101}%
  \BibitemOpen
  \bibfield  {author} {\bibinfo {author} {\bibfnamefont {B.~P.}\ \bibnamefont {Abbott}}\ and\ \bibinfo {author} {\bibfnamefont {R.}~\bibnamefont {Abbott~{\textit{et al.}}}},\ }\href {\doibase 10.1103/PhysRevLett.119.161101} {\bibfield  {journal} {\bibinfo  {journal} {Phys. Rev. Lett.}\ }\textbf {\bibinfo {volume} {119}},\ \bibinfo {pages} {161101} (\bibinfo {year} {2017})}\BibitemShut {NoStop}%
\bibitem [{\citenamefont {Abbott}\ \emph {et~al.}(2017)\citenamefont {Abbott} \emph {et~al.}}]{LIGOScientific:2017ync}%
  \BibitemOpen
  \bibfield  {author} {\bibinfo {author} {\bibfnamefont {B.~P.}\ \bibnamefont {Abbott}} \emph {et~al.} (\bibinfo {collaboration} {LIGO Scientific, Virgo}),\ }\href {\doibase 10.3847/2041-8213/aa91c9} {\bibfield  {journal} {\bibinfo  {journal} {Astrophys. J. Lett.}\ }\textbf {\bibinfo {volume} {848}},\ \bibinfo {pages} {L12} (\bibinfo {year} {2017})}\BibitemShut {NoStop}%
\bibitem [{\citenamefont {Abbott}\ and\ \citenamefont {Abbott~{\textit{et al.}}}(2018)}]{PhysRevLett.121.161101}%
  \BibitemOpen
  \bibfield  {author} {\bibinfo {author} {\bibfnamefont {B.~P.}\ \bibnamefont {Abbott}}\ and\ \bibinfo {author} {\bibfnamefont {R.}~\bibnamefont {Abbott~{\textit{et al.}}}},\ }\href {\doibase 10.1103/PhysRevLett.121.161101} {\bibfield  {journal} {\bibinfo  {journal} {Phys. Rev. Lett.}\ }\textbf {\bibinfo {volume} {121}},\ \bibinfo {pages} {161101} (\bibinfo {year} {2018})}\BibitemShut {NoStop}%
\bibitem [{\citenamefont {Typel}\ \emph {et~al.}(2010{\natexlab{a}})\citenamefont {Typel}, \citenamefont {R\"opke}, \citenamefont {Kl\"ahn}, \citenamefont {Blaschke},\ and\ \citenamefont {Wolter}}]{PhysRevC.81.015803}%
  \BibitemOpen
  \bibfield  {author} {\bibinfo {author} {\bibfnamefont {S.}~\bibnamefont {Typel}}, \bibinfo {author} {\bibfnamefont {G.}~\bibnamefont {R\"opke}}, \bibinfo {author} {\bibfnamefont {T.}~\bibnamefont {Kl\"ahn}}, \bibinfo {author} {\bibfnamefont {D.}~\bibnamefont {Blaschke}}, \ and\ \bibinfo {author} {\bibfnamefont {H.~H.}\ \bibnamefont {Wolter}},\ }\href {\doibase 10.1103/PhysRevC.81.015803} {\bibfield  {journal} {\bibinfo  {journal} {Phys. Rev. C}\ }\textbf {\bibinfo {volume} {81}},\ \bibinfo {pages} {015803} (\bibinfo {year} {2010}{\natexlab{a}})}\BibitemShut {NoStop}%
\bibitem [{\citenamefont {Lalazissis}\ \emph {et~al.}(1997)\citenamefont {Lalazissis}, \citenamefont {K\"onig},\ and\ \citenamefont {Ring}}]{PhysRevC.55.540}%
  \BibitemOpen
  \bibfield  {author} {\bibinfo {author} {\bibfnamefont {G.~A.}\ \bibnamefont {Lalazissis}}, \bibinfo {author} {\bibfnamefont {J.}~\bibnamefont {K\"onig}}, \ and\ \bibinfo {author} {\bibfnamefont {P.}~\bibnamefont {Ring}},\ }\href {\doibase 10.1103/PhysRevC.55.540} {\bibfield  {journal} {\bibinfo  {journal} {Phys. Rev. C}\ }\textbf {\bibinfo {volume} {55}},\ \bibinfo {pages} {540} (\bibinfo {year} {1997})}\BibitemShut {NoStop}%
\bibitem [{\citenamefont {Riley}\ \emph {et~al.}(2021{\natexlab{a}})\citenamefont {Riley} \emph {et~al.}}]{Riley:2021pdl}%
  \BibitemOpen
  \bibfield  {author} {\bibinfo {author} {\bibfnamefont {T.~E.}\ \bibnamefont {Riley}} \emph {et~al.},\ }\href {\doibase 10.3847/2041-8213/ac0a81} {\bibfield  {journal} {\bibinfo  {journal} {Astrophys. J. Lett.}\ }\textbf {\bibinfo {volume} {918}},\ \bibinfo {pages} {L27} (\bibinfo {year} {2021}{\natexlab{a}})},\ \Eprint {http://arxiv.org/abs/2105.06980} {arXiv:2105.06980 [astro-ph.HE]} \BibitemShut {NoStop}%
\bibitem [{\citenamefont {Miller}\ \emph {et~al.}(2021)\citenamefont {Miller} \emph {et~al.}}]{miller2021}%
  \BibitemOpen
  \bibfield  {author} {\bibinfo {author} {\bibfnamefont {M.~C.}\ \bibnamefont {Miller}} \emph {et~al.},\ }\href {\doibase 10.3847/2041-8213/ac089b} {\bibfield  {journal} {\bibinfo  {journal} {The Astrophysical Journal Letters}\ }\textbf {\bibinfo {volume} {918}},\ \bibinfo {pages} {L28} (\bibinfo {year} {2021})}\BibitemShut {NoStop}%
\bibitem [{\citenamefont {Salmi}\ \emph {et~al.}(2024)\citenamefont {Salmi} \emph {et~al.}}]{Salmi:2024aum}%
  \BibitemOpen
  \bibfield  {author} {\bibinfo {author} {\bibfnamefont {T.}~\bibnamefont {Salmi}} \emph {et~al.},\ }\href {\doibase 10.3847/1538-4357/ad5f1f} {\bibfield  {journal} {\bibinfo  {journal} {Astrophys. J.}\ }\textbf {\bibinfo {volume} {974}},\ \bibinfo {pages} {294} (\bibinfo {year} {2024})},\ \Eprint {http://arxiv.org/abs/2406.14466} {arXiv:2406.14466 [astro-ph.HE]} \BibitemShut {NoStop}%
\bibitem [{\citenamefont {Dittmann}\ \emph {et~al.}(2024)\citenamefont {Dittmann} \emph {et~al.}}]{dittmann2024precisemeasurementradiuspsr}%
  \BibitemOpen
  \bibfield  {author} {\bibinfo {author} {\bibfnamefont {A.~J.}\ \bibnamefont {Dittmann}} \emph {et~al.},\ }\href {https://arxiv.org/abs/2406.14467} {\  (\bibinfo {year} {2024})},\ \Eprint {http://arxiv.org/abs/2406.14467} {arXiv:2406.14467 [astro-ph.HE]} \BibitemShut {NoStop}%
\bibitem [{\citenamefont {Choudhury}\ \emph {et~al.}(2024)\citenamefont {Choudhury} \emph {et~al.}}]{Choudhury:2024xbk}%
  \BibitemOpen
  \bibfield  {author} {\bibinfo {author} {\bibfnamefont {D.}~\bibnamefont {Choudhury}} \emph {et~al.},\ }\href {\doibase 10.3847/2041-8213/ad5a6f} {\bibfield  {journal} {\bibinfo  {journal} {Astrophys. J. Lett.}\ }\textbf {\bibinfo {volume} {971}},\ \bibinfo {pages} {L20} (\bibinfo {year} {2024})}\BibitemShut {NoStop}%
\bibitem [{\citenamefont {Komoltsev}\ and\ \citenamefont {Kurkela}(2022)}]{Komoltsev:2021jzg}%
  \BibitemOpen
  \bibfield  {author} {\bibinfo {author} {\bibfnamefont {O.}~\bibnamefont {Komoltsev}}\ and\ \bibinfo {author} {\bibfnamefont {A.}~\bibnamefont {Kurkela}},\ }\href {\doibase 10.1103/PhysRevLett.128.202701} {\bibfield  {journal} {\bibinfo  {journal} {Phys. Rev. Lett.}\ }\textbf {\bibinfo {volume} {128}},\ \bibinfo {pages} {202701} (\bibinfo {year} {2022})},\ \Eprint {http://arxiv.org/abs/2111.05350} {arXiv:2111.05350 [nucl-th]} \BibitemShut {NoStop}%
\bibitem [{\citenamefont {Somasundaram}\ \emph {et~al.}(2023)\citenamefont {Somasundaram}, \citenamefont {Tews},\ and\ \citenamefont {Margueron}}]{Somasundaram:2022ztm}%
  \BibitemOpen
  \bibfield  {author} {\bibinfo {author} {\bibfnamefont {R.}~\bibnamefont {Somasundaram}}, \bibinfo {author} {\bibfnamefont {I.}~\bibnamefont {Tews}}, \ and\ \bibinfo {author} {\bibfnamefont {J.}~\bibnamefont {Margueron}},\ }\href {\doibase 10.1103/PhysRevC.107.L052801} {\bibfield  {journal} {\bibinfo  {journal} {Phys. Rev. C}\ }\textbf {\bibinfo {volume} {107}},\ \bibinfo {pages} {L052801} (\bibinfo {year} {2023})},\ \Eprint {http://arxiv.org/abs/2204.14039} {arXiv:2204.14039 [nucl-th]} \BibitemShut {NoStop}%
\bibitem [{\citenamefont {Gorda}\ \emph {et~al.}(2023{\natexlab{a}})\citenamefont {Gorda}, \citenamefont {Komoltsev}, \citenamefont {Kurkela},\ and\ \citenamefont {Mazeliauskas}}]{Gorda:2023usm}%
  \BibitemOpen
  \bibfield  {author} {\bibinfo {author} {\bibfnamefont {T.}~\bibnamefont {Gorda}}, \bibinfo {author} {\bibfnamefont {O.}~\bibnamefont {Komoltsev}}, \bibinfo {author} {\bibfnamefont {A.}~\bibnamefont {Kurkela}}, \ and\ \bibinfo {author} {\bibfnamefont {A.}~\bibnamefont {Mazeliauskas}},\ }\href {\doibase 10.1007/JHEP06(2023)002} {\bibfield  {journal} {\bibinfo  {journal} {JHEP}\ }\textbf {\bibinfo {volume} {06}},\ \bibinfo {pages} {002} (\bibinfo {year} {2023}{\natexlab{a}})},\ \Eprint {http://arxiv.org/abs/2303.02175} {arXiv:2303.02175 [hep-ph]} \BibitemShut {NoStop}%
\bibitem [{\citenamefont {Komoltsev}\ \emph {et~al.}(2024)\citenamefont {Komoltsev}, \citenamefont {Somasundaram}, \citenamefont {Gorda}, \citenamefont {Kurkela}, \citenamefont {Margueron},\ and\ \citenamefont {Tews}}]{Komoltsev:2023zor}%
  \BibitemOpen
  \bibfield  {author} {\bibinfo {author} {\bibfnamefont {O.}~\bibnamefont {Komoltsev}}, \bibinfo {author} {\bibfnamefont {R.}~\bibnamefont {Somasundaram}}, \bibinfo {author} {\bibfnamefont {T.}~\bibnamefont {Gorda}}, \bibinfo {author} {\bibfnamefont {A.}~\bibnamefont {Kurkela}}, \bibinfo {author} {\bibfnamefont {J.}~\bibnamefont {Margueron}}, \ and\ \bibinfo {author} {\bibfnamefont {I.}~\bibnamefont {Tews}},\ }\href {\doibase 10.1103/PhysRevD.109.094030} {\bibfield  {journal} {\bibinfo  {journal} {Phys. Rev. D}\ }\textbf {\bibinfo {volume} {109}},\ \bibinfo {pages} {094030} (\bibinfo {year} {2024})},\ \Eprint {http://arxiv.org/abs/2312.14127} {arXiv:2312.14127 [nucl-th]} \BibitemShut {NoStop}%
\bibitem [{\citenamefont {Gorda}\ \emph {et~al.}(2023{\natexlab{b}})\citenamefont {Gorda}, \citenamefont {Komoltsev},\ and\ \citenamefont {Kurkela}}]{Gorda:2022jvk}%
  \BibitemOpen
  \bibfield  {author} {\bibinfo {author} {\bibfnamefont {T.}~\bibnamefont {Gorda}}, \bibinfo {author} {\bibfnamefont {O.}~\bibnamefont {Komoltsev}}, \ and\ \bibinfo {author} {\bibfnamefont {A.}~\bibnamefont {Kurkela}},\ }\href {\doibase 10.3847/1538-4357/acce3a} {\bibfield  {journal} {\bibinfo  {journal} {Astrophys. J.}\ }\textbf {\bibinfo {volume} {950}},\ \bibinfo {pages} {107} (\bibinfo {year} {2023}{\natexlab{b}})},\ \Eprint {http://arxiv.org/abs/2204.11877} {arXiv:2204.11877 [nucl-th]} \BibitemShut {NoStop}%
\bibitem [{\citenamefont {Gorda}\ \emph {et~al.}(2021)\citenamefont {Gorda}, \citenamefont {Kurkela}, \citenamefont {Paatelainen}, \citenamefont {S\"appi},\ and\ \citenamefont {Vuorinen}}]{PhysRevLett.127.162003}%
  \BibitemOpen
  \bibfield  {author} {\bibinfo {author} {\bibfnamefont {T.}~\bibnamefont {Gorda}}, \bibinfo {author} {\bibfnamefont {A.}~\bibnamefont {Kurkela}}, \bibinfo {author} {\bibfnamefont {R.}~\bibnamefont {Paatelainen}}, \bibinfo {author} {\bibfnamefont {S.}~\bibnamefont {S\"appi}}, \ and\ \bibinfo {author} {\bibfnamefont {A.}~\bibnamefont {Vuorinen}},\ }\href {\doibase 10.1103/PhysRevLett.127.162003} {\bibfield  {journal} {\bibinfo  {journal} {Phys. Rev. Lett.}\ }\textbf {\bibinfo {volume} {127}},\ \bibinfo {pages} {162003} (\bibinfo {year} {2021})}\BibitemShut {NoStop}%
\bibitem [{\citenamefont {Hebeler}\ \emph {et~al.}(2013)\citenamefont {Hebeler}, \citenamefont {Lattimer}, \citenamefont {Pethick},\ and\ \citenamefont {Schwenk}}]{Hebeler:2013nza}%
  \BibitemOpen
  \bibfield  {author} {\bibinfo {author} {\bibfnamefont {K.}~\bibnamefont {Hebeler}}, \bibinfo {author} {\bibfnamefont {J.}~\bibnamefont {Lattimer}}, \bibinfo {author} {\bibfnamefont {C.}~\bibnamefont {Pethick}}, \ and\ \bibinfo {author} {\bibfnamefont {A.}~\bibnamefont {Schwenk}},\ }\href {\doibase 10.1088/0004-637X/773/1/11} {\bibfield  {journal} {\bibinfo  {journal} {Astrophys. J.}\ }\textbf {\bibinfo {volume} {773}},\ \bibinfo {pages} {11} (\bibinfo {year} {2013})},\ \Eprint {http://arxiv.org/abs/1303.4662} {arXiv:1303.4662 [astro-ph.SR]} \BibitemShut {NoStop}%
\bibitem [{\citenamefont {Drischler}\ \emph {et~al.}(2016{\natexlab{a}})\citenamefont {Drischler}, \citenamefont {Carbone}, \citenamefont {Hebeler},\ and\ \citenamefont {Schwenk}}]{Drischler:2016djf}%
  \BibitemOpen
  \bibfield  {author} {\bibinfo {author} {\bibfnamefont {C.}~\bibnamefont {Drischler}}, \bibinfo {author} {\bibfnamefont {A.}~\bibnamefont {Carbone}}, \bibinfo {author} {\bibfnamefont {K.}~\bibnamefont {Hebeler}}, \ and\ \bibinfo {author} {\bibfnamefont {A.}~\bibnamefont {Schwenk}},\ }\href {\doibase 10.1103/PhysRevC.94.054307} {\bibfield  {journal} {\bibinfo  {journal} {Phys. Rev. C}\ }\textbf {\bibinfo {volume} {94}},\ \bibinfo {pages} {054307} (\bibinfo {year} {2016}{\natexlab{a}})},\ \Eprint {http://arxiv.org/abs/1608.05615} {arXiv:1608.05615 [nucl-th]} \BibitemShut {NoStop}%
\bibitem [{\citenamefont {Annala}\ \emph {et~al.}(2020)\citenamefont {Annala}, \citenamefont {Gorda}, \citenamefont {Kurkela} \emph {et~al.}}]{Annala_2020}%
  \BibitemOpen
  \bibfield  {author} {\bibinfo {author} {\bibfnamefont {E.}~\bibnamefont {Annala}}, \bibinfo {author} {\bibfnamefont {T.}~\bibnamefont {Gorda}}, \bibinfo {author} {\bibfnamefont {A.}~\bibnamefont {Kurkela}},  \emph {et~al.},\ }\href {\doibase 10.1038/s41567-020-0914-9} {\bibfield  {journal} {\bibinfo  {journal} {Nature Physics}\ }\textbf {\bibinfo {volume} {16}},\ \bibinfo {pages} {907} (\bibinfo {year} {2020})}\BibitemShut {NoStop}%
\bibitem [{\citenamefont {Altiparmak}\ \emph {et~al.}(2022)\citenamefont {Altiparmak}, \citenamefont {Ecker},\ and\ \citenamefont {Rezzolla}}]{Altiparmak:2022bke}%
  \BibitemOpen
  \bibfield  {author} {\bibinfo {author} {\bibfnamefont {S.}~\bibnamefont {Altiparmak}}, \bibinfo {author} {\bibfnamefont {C.}~\bibnamefont {Ecker}}, \ and\ \bibinfo {author} {\bibfnamefont {L.}~\bibnamefont {Rezzolla}},\ }\href {\doibase 10.3847/2041-8213/ac9b2a} {\bibfield  {journal} {\bibinfo  {journal} {Astrophys. J. Lett.}\ }\textbf {\bibinfo {volume} {939}},\ \bibinfo {pages} {L34} (\bibinfo {year} {2022})},\ \Eprint {http://arxiv.org/abs/2203.14974} {arXiv:2203.14974 [astro-ph.HE]} \BibitemShut {NoStop}%
\bibitem [{\citenamefont {Semposki}\ \emph {et~al.}(2024)\citenamefont {Semposki}, \citenamefont {Drischler}, \citenamefont {Furnstahl}, \citenamefont {Melendez},\ and\ \citenamefont {Phillips}}]{Semposki:2024vnp}%
  \BibitemOpen
  \bibfield  {author} {\bibinfo {author} {\bibfnamefont {A.~C.}\ \bibnamefont {Semposki}}, \bibinfo {author} {\bibfnamefont {C.}~\bibnamefont {Drischler}}, \bibinfo {author} {\bibfnamefont {R.~J.}\ \bibnamefont {Furnstahl}}, \bibinfo {author} {\bibfnamefont {J.~A.}\ \bibnamefont {Melendez}}, \ and\ \bibinfo {author} {\bibfnamefont {D.~R.}\ \bibnamefont {Phillips}},\ }\href@noop {} {\  (\bibinfo {year} {2024})},\ \Eprint {http://arxiv.org/abs/2404.06323} {arXiv:2404.06323 [nucl-th]} \BibitemShut {NoStop}%
\bibitem [{\citenamefont {Albino}\ \emph {et~al.}(2024)\citenamefont {Albino}, \citenamefont {Malik}, \citenamefont {Ferreira},\ and\ \citenamefont {Providência}}]{albino2024hybridstarpropertiesnjl}%
  \BibitemOpen
  \bibfield  {author} {\bibinfo {author} {\bibfnamefont {M.}~\bibnamefont {Albino}}, \bibinfo {author} {\bibfnamefont {T.}~\bibnamefont {Malik}}, \bibinfo {author} {\bibfnamefont {M.}~\bibnamefont {Ferreira}}, \ and\ \bibinfo {author} {\bibfnamefont {C.}~\bibnamefont {Providência}},\ }\href {https://arxiv.org/abs/2406.15337} {\enquote {\bibinfo {title} {Hybrid star properties with njl and mftqcd model: A bayesian approach},}\ } (\bibinfo {year} {2024}),\ \Eprint {http://arxiv.org/abs/2406.15337} {arXiv:2406.15337 [nucl-th]} \BibitemShut {NoStop}%
\bibitem [{\citenamefont {Son}(1999)}]{Son:1998uk}%
  \BibitemOpen
  \bibfield  {author} {\bibinfo {author} {\bibfnamefont {D.~T.}\ \bibnamefont {Son}},\ }\href {\doibase 10.1103/PhysRevD.59.094019} {\bibfield  {journal} {\bibinfo  {journal} {Phys. Rev. D}\ }\textbf {\bibinfo {volume} {59}},\ \bibinfo {pages} {094019} (\bibinfo {year} {1999})},\ \Eprint {http://arxiv.org/abs/hep-ph/9812287} {arXiv:hep-ph/9812287} \BibitemShut {NoStop}%
\bibitem [{\citenamefont {Sch\"afer}\ and\ \citenamefont {Wilczek}(1999)}]{Schafer:1999jg}%
  \BibitemOpen
  \bibfield  {author} {\bibinfo {author} {\bibfnamefont {T.}~\bibnamefont {Sch\"afer}}\ and\ \bibinfo {author} {\bibfnamefont {F.}~\bibnamefont {Wilczek}},\ }\href {\doibase 10.1103/PhysRevD.60.114033} {\bibfield  {journal} {\bibinfo  {journal} {Phys. Rev. D}\ }\textbf {\bibinfo {volume} {60}},\ \bibinfo {pages} {114033} (\bibinfo {year} {1999})},\ \Eprint {http://arxiv.org/abs/hep-ph/9906512} {arXiv:hep-ph/9906512} \BibitemShut {NoStop}%
\bibitem [{\citenamefont {Pisarski}\ and\ \citenamefont {Rischke}(2000)}]{Pisarski:1999tv}%
  \BibitemOpen
  \bibfield  {author} {\bibinfo {author} {\bibfnamefont {R.~D.}\ \bibnamefont {Pisarski}}\ and\ \bibinfo {author} {\bibfnamefont {D.~H.}\ \bibnamefont {Rischke}},\ }\href {\doibase 10.1103/PhysRevD.61.074017} {\bibfield  {journal} {\bibinfo  {journal} {Phys. Rev. D}\ }\textbf {\bibinfo {volume} {61}},\ \bibinfo {pages} {074017} (\bibinfo {year} {2000})},\ \Eprint {http://arxiv.org/abs/nucl-th/9910056} {arXiv:nucl-th/9910056} \BibitemShut {NoStop}%
\bibitem [{\citenamefont {Alford}\ \emph {et~al.}(2008)\citenamefont {Alford}, \citenamefont {Schmitt}, \citenamefont {Rajagopal},\ and\ \citenamefont {Sch\"afer}}]{Alford:2007xm}%
  \BibitemOpen
  \bibfield  {author} {\bibinfo {author} {\bibfnamefont {M.~G.}\ \bibnamefont {Alford}}, \bibinfo {author} {\bibfnamefont {A.}~\bibnamefont {Schmitt}}, \bibinfo {author} {\bibfnamefont {K.}~\bibnamefont {Rajagopal}}, \ and\ \bibinfo {author} {\bibfnamefont {T.}~\bibnamefont {Sch\"afer}},\ }\href {\doibase 10.1103/RevModPhys.80.1455} {\bibfield  {journal} {\bibinfo  {journal} {Rev. Mod. Phys.}\ }\textbf {\bibinfo {volume} {80}},\ \bibinfo {pages} {1455} (\bibinfo {year} {2008})}\BibitemShut {NoStop}%
\bibitem [{\citenamefont {Alford}\ \emph {et~al.}(1999{\natexlab{a}})\citenamefont {Alford}, \citenamefont {Rajagopal},\ and\ \citenamefont {Wilczek}}]{Alford:1998mk}%
  \BibitemOpen
  \bibfield  {author} {\bibinfo {author} {\bibfnamefont {M.~G.}\ \bibnamefont {Alford}}, \bibinfo {author} {\bibfnamefont {K.}~\bibnamefont {Rajagopal}}, \ and\ \bibinfo {author} {\bibfnamefont {F.}~\bibnamefont {Wilczek}},\ }\href {\doibase 10.1016/S0550-3213(98)00668-3} {\bibfield  {journal} {\bibinfo  {journal} {Nucl. Phys. B}\ }\textbf {\bibinfo {volume} {537}},\ \bibinfo {pages} {443} (\bibinfo {year} {1999}{\natexlab{a}})}\BibitemShut {NoStop}%
\bibitem [{\citenamefont {Sch\"afer}(2000)}]{Schafer:1999fe}%
  \BibitemOpen
  \bibfield  {author} {\bibinfo {author} {\bibfnamefont {T.}~\bibnamefont {Sch\"afer}},\ }\href {\doibase 10.1016/S0550-3213(00)00063-8} {\bibfield  {journal} {\bibinfo  {journal} {Nucl. Phys. B}\ }\textbf {\bibinfo {volume} {575}},\ \bibinfo {pages} {269} (\bibinfo {year} {2000})},\ \Eprint {http://arxiv.org/abs/hep-ph/9909574} {arXiv:hep-ph/9909574} \BibitemShut {NoStop}%
\bibitem [{\citenamefont {Shovkovy}\ and\ \citenamefont {Wijewardhana}(1999)}]{Shovkovy:1999mr}%
  \BibitemOpen
  \bibfield  {author} {\bibinfo {author} {\bibfnamefont {I.~A.}\ \bibnamefont {Shovkovy}}\ and\ \bibinfo {author} {\bibfnamefont {L.~C.~R.}\ \bibnamefont {Wijewardhana}},\ }\href {\doibase 10.1016/S0370-2693(99)01297-6} {\bibfield  {journal} {\bibinfo  {journal} {Phys. Lett. B}\ }\textbf {\bibinfo {volume} {470}},\ \bibinfo {pages} {189} (\bibinfo {year} {1999})},\ \Eprint {http://arxiv.org/abs/hep-ph/9910225} {arXiv:hep-ph/9910225} \BibitemShut {NoStop}%
\bibitem [{\citenamefont {Alford}\ \emph {et~al.}(1999{\natexlab{b}})\citenamefont {Alford}, \citenamefont {Berges},\ and\ \citenamefont {Rajagopal}}]{Alford:1999pa}%
  \BibitemOpen
  \bibfield  {author} {\bibinfo {author} {\bibfnamefont {M.~G.}\ \bibnamefont {Alford}}, \bibinfo {author} {\bibfnamefont {J.}~\bibnamefont {Berges}}, \ and\ \bibinfo {author} {\bibfnamefont {K.}~\bibnamefont {Rajagopal}},\ }\href {\doibase 10.1016/S0550-3213(99)00410-1} {\bibfield  {journal} {\bibinfo  {journal} {Nucl. Phys. B}\ }\textbf {\bibinfo {volume} {558}},\ \bibinfo {pages} {219} (\bibinfo {year} {1999}{\natexlab{b}})},\ \Eprint {http://arxiv.org/abs/hep-ph/9903502} {arXiv:hep-ph/9903502} \BibitemShut {NoStop}%
\bibitem [{\citenamefont {Alford}\ and\ \citenamefont {Rajagopal}(2002)}]{Alford:2002kj}%
  \BibitemOpen
  \bibfield  {author} {\bibinfo {author} {\bibfnamefont {M.}~\bibnamefont {Alford}}\ and\ \bibinfo {author} {\bibfnamefont {K.}~\bibnamefont {Rajagopal}},\ }\href {\doibase 10.1088/1126-6708/2002/06/031} {\bibfield  {journal} {\bibinfo  {journal} {JHEP}\ }\textbf {\bibinfo {volume} {06}},\ \bibinfo {pages} {031} (\bibinfo {year} {2002})},\ \Eprint {http://arxiv.org/abs/hep-ph/0204001} {arXiv:hep-ph/0204001} \BibitemShut {NoStop}%
\bibitem [{\citenamefont {Yuan}\ and\ \citenamefont {Li}(2024)}]{Yuan:2023dxl}%
  \BibitemOpen
  \bibfield  {author} {\bibinfo {author} {\bibfnamefont {W.-L.}\ \bibnamefont {Yuan}}\ and\ \bibinfo {author} {\bibfnamefont {A.}~\bibnamefont {Li}},\ }\href {\doibase 10.3847/1538-4357/ad354f} {\bibfield  {journal} {\bibinfo  {journal} {Astrophys. J.}\ }\textbf {\bibinfo {volume} {966}},\ \bibinfo {pages} {3} (\bibinfo {year} {2024})},\ \Eprint {http://arxiv.org/abs/2312.17102} {arXiv:2312.17102 [nucl-th]} \BibitemShut {NoStop}%
\bibitem [{\citenamefont {Alford}\ \emph {et~al.}(2005)\citenamefont {Alford}, \citenamefont {Braby}, \citenamefont {Paris},\ and\ \citenamefont {Reddy}}]{Alford:2004pf}%
  \BibitemOpen
  \bibfield  {author} {\bibinfo {author} {\bibfnamefont {M.}~\bibnamefont {Alford}}, \bibinfo {author} {\bibfnamefont {M.}~\bibnamefont {Braby}}, \bibinfo {author} {\bibfnamefont {M.~W.}\ \bibnamefont {Paris}}, \ and\ \bibinfo {author} {\bibfnamefont {S.}~\bibnamefont {Reddy}},\ }\href {\doibase 10.1086/430902} {\bibfield  {journal} {\bibinfo  {journal} {Astrophys. J.}\ }\textbf {\bibinfo {volume} {629}},\ \bibinfo {pages} {969} (\bibinfo {year} {2005})},\ \Eprint {http://arxiv.org/abs/nucl-th/0411016} {arXiv:nucl-th/0411016} \BibitemShut {NoStop}%
\bibitem [{\citenamefont {Alford}\ \emph {et~al.}(2007)\citenamefont {Alford}, \citenamefont {Blaschke}, \citenamefont {Drago}, \citenamefont {Klahn}, \citenamefont {Pagliara},\ and\ \citenamefont {Schaffner-Bielich}}]{Alford:2006vz}%
  \BibitemOpen
  \bibfield  {author} {\bibinfo {author} {\bibfnamefont {M.}~\bibnamefont {Alford}}, \bibinfo {author} {\bibfnamefont {D.}~\bibnamefont {Blaschke}}, \bibinfo {author} {\bibfnamefont {A.}~\bibnamefont {Drago}}, \bibinfo {author} {\bibfnamefont {T.}~\bibnamefont {Klahn}}, \bibinfo {author} {\bibfnamefont {G.}~\bibnamefont {Pagliara}}, \ and\ \bibinfo {author} {\bibfnamefont {J.}~\bibnamefont {Schaffner-Bielich}},\ }\href {\doibase 10.1038/nature05582} {\bibfield  {journal} {\bibinfo  {journal} {Nature}\ }\textbf {\bibinfo {volume} {445}},\ \bibinfo {pages} {E7} (\bibinfo {year} {2007})},\ \Eprint {http://arxiv.org/abs/astro-ph/0606524} {arXiv:astro-ph/0606524} \BibitemShut {NoStop}%
\bibitem [{\citenamefont {Alford}\ \emph {et~al.}(2013)\citenamefont {Alford}, \citenamefont {Han},\ and\ \citenamefont {Prakash}}]{Alford:2013aca}%
  \BibitemOpen
  \bibfield  {author} {\bibinfo {author} {\bibfnamefont {M.~G.}\ \bibnamefont {Alford}}, \bibinfo {author} {\bibfnamefont {S.}~\bibnamefont {Han}}, \ and\ \bibinfo {author} {\bibfnamefont {M.}~\bibnamefont {Prakash}},\ }\href {\doibase 10.1103/PhysRevD.88.083013} {\bibfield  {journal} {\bibinfo  {journal} {Phys. Rev. D}\ }\textbf {\bibinfo {volume} {88}},\ \bibinfo {pages} {083013} (\bibinfo {year} {2013})},\ \Eprint {http://arxiv.org/abs/1302.4732} {arXiv:1302.4732 [astro-ph.SR]} \BibitemShut {NoStop}%
\bibitem [{\citenamefont {Contrera}\ \emph {et~al.}(2022)\citenamefont {Contrera}, \citenamefont {Blaschke}, \citenamefont {Carlomagno}, \citenamefont {Grunfeld},\ and\ \citenamefont {Liebing}}]{Contrera:2022tqh}%
  \BibitemOpen
  \bibfield  {author} {\bibinfo {author} {\bibfnamefont {G.~A.}\ \bibnamefont {Contrera}}, \bibinfo {author} {\bibfnamefont {D.}~\bibnamefont {Blaschke}}, \bibinfo {author} {\bibfnamefont {J.~P.}\ \bibnamefont {Carlomagno}}, \bibinfo {author} {\bibfnamefont {A.~G.}\ \bibnamefont {Grunfeld}}, \ and\ \bibinfo {author} {\bibfnamefont {S.}~\bibnamefont {Liebing}},\ }\href {\doibase 10.1103/PhysRevC.105.045808} {\bibfield  {journal} {\bibinfo  {journal} {Phys. Rev. C}\ }\textbf {\bibinfo {volume} {105}},\ \bibinfo {pages} {045808} (\bibinfo {year} {2022})},\ \Eprint {http://arxiv.org/abs/2201.00477} {arXiv:2201.00477 [nucl-th]} \BibitemShut {NoStop}%
\bibitem [{\citenamefont {Annala}\ \emph {et~al.}(2023)\citenamefont {Annala}, \citenamefont {Gorda}, \citenamefont {Hirvonen}, \citenamefont {Komoltsev}, \citenamefont {Kurkela}, \citenamefont {N\"attil\"a},\ and\ \citenamefont {Vuorinen}}]{Annala:2023cwx}%
  \BibitemOpen
  \bibfield  {author} {\bibinfo {author} {\bibfnamefont {E.}~\bibnamefont {Annala}}, \bibinfo {author} {\bibfnamefont {T.}~\bibnamefont {Gorda}}, \bibinfo {author} {\bibfnamefont {J.}~\bibnamefont {Hirvonen}}, \bibinfo {author} {\bibfnamefont {O.}~\bibnamefont {Komoltsev}}, \bibinfo {author} {\bibfnamefont {A.}~\bibnamefont {Kurkela}}, \bibinfo {author} {\bibfnamefont {J.}~\bibnamefont {N\"attil\"a}}, \ and\ \bibinfo {author} {\bibfnamefont {A.}~\bibnamefont {Vuorinen}},\ }\href {\doibase 10.1038/s41467-023-44051-y} {\bibfield  {journal} {\bibinfo  {journal} {Nature Commun.}\ }\textbf {\bibinfo {volume} {14}},\ \bibinfo {pages} {8451} (\bibinfo {year} {2023})},\ \Eprint {http://arxiv.org/abs/2303.11356} {arXiv:2303.11356 [astro-ph.HE]} \BibitemShut {NoStop}%
\bibitem [{\citenamefont {Christian}\ \emph {et~al.}(2024{\natexlab{a}})\citenamefont {Christian}, \citenamefont {Schaffner-Bielich},\ and\ \citenamefont {Rosswog}}]{Christian:2023hez}%
  \BibitemOpen
  \bibfield  {author} {\bibinfo {author} {\bibfnamefont {J.-E.}\ \bibnamefont {Christian}}, \bibinfo {author} {\bibfnamefont {J.}~\bibnamefont {Schaffner-Bielich}}, \ and\ \bibinfo {author} {\bibfnamefont {S.}~\bibnamefont {Rosswog}},\ }\href {\doibase 10.1103/PhysRevD.109.063035} {\bibfield  {journal} {\bibinfo  {journal} {Phys. Rev. D}\ }\textbf {\bibinfo {volume} {109}},\ \bibinfo {pages} {063035} (\bibinfo {year} {2024}{\natexlab{a}})},\ \Eprint {http://arxiv.org/abs/2312.10148} {arXiv:2312.10148 [nucl-th]} \BibitemShut {NoStop}%
\bibitem [{\citenamefont {Glendenning}(1992)}]{Glendenning:1992vb}%
  \BibitemOpen
  \bibfield  {author} {\bibinfo {author} {\bibfnamefont {N.~K.}\ \bibnamefont {Glendenning}},\ }\href {\doibase 10.1103/PhysRevD.46.1274} {\bibfield  {journal} {\bibinfo  {journal} {Phys. Rev. D}\ }\textbf {\bibinfo {volume} {46}},\ \bibinfo {pages} {1274} (\bibinfo {year} {1992})}\BibitemShut {NoStop}%
\bibitem [{\citenamefont {Baym}\ \emph {et~al.}(2019)\citenamefont {Baym}, \citenamefont {Furusawa}, \citenamefont {Hatsuda}, \citenamefont {Kojo},\ and\ \citenamefont {Togashi}}]{Baym:2019iky}%
  \BibitemOpen
  \bibfield  {author} {\bibinfo {author} {\bibfnamefont {G.}~\bibnamefont {Baym}}, \bibinfo {author} {\bibfnamefont {S.}~\bibnamefont {Furusawa}}, \bibinfo {author} {\bibfnamefont {T.}~\bibnamefont {Hatsuda}}, \bibinfo {author} {\bibfnamefont {T.}~\bibnamefont {Kojo}}, \ and\ \bibinfo {author} {\bibfnamefont {H.}~\bibnamefont {Togashi}},\ }\href {\doibase 10.3847/1538-4357/ab441e} {\bibfield  {journal} {\bibinfo  {journal} {Astrophys. J.}\ }\textbf {\bibinfo {volume} {885}},\ \bibinfo {pages} {42} (\bibinfo {year} {2019})},\ \Eprint {http://arxiv.org/abs/1903.08963} {arXiv:1903.08963 [astro-ph.HE]} \BibitemShut {NoStop}%
\bibitem [{\citenamefont {Huang}\ \emph {et~al.}(2022)\citenamefont {Huang}, \citenamefont {Hu}, \citenamefont {Zhang},\ and\ \citenamefont {Shen}}]{Huang:2022jiu}%
  \BibitemOpen
  \bibfield  {author} {\bibinfo {author} {\bibfnamefont {K.}~\bibnamefont {Huang}}, \bibinfo {author} {\bibfnamefont {J.}~\bibnamefont {Hu}}, \bibinfo {author} {\bibfnamefont {Y.}~\bibnamefont {Zhang}}, \ and\ \bibinfo {author} {\bibfnamefont {H.}~\bibnamefont {Shen}},\ }\href {\doibase 10.3847/1538-4357/ac7f3c} {\bibfield  {journal} {\bibinfo  {journal} {Astrophys. J.}\ }\textbf {\bibinfo {volume} {935}},\ \bibinfo {pages} {88} (\bibinfo {year} {2022})},\ \Eprint {http://arxiv.org/abs/2206.12760} {arXiv:2206.12760 [nucl-th]} \BibitemShut {NoStop}%
\bibitem [{\citenamefont {Christian}\ \emph {et~al.}(2019)\citenamefont {Christian}, \citenamefont {Zacchi},\ and\ \citenamefont {Schaffner-Bielich}}]{PhysRevD.99.023009}%
  \BibitemOpen
  \bibfield  {author} {\bibinfo {author} {\bibfnamefont {J.-E.}\ \bibnamefont {Christian}}, \bibinfo {author} {\bibfnamefont {A.}~\bibnamefont {Zacchi}}, \ and\ \bibinfo {author} {\bibfnamefont {J.}~\bibnamefont {Schaffner-Bielich}},\ }\href {\doibase 10.1103/PhysRevD.99.023009} {\bibfield  {journal} {\bibinfo  {journal} {Phys. Rev. D}\ }\textbf {\bibinfo {volume} {99}},\ \bibinfo {pages} {023009} (\bibinfo {year} {2019})}\BibitemShut {NoStop}%
\bibitem [{\citenamefont {Tews}\ \emph {et~al.}(2018)\citenamefont {Tews}, \citenamefont {Carlson}, \citenamefont {Gandolfi},\ and\ \citenamefont {Reddy}}]{Tews:2018kmu}%
  \BibitemOpen
  \bibfield  {author} {\bibinfo {author} {\bibfnamefont {I.}~\bibnamefont {Tews}}, \bibinfo {author} {\bibfnamefont {J.}~\bibnamefont {Carlson}}, \bibinfo {author} {\bibfnamefont {S.}~\bibnamefont {Gandolfi}}, \ and\ \bibinfo {author} {\bibfnamefont {S.}~\bibnamefont {Reddy}},\ }\href {\doibase 10.3847/1538-4357/aac267} {\bibfield  {journal} {\bibinfo  {journal} {Astrophys. J.}\ }\textbf {\bibinfo {volume} {860}},\ \bibinfo {pages} {149} (\bibinfo {year} {2018})},\ \Eprint {http://arxiv.org/abs/1801.01923} {arXiv:1801.01923 [nucl-th]} \BibitemShut {NoStop}%
\bibitem [{\citenamefont {Rutherford}\ \emph {et~al.}(2024)\citenamefont {Rutherford} \emph {et~al.}}]{Rutherford:2024srk}%
  \BibitemOpen
  \bibfield  {author} {\bibinfo {author} {\bibfnamefont {N.}~\bibnamefont {Rutherford}} \emph {et~al.},\ }\href {\doibase 10.3847/2041-8213/ad5f02} {\bibfield  {journal} {\bibinfo  {journal} {Astrophys. J. Lett.}\ }\textbf {\bibinfo {volume} {971}},\ \bibinfo {pages} {L19} (\bibinfo {year} {2024})},\ \Eprint {http://arxiv.org/abs/2407.06790} {arXiv:2407.06790 [astro-ph.HE]} \BibitemShut {NoStop}%
\bibitem [{\citenamefont {Fischer}(2019)}]{Fischer:2018sdj}%
  \BibitemOpen
  \bibfield  {author} {\bibinfo {author} {\bibfnamefont {C.~S.}\ \bibnamefont {Fischer}},\ }\href {\doibase 10.1016/j.ppnp.2019.01.002} {\bibfield  {journal} {\bibinfo  {journal} {Prog. Part. Nucl. Phys.}\ }\textbf {\bibinfo {volume} {105}},\ \bibinfo {pages} {1} (\bibinfo {year} {2019})},\ \Eprint {http://arxiv.org/abs/1810.12938} {arXiv:1810.12938 [hep-ph]} \BibitemShut {NoStop}%
\bibitem [{\citenamefont {Fu}\ \emph {et~al.}(2020)\citenamefont {Fu}, \citenamefont {Pawlowski},\ and\ \citenamefont {Rennecke}}]{Fu:2019hdw}%
  \BibitemOpen
  \bibfield  {author} {\bibinfo {author} {\bibfnamefont {W.-j.}\ \bibnamefont {Fu}}, \bibinfo {author} {\bibfnamefont {J.~M.}\ \bibnamefont {Pawlowski}}, \ and\ \bibinfo {author} {\bibfnamefont {F.}~\bibnamefont {Rennecke}},\ }\href {\doibase 10.1103/PhysRevD.101.054032} {\bibfield  {journal} {\bibinfo  {journal} {Phys. Rev. D}\ }\textbf {\bibinfo {volume} {101}},\ \bibinfo {pages} {054032} (\bibinfo {year} {2020})},\ \Eprint {http://arxiv.org/abs/1909.02991} {arXiv:1909.02991 [hep-ph]} \BibitemShut {NoStop}%
\bibitem [{\citenamefont {Nickel}\ \emph {et~al.}(2006{\natexlab{a}})\citenamefont {Nickel}, \citenamefont {Wambach},\ and\ \citenamefont {Alkofer}}]{Nickel:2006vf}%
  \BibitemOpen
  \bibfield  {author} {\bibinfo {author} {\bibfnamefont {D.}~\bibnamefont {Nickel}}, \bibinfo {author} {\bibfnamefont {J.}~\bibnamefont {Wambach}}, \ and\ \bibinfo {author} {\bibfnamefont {R.}~\bibnamefont {Alkofer}},\ }\href {\doibase 10.1103/PhysRevD.73.114028} {\bibfield  {journal} {\bibinfo  {journal} {Phys. Rev. D}\ }\textbf {\bibinfo {volume} {73}},\ \bibinfo {pages} {114028} (\bibinfo {year} {2006}{\natexlab{a}})},\ \Eprint {http://arxiv.org/abs/hep-ph/0603163} {arXiv:hep-ph/0603163} \BibitemShut {NoStop}%
\bibitem [{\citenamefont {Nickel}\ \emph {et~al.}(2006{\natexlab{b}})\citenamefont {Nickel}, \citenamefont {Alkofer},\ and\ \citenamefont {Wambach}}]{Nickel:2006kc}%
  \BibitemOpen
  \bibfield  {author} {\bibinfo {author} {\bibfnamefont {D.}~\bibnamefont {Nickel}}, \bibinfo {author} {\bibfnamefont {R.}~\bibnamefont {Alkofer}}, \ and\ \bibinfo {author} {\bibfnamefont {J.}~\bibnamefont {Wambach}},\ }\href {\doibase 10.1103/PhysRevD.74.114015} {\bibfield  {journal} {\bibinfo  {journal} {Phys. Rev. D}\ }\textbf {\bibinfo {volume} {74}},\ \bibinfo {pages} {114015} (\bibinfo {year} {2006}{\natexlab{b}})},\ \Eprint {http://arxiv.org/abs/hep-ph/0609198} {arXiv:hep-ph/0609198} \BibitemShut {NoStop}%
\bibitem [{\citenamefont {Marhauser}\ \emph {et~al.}(2007)\citenamefont {Marhauser}, \citenamefont {Nickel}, \citenamefont {Buballa},\ and\ \citenamefont {Wambach}}]{Marhauser:2006hy}%
  \BibitemOpen
  \bibfield  {author} {\bibinfo {author} {\bibfnamefont {F.}~\bibnamefont {Marhauser}}, \bibinfo {author} {\bibfnamefont {D.}~\bibnamefont {Nickel}}, \bibinfo {author} {\bibfnamefont {M.}~\bibnamefont {Buballa}}, \ and\ \bibinfo {author} {\bibfnamefont {J.}~\bibnamefont {Wambach}},\ }\href {\doibase 10.1103/PhysRevD.75.054022} {\bibfield  {journal} {\bibinfo  {journal} {Phys. Rev. D}\ }\textbf {\bibinfo {volume} {75}},\ \bibinfo {pages} {054022} (\bibinfo {year} {2007})},\ \Eprint {http://arxiv.org/abs/hep-ph/0612027} {arXiv:hep-ph/0612027} \BibitemShut {NoStop}%
\bibitem [{\citenamefont {Nickel}\ \emph {et~al.}(2008)\citenamefont {Nickel}, \citenamefont {Alkofer},\ and\ \citenamefont {Wambach}}]{Nickel:2008ef}%
  \BibitemOpen
  \bibfield  {author} {\bibinfo {author} {\bibfnamefont {D.}~\bibnamefont {Nickel}}, \bibinfo {author} {\bibfnamefont {R.}~\bibnamefont {Alkofer}}, \ and\ \bibinfo {author} {\bibfnamefont {J.}~\bibnamefont {Wambach}},\ }\href {\doibase 10.1103/PhysRevD.77.114010} {\bibfield  {journal} {\bibinfo  {journal} {Phys. Rev. D}\ }\textbf {\bibinfo {volume} {77}},\ \bibinfo {pages} {114010} (\bibinfo {year} {2008})},\ \Eprint {http://arxiv.org/abs/0802.3187} {arXiv:0802.3187 [hep-ph]} \BibitemShut {NoStop}%
\bibitem [{\citenamefont {M\"uller}\ \emph {et~al.}(2013)\citenamefont {M\"uller}, \citenamefont {Buballa},\ and\ \citenamefont {Wambach}}]{Muller:2013pya}%
  \BibitemOpen
  \bibfield  {author} {\bibinfo {author} {\bibfnamefont {D.}~\bibnamefont {M\"uller}}, \bibinfo {author} {\bibfnamefont {M.}~\bibnamefont {Buballa}}, \ and\ \bibinfo {author} {\bibfnamefont {J.}~\bibnamefont {Wambach}},\ }\href {\doibase 10.1140/epja/i2013-13096-5} {\bibfield  {journal} {\bibinfo  {journal} {Eur. Phys. J. A}\ }\textbf {\bibinfo {volume} {49}},\ \bibinfo {pages} {96} (\bibinfo {year} {2013})},\ \Eprint {http://arxiv.org/abs/1303.2693} {arXiv:1303.2693 [hep-ph]} \BibitemShut {NoStop}%
\bibitem [{\citenamefont {M\"uller}\ \emph {et~al.}(2016)\citenamefont {M\"uller}, \citenamefont {Buballa},\ and\ \citenamefont {Wambach}}]{Muller:2016fdr}%
  \BibitemOpen
  \bibfield  {author} {\bibinfo {author} {\bibfnamefont {D.}~\bibnamefont {M\"uller}}, \bibinfo {author} {\bibfnamefont {M.}~\bibnamefont {Buballa}}, \ and\ \bibinfo {author} {\bibfnamefont {J.}~\bibnamefont {Wambach}},\ }\href@noop {} {\  (\bibinfo {year} {2016})},\ \Eprint {http://arxiv.org/abs/1603.02865} {arXiv:1603.02865 [hep-ph]} \BibitemShut {NoStop}%
\bibitem [{\citenamefont {Braun}\ and\ \citenamefont {Schallmo}(2022{\natexlab{a}})}]{Braun:2021uua}%
  \BibitemOpen
  \bibfield  {author} {\bibinfo {author} {\bibfnamefont {J.}~\bibnamefont {Braun}}\ and\ \bibinfo {author} {\bibfnamefont {B.}~\bibnamefont {Schallmo}},\ }\href {\doibase 10.1103/PhysRevD.105.036003} {\bibfield  {journal} {\bibinfo  {journal} {Phys. Rev. D}\ }\textbf {\bibinfo {volume} {105}},\ \bibinfo {pages} {036003} (\bibinfo {year} {2022}{\natexlab{a}})},\ \Eprint {http://arxiv.org/abs/2106.04198} {arXiv:2106.04198 [hep-ph]} \BibitemShut {NoStop}%
\bibitem [{\citenamefont {Braun}\ \emph {et~al.}(2024)\citenamefont {Braun}, \citenamefont {Gei\ss{}el},\ and\ \citenamefont {Schallmo}}]{Braun:2022jme}%
  \BibitemOpen
  \bibfield  {author} {\bibinfo {author} {\bibfnamefont {J.}~\bibnamefont {Braun}}, \bibinfo {author} {\bibfnamefont {A.}~\bibnamefont {Gei\ss{}el}}, \ and\ \bibinfo {author} {\bibfnamefont {B.}~\bibnamefont {Schallmo}},\ }\href {\doibase 10.21468/SciPostPhysCore.7.2.015} {\bibfield  {journal} {\bibinfo  {journal} {SciPost Phys. Core}\ }\textbf {\bibinfo {volume} {7}},\ \bibinfo {pages} {015} (\bibinfo {year} {2024})},\ \Eprint {http://arxiv.org/abs/2206.06328} {arXiv:2206.06328 [nucl-th]} \BibitemShut {NoStop}%
\bibitem [{\citenamefont {Braun}\ and\ \citenamefont {Schallmo}(2022{\natexlab{b}})}]{Braun:2022olp}%
  \BibitemOpen
  \bibfield  {author} {\bibinfo {author} {\bibfnamefont {J.}~\bibnamefont {Braun}}\ and\ \bibinfo {author} {\bibfnamefont {B.}~\bibnamefont {Schallmo}},\ }\href {\doibase 10.1103/PhysRevD.106.076010} {\bibfield  {journal} {\bibinfo  {journal} {Phys. Rev. D}\ }\textbf {\bibinfo {volume} {106}},\ \bibinfo {pages} {076010} (\bibinfo {year} {2022}{\natexlab{b}})},\ \Eprint {http://arxiv.org/abs/2204.00358} {arXiv:2204.00358 [nucl-th]} \BibitemShut {NoStop}%
\bibitem [{\citenamefont {Gei\ss{}el}\ \emph {et~al.}(2024)\citenamefont {Gei\ss{}el}, \citenamefont {Gorda},\ and\ \citenamefont {Braun}}]{Geissel:2024nmx}%
  \BibitemOpen
  \bibfield  {author} {\bibinfo {author} {\bibfnamefont {A.}~\bibnamefont {Gei\ss{}el}}, \bibinfo {author} {\bibfnamefont {T.}~\bibnamefont {Gorda}}, \ and\ \bibinfo {author} {\bibfnamefont {J.}~\bibnamefont {Braun}},\ }\href {\doibase 10.1103/PhysRevD.110.014034} {\bibfield  {journal} {\bibinfo  {journal} {Phys. Rev. D}\ }\textbf {\bibinfo {volume} {110}},\ \bibinfo {pages} {014034} (\bibinfo {year} {2024})},\ \Eprint {http://arxiv.org/abs/2403.18010} {arXiv:2403.18010 [hep-ph]} \BibitemShut {NoStop}%
\bibitem [{\citenamefont {Schertler}\ \emph {et~al.}(1999)\citenamefont {Schertler}, \citenamefont {Leupold},\ and\ \citenamefont {Schaffner-Bielich}}]{PhysRevC.60.025801}%
  \BibitemOpen
  \bibfield  {author} {\bibinfo {author} {\bibfnamefont {K.}~\bibnamefont {Schertler}}, \bibinfo {author} {\bibfnamefont {S.}~\bibnamefont {Leupold}}, \ and\ \bibinfo {author} {\bibfnamefont {J.}~\bibnamefont {Schaffner-Bielich}},\ }\href {\doibase 10.1103/PhysRevC.60.025801} {\bibfield  {journal} {\bibinfo  {journal} {Phys. Rev. C}\ }\textbf {\bibinfo {volume} {\textbf{60}}},\ \bibinfo {pages} {025801} (\bibinfo {year} {1999})}\BibitemShut {NoStop}%
\bibitem [{\citenamefont {Baldo}\ \emph {et~al.}(2003)\citenamefont {Baldo}, \citenamefont {Buballa}, \citenamefont {Burgio}, \citenamefont {Neumann}, \citenamefont {Oertel},\ and\ \citenamefont {Schulze}}]{Baldo:2002ju}%
  \BibitemOpen
  \bibfield  {author} {\bibinfo {author} {\bibfnamefont {M.}~\bibnamefont {Baldo}}, \bibinfo {author} {\bibfnamefont {M.}~\bibnamefont {Buballa}}, \bibinfo {author} {\bibfnamefont {F.}~\bibnamefont {Burgio}}, \bibinfo {author} {\bibfnamefont {F.}~\bibnamefont {Neumann}}, \bibinfo {author} {\bibfnamefont {M.}~\bibnamefont {Oertel}}, \ and\ \bibinfo {author} {\bibfnamefont {H.~J.}\ \bibnamefont {Schulze}},\ }\href {\doibase 10.1016/S0370-2693(03)00556-2} {\bibfield  {journal} {\bibinfo  {journal} {Phys. Lett. B}\ }\textbf {\bibinfo {volume} {562}},\ \bibinfo {pages} {153} (\bibinfo {year} {2003})},\ \Eprint {http://arxiv.org/abs/nucl-th/0212096} {arXiv:nucl-th/0212096} \BibitemShut {NoStop}%
\bibitem [{\citenamefont {Buballa}\ \emph {et~al.}(2004)\citenamefont {Buballa}, \citenamefont {Neumann}, \citenamefont {Oertel},\ and\ \citenamefont {Shovkovy}}]{Buballa:2003et}%
  \BibitemOpen
  \bibfield  {author} {\bibinfo {author} {\bibfnamefont {M.}~\bibnamefont {Buballa}}, \bibinfo {author} {\bibfnamefont {F.}~\bibnamefont {Neumann}}, \bibinfo {author} {\bibfnamefont {M.}~\bibnamefont {Oertel}}, \ and\ \bibinfo {author} {\bibfnamefont {I.}~\bibnamefont {Shovkovy}},\ }\href {\doibase 10.1016/j.physletb.2004.05.064} {\bibfield  {journal} {\bibinfo  {journal} {Phys. Lett. B}\ }\textbf {\bibinfo {volume} {595}},\ \bibinfo {pages} {36} (\bibinfo {year} {2004})},\ \Eprint {http://arxiv.org/abs/nucl-th/0312078} {arXiv:nucl-th/0312078} \BibitemShut {NoStop}%
\bibitem [{\citenamefont {Klahn}\ \emph {et~al.}(2007)\citenamefont {Klahn}, \citenamefont {Blaschke}, \citenamefont {Sandin}, \citenamefont {Fuchs}, \citenamefont {Faessler}, \citenamefont {Grigorian}, \citenamefont {Ropke},\ and\ \citenamefont {Trumper}}]{Klahn:2006iw}%
  \BibitemOpen
  \bibfield  {author} {\bibinfo {author} {\bibfnamefont {T.}~\bibnamefont {Klahn}}, \bibinfo {author} {\bibfnamefont {D.}~\bibnamefont {Blaschke}}, \bibinfo {author} {\bibfnamefont {F.}~\bibnamefont {Sandin}}, \bibinfo {author} {\bibfnamefont {C.}~\bibnamefont {Fuchs}}, \bibinfo {author} {\bibfnamefont {A.}~\bibnamefont {Faessler}}, \bibinfo {author} {\bibfnamefont {H.}~\bibnamefont {Grigorian}}, \bibinfo {author} {\bibfnamefont {G.}~\bibnamefont {Ropke}}, \ and\ \bibinfo {author} {\bibfnamefont {J.}~\bibnamefont {Trumper}},\ }\href {\doibase 10.1016/j.physletb.2007.08.048} {\bibfield  {journal} {\bibinfo  {journal} {Phys. Lett. B}\ }\textbf {\bibinfo {volume} {654}},\ \bibinfo {pages} {170} (\bibinfo {year} {2007})},\ \Eprint {http://arxiv.org/abs/nucl-th/0609067} {arXiv:nucl-th/0609067} \BibitemShut {NoStop}%
\bibitem [{\citenamefont {Pagliara}\ and\ \citenamefont {Schaffner-Bielich}(2008)}]{Pagliara:2007ph}%
  \BibitemOpen
  \bibfield  {author} {\bibinfo {author} {\bibfnamefont {G.}~\bibnamefont {Pagliara}}\ and\ \bibinfo {author} {\bibfnamefont {J.}~\bibnamefont {Schaffner-Bielich}},\ }\href {\doibase 10.1103/PhysRevD.77.063004} {\bibfield  {journal} {\bibinfo  {journal} {Phys. Rev. D}\ }\textbf {\bibinfo {volume} {77}},\ \bibinfo {pages} {063004} (\bibinfo {year} {2008})},\ \Eprint {http://arxiv.org/abs/0711.1119} {arXiv:0711.1119 [astro-ph]} \BibitemShut {NoStop}%
\bibitem [{\citenamefont {Bonanno}\ and\ \citenamefont {Sedrakian}(2012)}]{Bonanno:2011ch}%
  \BibitemOpen
  \bibfield  {author} {\bibinfo {author} {\bibfnamefont {L.}~\bibnamefont {Bonanno}}\ and\ \bibinfo {author} {\bibfnamefont {A.}~\bibnamefont {Sedrakian}},\ }\href {\doibase 10.1051/0004-6361/201117832} {\bibfield  {journal} {\bibinfo  {journal} {Astron. Astrophys.}\ }\textbf {\bibinfo {volume} {539}},\ \bibinfo {pages} {A16} (\bibinfo {year} {2012})},\ \Eprint {http://arxiv.org/abs/1108.0559} {arXiv:1108.0559 [astro-ph.SR]} \BibitemShut {NoStop}%
\bibitem [{\citenamefont {Kl\"ahn}\ \emph {et~al.}(2013)\citenamefont {Kl\"ahn}, \citenamefont {\L{}astowiecki},\ and\ \citenamefont {Blaschke}}]{Klahn:2013kga}%
  \BibitemOpen
  \bibfield  {author} {\bibinfo {author} {\bibfnamefont {T.}~\bibnamefont {Kl\"ahn}}, \bibinfo {author} {\bibfnamefont {R.}~\bibnamefont {\L{}astowiecki}}, \ and\ \bibinfo {author} {\bibfnamefont {D.~B.}\ \bibnamefont {Blaschke}},\ }\href {\doibase 10.1103/PhysRevD.88.085001} {\bibfield  {journal} {\bibinfo  {journal} {Phys. Rev. D}\ }\textbf {\bibinfo {volume} {88}},\ \bibinfo {pages} {085001} (\bibinfo {year} {2013})},\ \Eprint {http://arxiv.org/abs/1307.6996} {arXiv:1307.6996 [nucl-th]} \BibitemShut {NoStop}%
\bibitem [{\citenamefont {Baym}\ \emph {et~al.}(2018)\citenamefont {Baym}, \citenamefont {Hatsuda}, \citenamefont {Kojo}, \citenamefont {Powell}, \citenamefont {Song},\ and\ \citenamefont {Takatsuka}}]{Baym:2017whm}%
  \BibitemOpen
  \bibfield  {author} {\bibinfo {author} {\bibfnamefont {G.}~\bibnamefont {Baym}}, \bibinfo {author} {\bibfnamefont {T.}~\bibnamefont {Hatsuda}}, \bibinfo {author} {\bibfnamefont {T.}~\bibnamefont {Kojo}}, \bibinfo {author} {\bibfnamefont {P.~D.}\ \bibnamefont {Powell}}, \bibinfo {author} {\bibfnamefont {Y.}~\bibnamefont {Song}}, \ and\ \bibinfo {author} {\bibfnamefont {T.}~\bibnamefont {Takatsuka}},\ }\href {\doibase 10.1088/1361-6633/aaae14} {\bibfield  {journal} {\bibinfo  {journal} {Rept. Prog. Phys.}\ }\textbf {\bibinfo {volume} {81}},\ \bibinfo {pages} {056902} (\bibinfo {year} {2018})},\ \Eprint {http://arxiv.org/abs/1707.04966} {arXiv:1707.04966 [astro-ph.HE]} \BibitemShut {NoStop}%
\bibitem [{\citenamefont {Alaverdyan}(2020)}]{Alaverdyan:2020xnv}%
  \BibitemOpen
  \bibfield  {author} {\bibinfo {author} {\bibfnamefont {G.}~\bibnamefont {Alaverdyan}},\ }\href@noop {} {\  (\bibinfo {year} {2020})},\ \Eprint {http://arxiv.org/abs/2011.12593} {arXiv:2011.12593 [nucl-th]} \BibitemShut {NoStop}%
\bibitem [{\citenamefont {Alaverdyan}(2022)}]{Alaverdyan:2022foz}%
  \BibitemOpen
  \bibfield  {author} {\bibinfo {author} {\bibfnamefont {G.~B.}\ \bibnamefont {Alaverdyan}},\ }\href {\doibase 10.1007/s10511-022-09737-z} {\bibfield  {journal} {\bibinfo  {journal} {Astrophysics}\ }\textbf {\bibinfo {volume} {65}},\ \bibinfo {pages} {278} (\bibinfo {year} {2022})},\ \Eprint {http://arxiv.org/abs/2208.00466} {arXiv:2208.00466 [nucl-th]} \BibitemShut {NoStop}%
\bibitem [{\citenamefont {Gao}\ \emph {et~al.}(2024)\citenamefont {Gao}, \citenamefont {Yuan}, \citenamefont {Harada},\ and\ \citenamefont {Ma}}]{PhysRevC.110.045802}%
  \BibitemOpen
  \bibfield  {author} {\bibinfo {author} {\bibfnamefont {B.}~\bibnamefont {Gao}}, \bibinfo {author} {\bibfnamefont {W.-L.}\ \bibnamefont {Yuan}}, \bibinfo {author} {\bibfnamefont {M.}~\bibnamefont {Harada}}, \ and\ \bibinfo {author} {\bibfnamefont {Y.-L.}\ \bibnamefont {Ma}},\ }\href {\doibase 10.1103/PhysRevC.110.045802} {\bibfield  {journal} {\bibinfo  {journal} {Phys. Rev. C}\ }\textbf {\bibinfo {volume} {110}},\ \bibinfo {pages} {045802} (\bibinfo {year} {2024})}\BibitemShut {NoStop}%
\bibitem [{\citenamefont {Tanimoto}\ \emph {et~al.}(2020)\citenamefont {Tanimoto}, \citenamefont {Bentz},\ and\ \citenamefont {Clo\"et}}]{Tanimoto:2019tsl}%
  \BibitemOpen
  \bibfield  {author} {\bibinfo {author} {\bibfnamefont {T.}~\bibnamefont {Tanimoto}}, \bibinfo {author} {\bibfnamefont {W.}~\bibnamefont {Bentz}}, \ and\ \bibinfo {author} {\bibfnamefont {I.~C.}\ \bibnamefont {Clo\"et}},\ }\href {\doibase 10.1103/PhysRevC.101.055204} {\bibfield  {journal} {\bibinfo  {journal} {Phys. Rev. C}\ }\textbf {\bibinfo {volume} {101}},\ \bibinfo {pages} {055204} (\bibinfo {year} {2020})},\ \Eprint {http://arxiv.org/abs/1903.06851} {arXiv:1903.06851 [nucl-th]} \BibitemShut {NoStop}%
\bibitem [{\citenamefont {Shahrbaf}\ \emph {et~al.}(2023)\citenamefont {Shahrbaf}, \citenamefont {Anti\'c}, \citenamefont {Ayriyan}, \citenamefont {Blaschke},\ and\ \citenamefont {Grunfeld}}]{Shahrbaf:2021cjz}%
  \BibitemOpen
  \bibfield  {author} {\bibinfo {author} {\bibfnamefont {M.}~\bibnamefont {Shahrbaf}}, \bibinfo {author} {\bibfnamefont {S.}~\bibnamefont {Anti\'c}}, \bibinfo {author} {\bibfnamefont {A.}~\bibnamefont {Ayriyan}}, \bibinfo {author} {\bibfnamefont {D.}~\bibnamefont {Blaschke}}, \ and\ \bibinfo {author} {\bibfnamefont {A.~G.}\ \bibnamefont {Grunfeld}},\ }\href {\doibase 10.1103/PhysRevD.107.054011} {\bibfield  {journal} {\bibinfo  {journal} {Phys. Rev. D}\ }\textbf {\bibinfo {volume} {107}},\ \bibinfo {pages} {054011} (\bibinfo {year} {2023})},\ \Eprint {http://arxiv.org/abs/2105.00029} {arXiv:2105.00029 [nucl-th]} \BibitemShut {NoStop}%
\bibitem [{\citenamefont {Blaschke}\ \emph {et~al.}(2023)\citenamefont {Blaschke}, \citenamefont {Shukla}, \citenamefont {Ivanytskyi},\ and\ \citenamefont {Liebing}}]{Blaschke:2022egm}%
  \BibitemOpen
  \bibfield  {author} {\bibinfo {author} {\bibfnamefont {D.}~\bibnamefont {Blaschke}}, \bibinfo {author} {\bibfnamefont {U.}~\bibnamefont {Shukla}}, \bibinfo {author} {\bibfnamefont {O.}~\bibnamefont {Ivanytskyi}}, \ and\ \bibinfo {author} {\bibfnamefont {S.}~\bibnamefont {Liebing}},\ }\href {\doibase 10.1103/PhysRevD.107.063034} {\bibfield  {journal} {\bibinfo  {journal} {Phys. Rev. D}\ }\textbf {\bibinfo {volume} {107}},\ \bibinfo {pages} {063034} (\bibinfo {year} {2023})},\ \Eprint {http://arxiv.org/abs/2212.14856} {arXiv:2212.14856 [nucl-th]} \BibitemShut {NoStop}%
\bibitem [{\citenamefont {Ivanytskyi}(2024)}]{Ivanytskyi:2024zip}%
  \BibitemOpen
  \bibfield  {author} {\bibinfo {author} {\bibfnamefont {O.}~\bibnamefont {Ivanytskyi}},\ }\href@noop {} {\  (\bibinfo {year} {2024})},\ \Eprint {http://arxiv.org/abs/2409.05859} {arXiv:2409.05859 [hep-ph]} \BibitemShut {NoStop}%
\bibitem [{\citenamefont {Zhang}\ and\ \citenamefont {Mann}(2021)}]{Zhang:2020jmb}%
  \BibitemOpen
  \bibfield  {author} {\bibinfo {author} {\bibfnamefont {C.}~\bibnamefont {Zhang}}\ and\ \bibinfo {author} {\bibfnamefont {R.~B.}\ \bibnamefont {Mann}},\ }\href {\doibase 10.1103/PhysRevD.103.063018} {\bibfield  {journal} {\bibinfo  {journal} {Phys. Rev. D}\ }\textbf {\bibinfo {volume} {103}},\ \bibinfo {pages} {063018} (\bibinfo {year} {2021})},\ \Eprint {http://arxiv.org/abs/2009.07182} {arXiv:2009.07182 [astro-ph.HE]} \BibitemShut {NoStop}%
\bibitem [{\citenamefont {Christian}\ \emph {et~al.}(2024{\natexlab{b}})\citenamefont {Christian}, \citenamefont {Schaffner-Bielich},\ and\ \citenamefont {Rosswog}}]{christian2024orderphasetransitionsquark}%
  \BibitemOpen
  \bibfield  {author} {\bibinfo {author} {\bibfnamefont {J.-E.}\ \bibnamefont {Christian}}, \bibinfo {author} {\bibfnamefont {J.}~\bibnamefont {Schaffner-Bielich}}, \ and\ \bibinfo {author} {\bibfnamefont {S.}~\bibnamefont {Rosswog}},\ }\href {https://arxiv.org/abs/2312.10148} {\enquote {\bibinfo {title} {Which first order phase transitions to quark matter are possible in neutron stars?}}\ } (\bibinfo {year} {2024}{\natexlab{b}}),\ \Eprint {http://arxiv.org/abs/2312.10148} {arXiv:2312.10148 [nucl-th]} \BibitemShut {NoStop}%
\bibitem [{\citenamefont {Alford}\ and\ \citenamefont {Sedrakian}(2017)}]{Alford:2017qgh}%
  \BibitemOpen
  \bibfield  {author} {\bibinfo {author} {\bibfnamefont {M.~G.}\ \bibnamefont {Alford}}\ and\ \bibinfo {author} {\bibfnamefont {A.}~\bibnamefont {Sedrakian}},\ }\href {\doibase 10.1103/PhysRevLett.119.161104} {\bibfield  {journal} {\bibinfo  {journal} {Phys. Rev. Lett.}\ }\textbf {\bibinfo {volume} {119}},\ \bibinfo {pages} {161104} (\bibinfo {year} {2017})},\ \Eprint {http://arxiv.org/abs/1706.01592} {arXiv:1706.01592 [astro-ph.HE]} \BibitemShut {NoStop}%
\bibitem [{\citenamefont {Li}\ \emph {et~al.}(2020)\citenamefont {Li}, \citenamefont {Sedrakian},\ and\ \citenamefont {Alford}}]{Li:2019fqe}%
  \BibitemOpen
  \bibfield  {author} {\bibinfo {author} {\bibfnamefont {J.~J.}\ \bibnamefont {Li}}, \bibinfo {author} {\bibfnamefont {A.}~\bibnamefont {Sedrakian}}, \ and\ \bibinfo {author} {\bibfnamefont {M.}~\bibnamefont {Alford}},\ }\href {\doibase 10.1103/PhysRevD.101.063022} {\bibfield  {journal} {\bibinfo  {journal} {Phys. Rev. D}\ }\textbf {\bibinfo {volume} {101}},\ \bibinfo {pages} {063022} (\bibinfo {year} {2020})},\ \Eprint {http://arxiv.org/abs/1911.00276} {arXiv:1911.00276 [astro-ph.HE]} \BibitemShut {NoStop}%
\bibitem [{\citenamefont {Li}\ \emph {et~al.}(2023)\citenamefont {Li}, \citenamefont {Sedrakian},\ and\ \citenamefont {Alford}}]{Li:2023zty}%
  \BibitemOpen
  \bibfield  {author} {\bibinfo {author} {\bibfnamefont {J.~J.}\ \bibnamefont {Li}}, \bibinfo {author} {\bibfnamefont {A.}~\bibnamefont {Sedrakian}}, \ and\ \bibinfo {author} {\bibfnamefont {M.}~\bibnamefont {Alford}},\ }\href {\doibase 10.3847/1538-4357/acb688} {\bibfield  {journal} {\bibinfo  {journal} {Astrophys. J.}\ }\textbf {\bibinfo {volume} {944}},\ \bibinfo {pages} {206} (\bibinfo {year} {2023})},\ \Eprint {http://arxiv.org/abs/2301.10940} {arXiv:2301.10940 [astro-ph.HE]} \BibitemShut {NoStop}%
\bibitem [{\citenamefont {Ranea-Sandoval}\ \emph {et~al.}(2017)\citenamefont {Ranea-Sandoval}, \citenamefont {Orsaria}, \citenamefont {Han}, \citenamefont {Weber},\ and\ \citenamefont {Spinella}}]{Ranea-Sandoval:2017ort}%
  \BibitemOpen
  \bibfield  {author} {\bibinfo {author} {\bibfnamefont {I.~F.}\ \bibnamefont {Ranea-Sandoval}}, \bibinfo {author} {\bibfnamefont {M.~G.}\ \bibnamefont {Orsaria}}, \bibinfo {author} {\bibfnamefont {S.}~\bibnamefont {Han}}, \bibinfo {author} {\bibfnamefont {F.}~\bibnamefont {Weber}}, \ and\ \bibinfo {author} {\bibfnamefont {W.~M.}\ \bibnamefont {Spinella}},\ }\href {\doibase 10.1103/PhysRevC.96.065807} {\bibfield  {journal} {\bibinfo  {journal} {Phys. Rev. C}\ }\textbf {\bibinfo {volume} {96}},\ \bibinfo {pages} {065807} (\bibinfo {year} {2017})}\BibitemShut {NoStop}%
\bibitem [{\citenamefont {Rehberg}\ \emph {et~al.}(1996)\citenamefont {Rehberg}, \citenamefont {Klevansky},\ and\ \citenamefont {Hufner}}]{Rehberg:1995kh}%
  \BibitemOpen
  \bibfield  {author} {\bibinfo {author} {\bibfnamefont {P.}~\bibnamefont {Rehberg}}, \bibinfo {author} {\bibfnamefont {S.~P.}\ \bibnamefont {Klevansky}}, \ and\ \bibinfo {author} {\bibfnamefont {J.}~\bibnamefont {Hufner}},\ }\href {\doibase 10.1103/PhysRevC.53.410} {\bibfield  {journal} {\bibinfo  {journal} {Phys. Rev. C}\ }\textbf {\bibinfo {volume} {53}},\ \bibinfo {pages} {410} (\bibinfo {year} {1996})},\ \Eprint {http://arxiv.org/abs/hep-ph/9506436} {arXiv:hep-ph/9506436} \BibitemShut {NoStop}%
\bibitem [{\citenamefont {Gholami}\ \emph {et~al.}(2024)\citenamefont {Gholami}, \citenamefont {Hofmann},\ and\ \citenamefont {Buballa}}]{Gholami:2024diy}%
  \BibitemOpen
  \bibfield  {author} {\bibinfo {author} {\bibfnamefont {H.}~\bibnamefont {Gholami}}, \bibinfo {author} {\bibfnamefont {M.}~\bibnamefont {Hofmann}}, \ and\ \bibinfo {author} {\bibfnamefont {M.}~\bibnamefont {Buballa}},\ }\href@noop {} {\  (\bibinfo {year} {2024})},\ \Eprint {http://arxiv.org/abs/2408.06704} {arXiv:2408.06704 [hep-ph]} \BibitemShut {NoStop}%
\bibitem [{\citenamefont {Braun}\ \emph {et~al.}(2019)\citenamefont {Braun}, \citenamefont {Leonhardt},\ and\ \citenamefont {Pawlowski}}]{Braun:2018svj}%
  \BibitemOpen
  \bibfield  {author} {\bibinfo {author} {\bibfnamefont {J.}~\bibnamefont {Braun}}, \bibinfo {author} {\bibfnamefont {M.}~\bibnamefont {Leonhardt}}, \ and\ \bibinfo {author} {\bibfnamefont {J.~M.}\ \bibnamefont {Pawlowski}},\ }\href {\doibase 10.21468/SciPostPhys.6.5.056} {\bibfield  {journal} {\bibinfo  {journal} {SciPost Phys.}\ }\textbf {\bibinfo {volume} {6}},\ \bibinfo {pages} {056} (\bibinfo {year} {2019})},\ \Eprint {http://arxiv.org/abs/1806.04432} {arXiv:1806.04432 [hep-ph]} \BibitemShut {NoStop}%
\bibitem [{\citenamefont {Kobayashi}\ and\ \citenamefont {Maskawa}(1970)}]{Kobayashi:1970ji}%
  \BibitemOpen
  \bibfield  {author} {\bibinfo {author} {\bibfnamefont {M.}~\bibnamefont {Kobayashi}}\ and\ \bibinfo {author} {\bibfnamefont {T.}~\bibnamefont {Maskawa}},\ }\href {\doibase 10.1143/PTP.44.1422} {\bibfield  {journal} {\bibinfo  {journal} {Prog. Theor. Phys.}\ }\textbf {\bibinfo {volume} {44}},\ \bibinfo {pages} {1422} (\bibinfo {year} {1970})}\BibitemShut {NoStop}%
\bibitem [{\citenamefont {'t~Hooft}(1976)}]{tHooft:1976rip}%
  \BibitemOpen
  \bibfield  {author} {\bibinfo {author} {\bibfnamefont {G.}~\bibnamefont {'t~Hooft}},\ }\href {\doibase 10.1103/PhysRevLett.37.8} {\bibfield  {journal} {\bibinfo  {journal} {Phys. Rev. Lett.}\ }\textbf {\bibinfo {volume} {37}},\ \bibinfo {pages} {8} (\bibinfo {year} {1976})}\BibitemShut {NoStop}%
\bibitem [{\citenamefont {Buballa}(2005)}]{BUBALLA2005205}%
  \BibitemOpen
  \bibfield  {author} {\bibinfo {author} {\bibfnamefont {M.}~\bibnamefont {Buballa}},\ }\href {\doibase https://doi.org/10.1016/j.physrep.2004.11.004} {\bibfield  {journal} {\bibinfo  {journal} {Physics Reports}\ }\textbf {\bibinfo {volume} {407}},\ \bibinfo {pages} {205 } (\bibinfo {year} {2005})}\BibitemShut {NoStop}%
\bibitem [{\citenamefont {Zel'dovich}(1961)}]{Zeldovich:1961sbr}%
  \BibitemOpen
  \bibfield  {author} {\bibinfo {author} {\bibfnamefont {Y.~B.}\ \bibnamefont {Zel'dovich}},\ }\href@noop {} {\bibfield  {journal} {\bibinfo  {journal} {Zh. Eksp. Teor. Fiz.}\ }\textbf {\bibinfo {volume} {41}},\ \bibinfo {pages} {1609} (\bibinfo {year} {1961})}\BibitemShut {NoStop}%
\bibitem [{\citenamefont {Riley}\ \emph {et~al.}(2021{\natexlab{b}})\citenamefont {Riley} \emph {et~al.}}]{2021ApJ...918L..27R}%
  \BibitemOpen
  \bibfield  {author} {\bibinfo {author} {\bibfnamefont {T.~E.}\ \bibnamefont {Riley}} \emph {et~al.},\ }\href {\doibase 10.3847/2041-8213/ac0a81} {\bibfield  {journal} {\bibinfo  {journal} {The Astrophysical Journal Letters}\ }\textbf {\bibinfo {volume} {918}},\ \bibinfo {pages} {L27} (\bibinfo {year} {2021}{\natexlab{b}})}\BibitemShut {NoStop}%
\bibitem [{\citenamefont {Miller~{\textit{et al.}}}(2019)}]{Miller_2019a}%
  \BibitemOpen
  \bibfield  {author} {\bibinfo {author} {\bibfnamefont {M.~C.}\ \bibnamefont {Miller~{\textit{et al.}}}},\ }\href {\doibase 10.3847/2041-8213/ab50c5} {\bibfield  {journal} {\bibinfo  {journal} {Astrophys. J.}\ }\textbf {\bibinfo {volume} {887}},\ \bibinfo {pages} {L24} (\bibinfo {year} {2019})}\BibitemShut {NoStop}%
\bibitem [{\citenamefont {Riley~{\textit{et al.}}}(2019)}]{Riley_2019}%
  \BibitemOpen
  \bibfield  {author} {\bibinfo {author} {\bibfnamefont {T.~E.}\ \bibnamefont {Riley~{\textit{et al.}}}},\ }\href {\doibase 10.3847/2041-8213/ab481c} {\bibfield  {journal} {\bibinfo  {journal} {Astrophys. J.}\ }\textbf {\bibinfo {volume} {887}},\ \bibinfo {pages} {L21} (\bibinfo {year} {2019})}\BibitemShut {NoStop}%
\bibitem [{\citenamefont {Doroshenko}\ \emph {et~al.}(2022)\citenamefont {Doroshenko}, \citenamefont {Suleimanov}, \citenamefont {P{\"u}hlhofer},\ and\ \citenamefont {Santangelo}}]{Doroshenko2022}%
  \BibitemOpen
  \bibfield  {author} {\bibinfo {author} {\bibfnamefont {V.}~\bibnamefont {Doroshenko}}, \bibinfo {author} {\bibfnamefont {V.}~\bibnamefont {Suleimanov}}, \bibinfo {author} {\bibfnamefont {G.}~\bibnamefont {P{\"u}hlhofer}}, \ and\ \bibinfo {author} {\bibfnamefont {A.}~\bibnamefont {Santangelo}},\ }\href {\doibase 10.1038/s41550-022-01800-1} {\bibfield  {journal} {\bibinfo  {journal} {Nature Astronomy}\ }\textbf {\bibinfo {volume} {6}},\ \bibinfo {pages} {1444} (\bibinfo {year} {2022})}\BibitemShut {NoStop}%
\bibitem [{\citenamefont {Oppenheimer}\ and\ \citenamefont {Volkoff}(1939)}]{PhysRev.55.374}%
  \BibitemOpen
  \bibfield  {author} {\bibinfo {author} {\bibfnamefont {J.~R.}\ \bibnamefont {Oppenheimer}}\ and\ \bibinfo {author} {\bibfnamefont {G.~M.}\ \bibnamefont {Volkoff}},\ }\href {\doibase 10.1103/PhysRev.55.374} {\bibfield  {journal} {\bibinfo  {journal} {Phys. Rev.}\ }\textbf {\bibinfo {volume} {55}},\ \bibinfo {pages} {374} (\bibinfo {year} {1939})}\BibitemShut {NoStop}%
\bibitem [{\citenamefont {Tolman}(1939)}]{PhysRev.55.364}%
  \BibitemOpen
  \bibfield  {author} {\bibinfo {author} {\bibfnamefont {R.~C.}\ \bibnamefont {Tolman}},\ }\href {\doibase 10.1103/PhysRev.55.364} {\bibfield  {journal} {\bibinfo  {journal} {Phys. Rev.}\ }\textbf {\bibinfo {volume} {55}},\ \bibinfo {pages} {364} (\bibinfo {year} {1939})}\BibitemShut {NoStop}%
\bibitem [{\citenamefont {Buballa}\ and\ \citenamefont {Oertel}(1999)}]{Buballa:1998pr}%
  \BibitemOpen
  \bibfield  {author} {\bibinfo {author} {\bibfnamefont {M.}~\bibnamefont {Buballa}}\ and\ \bibinfo {author} {\bibfnamefont {M.}~\bibnamefont {Oertel}},\ }\href {\doibase 10.1016/S0370-2693(99)00533-X} {\bibfield  {journal} {\bibinfo  {journal} {Phys. Lett. B}\ }\textbf {\bibinfo {volume} {457}},\ \bibinfo {pages} {261} (\bibinfo {year} {1999})},\ \Eprint {http://arxiv.org/abs/hep-ph/9810529} {arXiv:hep-ph/9810529} \BibitemShut {NoStop}%
\bibitem [{\citenamefont {Witten}(1984)}]{Witten1984}%
  \BibitemOpen
  \bibfield  {author} {\bibinfo {author} {\bibfnamefont {E.}~\bibnamefont {Witten}},\ }\href@noop {} {\bibfield  {journal} {\bibinfo  {journal} {Physical Review D}\ }\textbf {\bibinfo {volume} {30}},\ \bibinfo {pages} {272} (\bibinfo {year} {1984})}\BibitemShut {NoStop}%
\bibitem [{\citenamefont {Bodmer}(1971)}]{Bodmer1971}%
  \BibitemOpen
  \bibfield  {author} {\bibinfo {author} {\bibfnamefont {A.~R.}\ \bibnamefont {Bodmer}},\ }\href@noop {} {\bibfield  {journal} {\bibinfo  {journal} {Physical Review D}\ }\textbf {\bibinfo {volume} {4}},\ \bibinfo {pages} {1601} (\bibinfo {year} {1971})}\BibitemShut {NoStop}%
\bibitem [{\citenamefont {Farhi}\ and\ \citenamefont {Jaffe}(1984)}]{Farhi:1984qu}%
  \BibitemOpen
  \bibfield  {author} {\bibinfo {author} {\bibfnamefont {E.}~\bibnamefont {Farhi}}\ and\ \bibinfo {author} {\bibfnamefont {R.~L.}\ \bibnamefont {Jaffe}},\ }\href {\doibase 10.1103/PhysRevD.30.2379} {\bibfield  {journal} {\bibinfo  {journal} {Phys. Rev. D}\ }\textbf {\bibinfo {volume} {30}},\ \bibinfo {pages} {2379} (\bibinfo {year} {1984})}\BibitemShut {NoStop}%
\bibitem [{\citenamefont {Vinciguerra}\ \emph {et~al.}(2024)\citenamefont {Vinciguerra} \emph {et~al.}}]{Vinciguerra:2023qxq}%
  \BibitemOpen
  \bibfield  {author} {\bibinfo {author} {\bibfnamefont {S.}~\bibnamefont {Vinciguerra}} \emph {et~al.},\ }\href {\doibase 10.3847/1538-4357/acfb83} {\bibfield  {journal} {\bibinfo  {journal} {Astrophys. J.}\ }\textbf {\bibinfo {volume} {961}},\ \bibinfo {pages} {62} (\bibinfo {year} {2024})},\ \Eprint {http://arxiv.org/abs/2308.09469} {arXiv:2308.09469 [astro-ph.HE]} \BibitemShut {NoStop}%
\bibitem [{\citenamefont {Brodie}\ and\ \citenamefont {Haber}(2023)}]{Brodie:2023pjw}%
  \BibitemOpen
  \bibfield  {author} {\bibinfo {author} {\bibfnamefont {L.}~\bibnamefont {Brodie}}\ and\ \bibinfo {author} {\bibfnamefont {A.}~\bibnamefont {Haber}},\ }\href {\doibase 10.1103/PhysRevC.108.025806} {\bibfield  {journal} {\bibinfo  {journal} {Phys. Rev. C}\ }\textbf {\bibinfo {volume} {108}},\ \bibinfo {pages} {025806} (\bibinfo {year} {2023})},\ \Eprint {http://arxiv.org/abs/2302.02989} {arXiv:2302.02989 [nucl-th]} \BibitemShut {NoStop}%
\bibitem [{\citenamefont {Tewari}\ \emph {et~al.}(2024)\citenamefont {Tewari}, \citenamefont {Chatterjee}, \citenamefont {Kumar},\ and\ \citenamefont {Mallick}}]{Tewari:2024qit}%
  \BibitemOpen
  \bibfield  {author} {\bibinfo {author} {\bibfnamefont {S.}~\bibnamefont {Tewari}}, \bibinfo {author} {\bibfnamefont {S.}~\bibnamefont {Chatterjee}}, \bibinfo {author} {\bibfnamefont {D.}~\bibnamefont {Kumar}}, \ and\ \bibinfo {author} {\bibfnamefont {R.}~\bibnamefont {Mallick}},\ }\href@noop {} {\  (\bibinfo {year} {2024})},\ \Eprint {http://arxiv.org/abs/2410.20355} {arXiv:2410.20355 [astro-ph.HE]} \BibitemShut {NoStop}%
\bibitem [{\citenamefont {Mariani}\ \emph {et~al.}(2024)\citenamefont {Mariani}, \citenamefont {Ranea-Sandoval}, \citenamefont {Lugones},\ and\ \citenamefont {Orsaria}}]{Mariani:2024gqi}%
  \BibitemOpen
  \bibfield  {author} {\bibinfo {author} {\bibfnamefont {M.}~\bibnamefont {Mariani}}, \bibinfo {author} {\bibfnamefont {I.~F.}\ \bibnamefont {Ranea-Sandoval}}, \bibinfo {author} {\bibfnamefont {G.}~\bibnamefont {Lugones}}, \ and\ \bibinfo {author} {\bibfnamefont {M.~G.}\ \bibnamefont {Orsaria}},\ }\href {\doibase 10.1103/PhysRevD.110.043026} {\bibfield  {journal} {\bibinfo  {journal} {Phys. Rev. D}\ }\textbf {\bibinfo {volume} {110}},\ \bibinfo {pages} {043026} (\bibinfo {year} {2024})},\ \Eprint {http://arxiv.org/abs/2407.06347} {arXiv:2407.06347 [astro-ph.HE]} \BibitemShut {NoStop}%
\bibitem [{\citenamefont {{Seidov}}(1971)}]{1971SvA....15..347S}%
  \BibitemOpen
  \bibfield  {author} {\bibinfo {author} {\bibfnamefont {Z.~F.}\ \bibnamefont {{Seidov}}},\ }\href@noop {} {\bibfield  {journal} {\bibinfo  {journal} {Sov. Astron.}\ }\textbf {\bibinfo {volume} {15}},\ \bibinfo {pages} {347} (\bibinfo {year} {1971})}\BibitemShut {NoStop}%
\bibitem [{\citenamefont {K\"ampfer}(1981)}]{Kampfer:1981zmq}%
  \BibitemOpen
  \bibfield  {author} {\bibinfo {author} {\bibfnamefont {B.}~\bibnamefont {K\"ampfer}},\ }\href {\doibase 10.1016/0370-2693(81)90065-4} {\bibfield  {journal} {\bibinfo  {journal} {Phys. Lett. B}\ }\textbf {\bibinfo {volume} {101}},\ \bibinfo {pages} {366} (\bibinfo {year} {1981})}\BibitemShut {NoStop}%
\bibitem [{\citenamefont {Serot}\ and\ \citenamefont {Walecka}(1986)}]{Serot:1984ey}%
  \BibitemOpen
  \bibfield  {author} {\bibinfo {author} {\bibfnamefont {B.~D.}\ \bibnamefont {Serot}}\ and\ \bibinfo {author} {\bibfnamefont {J.~D.}\ \bibnamefont {Walecka}},\ }\href@noop {} {\bibfield  {journal} {\bibinfo  {journal} {Adv. Nucl. Phys.}\ }\textbf {\bibinfo {volume} {16}},\ \bibinfo {pages} {1} (\bibinfo {year} {1986})}\BibitemShut {NoStop}%
\bibitem [{\citenamefont {Glendenning}(1997)}]{Glendenning:1997wn}%
  \BibitemOpen
  \bibfield  {author} {\bibinfo {author} {\bibfnamefont {N.~K.}\ \bibnamefont {Glendenning}},\ }\href@noop {} {\emph {\bibinfo {title} {{Compact stars: Nuclear physics, particle physics, and general relativity}}}}\ (\bibinfo {year} {1997})\BibitemShut {NoStop}%
\bibitem [{\citenamefont {Horowitz}\ and\ \citenamefont {Piekarewicz}(2001)}]{PhysRevLett.86.5647}%
  \BibitemOpen
  \bibfield  {author} {\bibinfo {author} {\bibfnamefont {C.~J.}\ \bibnamefont {Horowitz}}\ and\ \bibinfo {author} {\bibfnamefont {J.}~\bibnamefont {Piekarewicz}},\ }\href {\doibase 10.1103/PhysRevLett.86.5647} {\bibfield  {journal} {\bibinfo  {journal} {Phys. Rev. Lett.}\ }\textbf {\bibinfo {volume} {86}},\ \bibinfo {pages} {5647} (\bibinfo {year} {2001})}\BibitemShut {NoStop}%
\bibitem [{\citenamefont {Walecka}(1974)}]{Walecka:1974qa}%
  \BibitemOpen
  \bibfield  {author} {\bibinfo {author} {\bibfnamefont {J.~D.}\ \bibnamefont {Walecka}},\ }\href {\doibase 10.1016/0003-4916(74)90208-5} {\bibfield  {journal} {\bibinfo  {journal} {Ann. Phys.}\ }\textbf {\bibinfo {volume} {\textbf{83}}},\ \bibinfo {pages} {491} (\bibinfo {year} {1974})}\BibitemShut {NoStop}%
\bibitem [{\citenamefont {Mueller}\ and\ \citenamefont {Serot}(1996)}]{Mueller:1996pm}%
  \BibitemOpen
  \bibfield  {author} {\bibinfo {author} {\bibfnamefont {H.}~\bibnamefont {Mueller}}\ and\ \bibinfo {author} {\bibfnamefont {B.~D.}\ \bibnamefont {Serot}},\ }\href {\doibase 10.1016/0375-9474(96)00187-X} {\bibfield  {journal} {\bibinfo  {journal} {Nucl. Phys. A}\ }\textbf {\bibinfo {volume} {606}},\ \bibinfo {pages} {508} (\bibinfo {year} {1996})},\ \Eprint {http://arxiv.org/abs/nucl-th/9603037} {arXiv:nucl-th/9603037} \BibitemShut {NoStop}%
\bibitem [{\citenamefont {Serot}\ and\ \citenamefont {Walecka}(1997)}]{Serot:1997xg}%
  \BibitemOpen
  \bibfield  {author} {\bibinfo {author} {\bibfnamefont {B.~D.}\ \bibnamefont {Serot}}\ and\ \bibinfo {author} {\bibfnamefont {J.~D.}\ \bibnamefont {Walecka}},\ }\href {\doibase 10.1142/S0218301397000299} {\bibfield  {journal} {\bibinfo  {journal} {Int. J. Mod. Phys. E}\ }\textbf {\bibinfo {volume} {6}},\ \bibinfo {pages} {515} (\bibinfo {year} {1997})},\ \Eprint {http://arxiv.org/abs/nucl-th/9701058} {arXiv:nucl-th/9701058} \BibitemShut {NoStop}%
\bibitem [{\citenamefont {Gambhir}\ \emph {et~al.}(1990)\citenamefont {Gambhir}, \citenamefont {Ring},\ and\ \citenamefont {Thimet}}]{Gambhir:1990uyn}%
  \BibitemOpen
  \bibfield  {author} {\bibinfo {author} {\bibfnamefont {Y.~K.}\ \bibnamefont {Gambhir}}, \bibinfo {author} {\bibfnamefont {P.}~\bibnamefont {Ring}}, \ and\ \bibinfo {author} {\bibfnamefont {A.}~\bibnamefont {Thimet}},\ }\href {\doibase 10.1016/0003-4916(90)90330-Q} {\bibfield  {journal} {\bibinfo  {journal} {Annals Phys.}\ }\textbf {\bibinfo {volume} {198}},\ \bibinfo {pages} {132} (\bibinfo {year} {1990})}\BibitemShut {NoStop}%
\bibitem [{\citenamefont {Hornick}\ \emph {et~al.}(2018)\citenamefont {Hornick}, \citenamefont {Tolos}, \citenamefont {Zacchi}, \citenamefont {Christian},\ and\ \citenamefont {Schaffner-Bielich}}]{PhysRevC.98.065804}%
  \BibitemOpen
  \bibfield  {author} {\bibinfo {author} {\bibfnamefont {N.}~\bibnamefont {Hornick}}, \bibinfo {author} {\bibfnamefont {L.}~\bibnamefont {Tolos}}, \bibinfo {author} {\bibfnamefont {A.}~\bibnamefont {Zacchi}}, \bibinfo {author} {\bibfnamefont {J.-E.}\ \bibnamefont {Christian}}, \ and\ \bibinfo {author} {\bibfnamefont {J.}~\bibnamefont {Schaffner-Bielich}},\ }\href {\doibase 10.1103/PhysRevC.98.065804} {\bibfield  {journal} {\bibinfo  {journal} {Phys. Rev. C}\ }\textbf {\bibinfo {volume} {98}},\ \bibinfo {pages} {065804} (\bibinfo {year} {2018})}\BibitemShut {NoStop}%
\bibitem [{\citenamefont {Drischler}\ \emph {et~al.}(2016{\natexlab{b}})\citenamefont {Drischler}, \citenamefont {Carbone}, \citenamefont {Hebeler},\ and\ \citenamefont {Schwenk}}]{PhysRevC.94.054307}%
  \BibitemOpen
  \bibfield  {author} {\bibinfo {author} {\bibfnamefont {C.}~\bibnamefont {Drischler}}, \bibinfo {author} {\bibfnamefont {A.}~\bibnamefont {Carbone}}, \bibinfo {author} {\bibfnamefont {K.}~\bibnamefont {Hebeler}}, \ and\ \bibinfo {author} {\bibfnamefont {A.}~\bibnamefont {Schwenk}},\ }\href {\doibase 10.1103/PhysRevC.94.054307} {\bibfield  {journal} {\bibinfo  {journal} {Phys. Rev. C}\ }\textbf {\bibinfo {volume} {94}},\ \bibinfo {pages} {054307} (\bibinfo {year} {2016}{\natexlab{b}})}\BibitemShut {NoStop}%
\bibitem [{\citenamefont {Kurkela}\ \emph {et~al.}(2024)\citenamefont {Kurkela}, \citenamefont {Rajagopal},\ and\ \citenamefont {Steinhorst}}]{PhysRevLett.132.262701}%
  \BibitemOpen
  \bibfield  {author} {\bibinfo {author} {\bibfnamefont {A.}~\bibnamefont {Kurkela}}, \bibinfo {author} {\bibfnamefont {K.}~\bibnamefont {Rajagopal}}, \ and\ \bibinfo {author} {\bibfnamefont {R.}~\bibnamefont {Steinhorst}},\ }\href {\doibase 10.1103/PhysRevLett.132.262701} {\bibfield  {journal} {\bibinfo  {journal} {Phys. Rev. Lett.}\ }\textbf {\bibinfo {volume} {132}},\ \bibinfo {pages} {262701} (\bibinfo {year} {2024})}\BibitemShut {NoStop}%
\bibitem [{\citenamefont {Hempel}\ and\ \citenamefont {Schaffner-Bielich}(2010)}]{Hempel:2009mc}%
  \BibitemOpen
  \bibfield  {author} {\bibinfo {author} {\bibfnamefont {M.}~\bibnamefont {Hempel}}\ and\ \bibinfo {author} {\bibfnamefont {J.}~\bibnamefont {Schaffner-Bielich}},\ }\href {\doibase 10.1016/j.nuclphysa.2010.02.010} {\bibfield  {journal} {\bibinfo  {journal} {Nucl. Phys. A}\ }\textbf {\bibinfo {volume} {837}},\ \bibinfo {pages} {210} (\bibinfo {year} {2010})},\ \Eprint {http://arxiv.org/abs/0911.4073} {arXiv:0911.4073 [nucl-th]} \BibitemShut {NoStop}%
\bibitem [{\citenamefont {Typel}\ \emph {et~al.}(2010{\natexlab{b}})\citenamefont {Typel}, \citenamefont {Ropke}, \citenamefont {Klahn}, \citenamefont {Blaschke},\ and\ \citenamefont {Wolter}}]{Typel:2009sy}%
  \BibitemOpen
  \bibfield  {author} {\bibinfo {author} {\bibfnamefont {S.}~\bibnamefont {Typel}}, \bibinfo {author} {\bibfnamefont {G.}~\bibnamefont {Ropke}}, \bibinfo {author} {\bibfnamefont {T.}~\bibnamefont {Klahn}}, \bibinfo {author} {\bibfnamefont {D.}~\bibnamefont {Blaschke}}, \ and\ \bibinfo {author} {\bibfnamefont {H.~H.}\ \bibnamefont {Wolter}},\ }\href {\doibase 10.1103/PhysRevC.81.015803} {\bibfield  {journal} {\bibinfo  {journal} {Phys. Rev. C}\ }\textbf {\bibinfo {volume} {81}},\ \bibinfo {pages} {015803} (\bibinfo {year} {2010}{\natexlab{b}})},\ \Eprint {http://arxiv.org/abs/0908.2344} {arXiv:0908.2344 [nucl-th]} \BibitemShut {NoStop}%
\bibitem [{\citenamefont {Alford}\ \emph {et~al.}(2022)\citenamefont {Alford}, \citenamefont {Brodie}, \citenamefont {Haber},\ and\ \citenamefont {Tews}}]{Alford:2022bpp}%
  \BibitemOpen
  \bibfield  {author} {\bibinfo {author} {\bibfnamefont {M.~G.}\ \bibnamefont {Alford}}, \bibinfo {author} {\bibfnamefont {L.}~\bibnamefont {Brodie}}, \bibinfo {author} {\bibfnamefont {A.}~\bibnamefont {Haber}}, \ and\ \bibinfo {author} {\bibfnamefont {I.}~\bibnamefont {Tews}},\ }\href {\doibase 10.1103/PhysRevC.106.055804} {\bibfield  {journal} {\bibinfo  {journal} {Phys. Rev. C}\ }\textbf {\bibinfo {volume} {106}},\ \bibinfo {pages} {055804} (\bibinfo {year} {2022})},\ \Eprint {http://arxiv.org/abs/2205.10283} {arXiv:2205.10283 [nucl-th]} \BibitemShut {NoStop}%
\bibitem [{\citenamefont {Alford}\ \emph {et~al.}(2023)\citenamefont {Alford}, \citenamefont {Brodie}, \citenamefont {Haber},\ and\ \citenamefont {Tews}}]{Alford:2023rgp}%
  \BibitemOpen
  \bibfield  {author} {\bibinfo {author} {\bibfnamefont {M.~G.}\ \bibnamefont {Alford}}, \bibinfo {author} {\bibfnamefont {L.}~\bibnamefont {Brodie}}, \bibinfo {author} {\bibfnamefont {A.}~\bibnamefont {Haber}}, \ and\ \bibinfo {author} {\bibfnamefont {I.}~\bibnamefont {Tews}},\ }\href {\doibase 10.1088/1402-4896/ad03c8} {\bibfield  {journal} {\bibinfo  {journal} {Phys. Scripta}\ }\textbf {\bibinfo {volume} {98}},\ \bibinfo {pages} {125302} (\bibinfo {year} {2023})},\ \Eprint {http://arxiv.org/abs/2304.07836} {arXiv:2304.07836 [nucl-th]} \BibitemShut {NoStop}%
\bibitem [{\citenamefont {Hanauske}\ \emph {et~al.}(2001)\citenamefont {Hanauske}, \citenamefont {Satarov}, \citenamefont {Mishustin}, \citenamefont {Stoecker},\ and\ \citenamefont {Greiner}}]{Hanauske:2001nc}%
  \BibitemOpen
  \bibfield  {author} {\bibinfo {author} {\bibfnamefont {M.}~\bibnamefont {Hanauske}}, \bibinfo {author} {\bibfnamefont {L.~M.}\ \bibnamefont {Satarov}}, \bibinfo {author} {\bibfnamefont {I.~N.}\ \bibnamefont {Mishustin}}, \bibinfo {author} {\bibfnamefont {H.}~\bibnamefont {Stoecker}}, \ and\ \bibinfo {author} {\bibfnamefont {W.}~\bibnamefont {Greiner}},\ }\href {\doibase 10.1103/PhysRevD.64.043005} {\bibfield  {journal} {\bibinfo  {journal} {Phys. Rev. D}\ }\textbf {\bibinfo {volume} {64}},\ \bibinfo {pages} {043005} (\bibinfo {year} {2001})},\ \Eprint {http://arxiv.org/abs/astro-ph/0101267} {arXiv:astro-ph/0101267} \BibitemShut {NoStop}%
\end{thebibliography}%
\bibliographystyle{apsrev4-1}

\end{document}